\documentclass[preprint,aps,superscriptaddress,nofootinbib,showkeys,11pt]{revtex4}
\pdfoutput=1
\usepackage{graphicx}
\usepackage{amsmath}
\usepackage{amsfonts}
\usepackage{amssymb}
\usepackage{epstopdf}
\usepackage{float}
\usepackage{subfigure}
\usepackage{capt-of}
\usepackage{hyperref}
\usepackage[dvipsnames]{xcolor}
\hypersetup{
     colorlinks   = true,
     citecolor    = red,
     linkcolor    = blue,
     urlcolor     = blue,
}
\definecolor{deeppink}{rgb}{0.9, 0.17, 0.31}
\def\beq{\begin{equation}}
\def\eeq{\end{equation}}
\def\bea{\begin{eqnarray}}
\def\eea{\end{eqnarray}}
\def\be{\begin{equation}}
\def\ee{\end{equation}}
\def\pr{\partial}
\def\nno{\nonumber}
\def\bse{\begin{subequations}}
\def\ese{\end{subequations}}

\graphicspath{{./figs/}}

\begin{document}

\title{Two phase reheating: CMB constraints on inflaton and dark matter phenomenology}
\author{Md Riajul Haque}%
\email{riaju176121018@iitg.ac.in}
\author{Debaprasad Maity}
\email{debu@iitg.ac.in}
\affiliation{%
	Department of Physics, Indian Institute of Technology Guwahati,\\
	Guwahati, Assam, India, 781039 
}%
\author{Pankaj Saha}%
\email{pankaj@physics.iitm.ac.in}
\affiliation{%
	Department of Physics, Indian Institute of Technology Madras,\\
	Chennai, Tamilnadu, India, 600036
}%

\date{\today}

\begin{abstract}
We propose a two-phase reheating scenario where the initial preheating dynamics is described by an effective dynamics followed by the standard perturbative reheating. Some of the important universal results of lattice simulation during preheating have been considered as crucial inputs in our two-phase dynamics. In this framework, detailed phenomenological constraints have been obtained on the inflaton couplings with reheating fields, and dark matter parameters in terms of CMB constrained inflationary scalar spectral index. It is observed that the conventional reheating scenario generically predicts the maximum reheating temperature $T_{re}^{max} \simeq 10^{15}$ GeV, corresponding to an almost instantaneous transition from the end of inflation to radiation domination. This fact will naturally lead to the problem of non-perturbative inflaton decay, which is in direct conflict with the perturbative reheating itself. Taking into account this by incorporating effective non-perturbative dynamics as the initial phase, our model of two-phase reheating scenarios also predicts model-independent maximum reheating temperature, which does not correspond to the instantaneous process. Further, $T_{re}^{max}$ is predicted to lie within $(10^{13}, 10^{10})$ GeV if CMB constraints on inflaton couplings with different reheating field are taken into account. We have further studied in detail the dark matter phenomenology in a model-independent manner and show how dark matter parameter space can be constrained through CMB parameters via the inflaton spectral index. Considering dark matter production during reheating via the Freeze-in mechanism, its parameter space has been observed to be highly constrained by our two-phase reheating than the constraints predicted by the conventional reheating scenarios, which are believed to theoretically incomplete. 
 
\end{abstract}
\maketitle
\tableofcontents
\section{Introduction}
The inflationary universe \cite{inflation_glass} is currently the leading paradigm to explain the inhomogeneities in the Cosmic Microwave Background (CMB)\cite{Akrami:2018odb}, which plays the crucial role of seed perturbations for the large scale structure of the universe \cite{Liddle:2000cg}. Within the present setup, the inflationary phase must be followed by a phase known as reheating\footnote{Particle production during inflation is also considered in the so-called warm inflationary scenario\cite{Berera:1995ie, Berera:1996fm}} when the energy stored in the inflaton field is released to defrost the universe. Unlike inflation, the reheating phase is not constrained by direct observables; however, the modified expansion history of the universe due to the presence of the reheating phase prior to the hot big bang evolution influences the relation between physical scales of the CMB mode today and that at the time of their Hubble exit during inflation (See, Fig.2 in this context). This was the basic idea of reheating constraints to inflationary models from CMB \cite{martin}. Despite, the thermalization process erasing many micro-physical details of this phase, a better understanding of this phase is necessary and can shed light on how inflationary mechanism is connected to the rest of the universe dynamics~\cite{Drewes:2015coa,Dalianis:2018afb,Drewes:2019rxn}, the production mechanism of baryonic asymmetry in the early universe called baryogenesis \cite{Kolb:1996jt,Dolgov:1996qq,Riotto:1999yt,GarciaBellido:1999sv,Allahverdi:2000zd,Davidson:2000dw,Megevand:2000da}, the origin of dark matter~\cite{Gorbunov:2020wfj}, and the generation of primordial gravitational waves~\cite{Turner:1990rc,Kosowsky:1992rz,Finelli:1998bu,Easther:2006vd,Easther:2006gt,GarciaBellido:2007dg,GarciaBellido:2007af,Dufaux:2007pt,Bernal:2019lpc,Lozanov:2019ylm} and constraining the etc. There exist two approaches that can constrain the reheating phase through the inflationary models. We can either model the expansion during the reheating phase using an effective equation of state parameter \cite{martin,Saha:2020bis}, or we can solve the Boltzmann equation system supplemented with the background expansion~\cite{Maity:2018dgy}. Both the description have their limitations and are not theoretically complete. However, the latter approach's advantage is that one can further generalize it by including matter components in addition to radiation, which could be physically motivated. For example, in the original work, the production of dark matter has been studied. The study revealed a very interesting link between the dark matter with CMB through inflation  \cite{Maity:2018dgy}. Later this formalism has been extended in considering various models of inflaton with general power-law types potential \cite{Maity:2018exj} and non-perturbative effect from numerical lattice simulation has also been considered~\cite{Maity:2018qhi}. In this paper, we take up this issue of the non-perturbative phase known as preheating and formulate an effective approach that will be shown to lead qualitatively different results than that of the usual reheating constraint analysis. Preheating is the phase when the occupation number of field quanta for both the inflaton and daughter field(s) grows exponentially due to parametric resonance \cite{reheating}. The equation of state during steady state of the preheating phase is crucial for model independent reheating constraint analysis. Further, some scenarios can lead to non-standard case such as sudden blocking reheating process due to Higgs field \cite{Freese:2017ace} or  breakdown of coherent oscillation without thermalization \cite{Easther:2010mr}. Considering those into our present scenario would interesting to consider. Although, the preheating phenomenon depends on the inflation model and its interaction with the daughter fields, certain {\it universal behaviors} have been observed to emerge irrespective of inflation models\cite{Figueroa:2016wxr,Maity:2018qhi,Antusch:2020iyq}. Namely, i) The preheating phase is episodic with at-least three distinct phases.
 	(iii) The equation of state (EoS) of the system does no reach to that of the radiation  for quadratic potentials  $V(\phi)\propto\phi^2$. However, for other form $V(\phi)\propto\phi^n$ with $n\geq4$, EoS reach $w\to1/3$ at the end of preheating\cite{Lozanov:2016hid,Lozanov:2017hjm,Maity:2018qhi,Antusch:2020iyq}.
 	\\
 	These results indicate that while non-perturbative processes dominate the initial stage, the inflaton decay in later stages should be described by perturbative channels. A systematic study of reheating constraints incorporating the non-perturbative phase for various models and interactions is missing in the literature.
 	\\
 	In this work, we will extend the formalism developed in \cite{Maity:2018dgy, Maity:2018qhi} to include the non-perturbative effects in reheating constrains analysis. As we just mentioned, reheating happens in multiple stages, with the initial non-perturbative stage followed by the perturbative one. Let us now briefly describe the main idea of the present work: endowed with the above few universal features, we will consider reheating as a two-stage process. We will model the initial non-perturbative stage (henceforth, phase-I) governed by an effective fluid with an effective equation of state ($w_{eff}$). One of the boundary conditions of the effective non-perturbative dynamics will be set by the inflation model potential expressed in terms of the scalar spectral index (in this regard, the reader may find this in parallel with the conventional works in \cite{martin}). This phase is assumed to be continued until the inflaton energy decaying into $50\%$ of its initial energy and the subsequent perturbative stage (henceforth, phase-II) follows. While evolving through phase-I and connecting to phase-II, we allow the system to satisfy an important {\it consistency relation} associated with the total energy conservation comprising inflaton and various daughter fields such as radiation, dark matter (which we considered separately). We will see that this consistency relation will restrict the possible values of $w_{eff}$ during phase-I as opposed to the conventional analysis \cite{martin}. Furthermore, the perturbative decay in phase-II will be restricted by the CMB constrains\cite{Maity:2018dgy}.\\
 	We have structured our paper as follows: In section, \ref{pert}, we discuss the general analysis of single-stage perturbative reheating, and in section \ref{valid}, we will try to specify the possible limits on perturbative reheating considering some specified form of the interaction between inflation and radiation field. Finally, In section \ref{section4}, we briefly describe our proposed two-phase reheating analysis and, in section \ref{sec5}, illustrate the strategy of our numerical study. After that, in section \ref{analytic}, we will try to find out an analytical estimation of the maximum radiation temperature and reheating temperature. Next, in section \ref{models}, we consider different inflationary models of and analyze in the context of the two-phase scenario and compare it with conventional reheating dynamics. In section \ref{bound}, we analyze the possible constraints on the coupling parameter correspond to different inflaton-radiation field interactions. Furthermore, in section \ref{dark}, we include additional dark matter components and discuss the viable restrictions on the dark matter parameter space.


\section{Reheating constraint  analysis for perturbatively decaying inflaton}\label{pert}
Before we directly jump into constructing  the two phase reheating model, let us first elaborate on widely studied single phase perturbatie reheating with decaying inflaton following \cite{Maity:2018dgy}. This not only explains the methodology of our analysis, but also helps us to identify the regime of its validity which will further motivate the reader, the need for considering physically more acceptable two phase reheating process mentioned in the introduction. While discussing this we will see one of the important results that is the existence of maximum reheating temperature. Subsequently the generalization to two phase reheating will show how the aforesaid maximum reheating temperature reduces depending upon the initial condition. Let us start with the following Einstein's equation for the cosmological scale factor and conservation of energy, 
\bea \label{con}
&&\ddot{n}_{re} = - 2 \dot{n}_{re}^2  + \frac{1-3 w}{6 M_p^2} \rho_{\phi} \\ \nno
&&\dot{\rho}_{\phi} + 3 \dot{n}_{re}(\rho_{\phi}+p_{\phi})
+ \dot{\rho}_{rad} + 4 \dot{n}_{re} \rho_{rad} = 0 ,
\eea
with the following Freedman-Roberson-Walker (FRW) form of the metric
\bea
ds^2 = -dt^2 + a^2 (dx^2 + dy^2 + dz^2)
\eea 
Where, "$\rho$"s are the energy densities of two different components. 
 At any instant of time during reheating, we parametrize the duration of reheating by e-folding number $n_{re}(t) = \ln (a/a_i)$,
 where "$a$" is the cosmological scale factor. The time derivative of $n_{re}$ is the Hubble expansion parameter $\dot{n}_{re}= H$ during reheating. During reheating we assume the effective equation of state of the inflaton $w =\langle p_{\phi}/\rho_{\phi}\rangle$ to be approximately constant.
 The fundamental difference between our present analysis followed from \cite{Maity:2018dgy} and that in \cite{martin} is the consideration of Eq.\ref{con}, where we consider the
 multiple dynamical components. Considering the evolution of $\rho_{\phi} + \rho_{rad} = \rho_{eff}$ together, the  effective equation of state during reheating can be defined as
 \bea
 w_{eff} = \left\langle \frac {3 p_{\phi}+\rho_{rad}}{3(\rho_{\phi} + \rho_{rad})}\right\rangle .
 \eea 
Hence, $w_{eff}$ will essentially interpolates between two values $(w,1/3)$ through non-trivial time dynamics for decaying inflaton $\rho_{\phi}$ and the growing radiation field $\rho_{rad}$. However, in \cite{martin}, the authors have taken it to be constant during their analysis. Therefore, we not only employ realistic decay dynamics into the reheating constraint analysis but also provides a new framework to go beyond which is our main purpose of the present paper. 

 Keeping the above points in mind, let us express the total energy density  as,  
\bea \label{rhoh}
\rho_{rad}+ \rho_{\phi} &=&  e^{-4 n_{re}}\left(\rho^{i}_{\phi} + (1-3 w)  \int_{t_i}^{t}  \rho_{\phi} e^{4 n_{re}}  dn_{re}\right).
\eea 
which is followed from the conservation eq.\ref{con}. The index "i" stands for the initial stage of reheating, which also marks the end of inflation.
At the beginning of reheating we set $\rho_{rad}(t_i) = 0$. For solving the above set of equations, the boundary condition is set by the inflaton energy density as $ \dot{n}_{re}(t_i) = H(t_i) = \sqrt{{\rho^i_{\phi}}/{3 M_p^2}}$.
 The physical quantity of our interest is the ratio of the radiation energy density and the inflaton energy density. From eq.\ref{rhoh}, one gets
\bea
\frac{\rho^f_{rad}}{\rho^i_{\phi}} = e^{-4 N_{re}} -
\frac {\rho^f_{\phi}} {\rho^i_{\phi}}
+ (1-3 w)e^{-4 N_{re}} \int_i^{f} \frac{\rho_{\phi}}{\rho^i_{\phi}} e^{4 n_r}  dn_{re} ~.
\eea
Where, "f" corresponds to the final value of radiation density. We define total e-folding number during reheating as
$N_{re} = n_{re}(t_f)$.

The main goal of this whole program of reheating constraint analysis is to understand the relation among early universe inflaton dynamics, the intermediate reheating dynamics and late time dynamics. 
A particular cosmological scale $k$ going out of the horizon during inflation will re-enter the horizon during
late time cosmological evolution. This fact will provide an important relation among different phases just mentioned as follows 
\bea
\ln{\left(\frac {a_k H_k}{a_0 H_0}\right)} =-N_k -N_{re} -
\ln{\left(\frac {a_{re} H_k}{a_0 H_0}\right)},
\label{scalek}
\eea  
where, a particular scale $k$ satisfies the relation $k = a_0 H_0 = a_k H_k$. $(a_{re}, a_0)$, are the cosmological scale factors at the end of reheating phase and at the present time respectively.
$(N_k,H_k)$ are the efolding number and the Hubble parameter respectively during inflation. $H_0$ is the present value of the Hubble constant.

The usual approach is to define the effective equation of state of the total energy density during reheating and study its evolution. However, we consider only the radiation part during reheating and try to understand the evolution of its temperature $T_{rad}$ as a function of scalar spectral index, and finally connect the temperature with CMB one on large scale \cite{Maity:2018dgy}.  
The reheating temperature $T_{re}$ is identified with radiation temperature $T_{rad}$
at thermal equilibrium between the decaying inflaton and the radiation. From the entropy conservation of thermal radiation, the relation among $T_{rad}=T_{re}$ at equilibrium, and $(T_0, T_{\nu 0} = (4/11)^{1/3} T_0)$, temperature of the CMB photon and neutrino background at the present day respectively, can be written as 
\bea \label{entropy}
g_{re} T_{re}^3 = \left(\frac {a_0}{a_{re}}\right)^3\left( 2 T_0^3 + 6 \frac 7 8 T_{\nu 0}^3\right).
\eea
Using eq.(\ref{scalek},\ref{entropy}), one arrives at the following well known relation
\bea \label{eqtre}
T_{re}= \left(\frac{43}{11 g_{re}}\right)^{\frac 1 3}\left(\frac{a_0T_0}{k}\right) H_k e^{-N_k} e^{-N_{re}} ={\cal G}_k e^{-N_{re}} .
\eea
Where, $g_{re} \sim 100$ is the effective number of relativistic degrees of freedom during radiation phase. In our subsequent study we identify the cosmological scale $k$ as the pivot scale set by PLANCK, $k/a_0 = 0.05 Mpc^{-1}$ and compare our result with the corresponding estimated scalar spectral index $n_s = 0.9682 \pm 0.0062$ 
\cite{PLANCK}.

\subsection{Example I: Exactly solvable case}
As has already been discussed in 
\cite{Maity:2018dgy}, one of the important outcome of our formalism is the existence of maximum possible reheating temperature. In this and next section, we will elaborate on this considering simple ansatz of decaying inflaton. We first consider an analytically solvable case where the inflaton is decaying as
\bea
&&{\dot \rho}_{\phi} + 3 H (1+w)\rho_{\phi} = -\bar{\varGamma_\phi} H \rho_{\phi} \nno\\  
\implies && \rho_{\phi}(t) =  \rho^i_{\phi}  e^{-3(1+w)n_{re}} e^{-\bar{\varGamma_\phi} n_{re}} .
\eea 
$\bar{\varGamma_\phi}$ is a dimensionless constant,
\begin{figure}[t!]
	\begin{center}
		\includegraphics[width=0011.0cm,height=05.00cm]{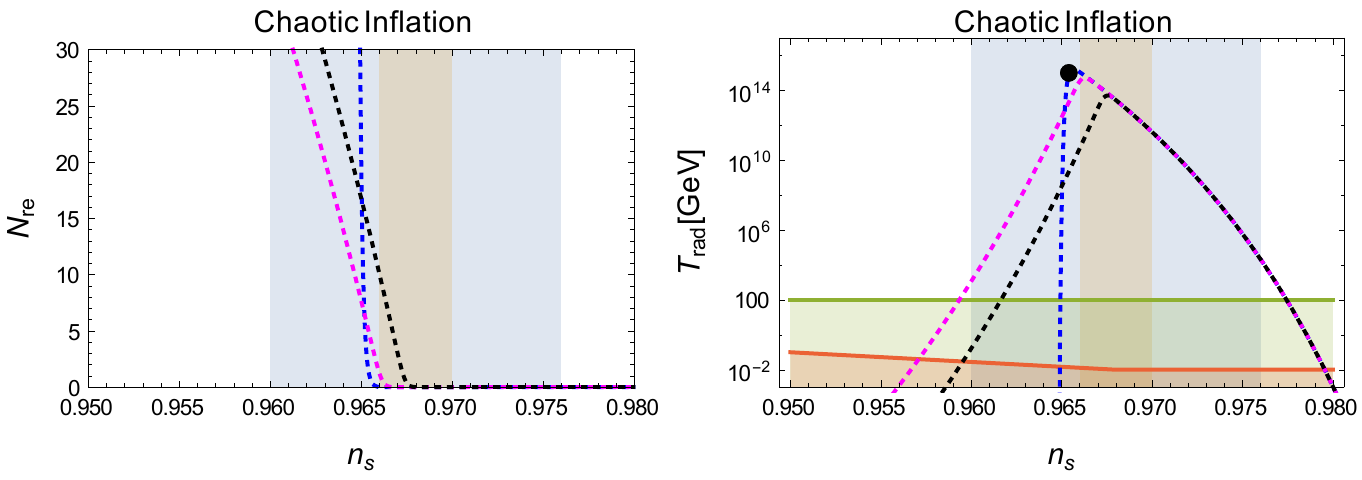}
		\caption{\scriptsize Variation of $(N_{re},T_{rad})$ as a function of $n_s$ have been plotted for $\bar{\varGamma_\phi} = (1.0,0.01, 0.000001)$ corresponding to \textcolor{blue}{\bf{blue}}, \textcolor{Rhodamine}{\bf{pink}}, \textcolor{black}{\bf{black}} curves respectively. The \textcolor{CadetBlue}{\bf{Light blue}} shaded region 
			corresponds to the $1 \sigma$ bounds on $n_s$ from Planck. The \textcolor{brown}{\bf{brown}} shaded region corresponds to the $1 \sigma$
			bounds of a further CMB experiment with sensitivity $\pm 10^{-3}$ \cite{limit1,limit2}, using the same central $n_s$ value as
			Planck. Temperatures below the horizontal \textcolor{red}{\bf{red line}} is ruled out by BBN. The \textcolor{Green}{\bf{deep green}} shaded region is below the 
			electroweak scale, assumed 100 GeV for reference.} 
		\label{plotexact}
	\end{center}
\end{figure} 
which parametrizes the decay of inflaton. This form of decay essentially modifies the Hubble friction term for the dynamics of inflaton during reheating. With aforementioned ansatz for the decaying inflaton, radiation density is analytically solved as 
\bea \label{eqnreexact}
\frac{\rho^f_{rad}}{\rho^i_{\phi}} &=& \frac{\bar{\Gamma_\phi} }{\bar{\Gamma_\phi}  + 3 w -1} \left(e^{-4 N_{re}} -
e^{-3(1+w)N_{re} -\bar{\Gamma_\phi} N_{re}} \right). 
\eea
The second term in the parenthesis is quantifying the fractional amount of inflaton energy
left after the reheating process is over. Expressing  $\rho_f$ in term of radiation temperature as $\rho_{rad}^f=\pi^2 (g_{re}/30) T_{rad}^4$, the Eq.\ref{eqnreexact} leads to the following maximum radiation temperature \cite{tmax} for a given $\varGamma_{\phi}$ as   
\bea \label{tmaxexact}
T_{rad}^{max} = \left(\frac{30 \rho^i_{\phi} P}{\pi^2 g_{re}} \right)^{\frac 1 4} \left[x^{\frac {4}{\bar{\varGamma_\phi} +3w-1}}-x^{\frac{3+3w+\bar{\varGamma_\phi}}{\bar{\varGamma_\phi} +3w-1}}
	\right]^{\frac 1 4} ,
\eea
where, $(x = 4/(3+3w+\bar{\varGamma_\phi}), P ={{\bar{\varGamma_\phi}} }/{(\bar{\varGamma_\phi}  + 3 w -1})$. This also can be clearly seen from the Fig.\ref{plotexact} for each value of $\bar{\varGamma_\phi}$. From the perturbative point of view, the value of $\varGamma_{\phi}$ should be $\leq 1$. However, if we naively extrapolate the above result for large $\bar{\varGamma_\phi}$ most important result turned out to be the existence of a maximum possible temperature, 
\bea
\lim_{\bar{\varGamma_\phi}\gg 1} T_{rad}^{max}  =    \left(\frac{30 \rho^i_{\phi}}{\pi^2 g_{re}}\right)^{\frac 1 4} \simeq 2.9 \times 10^{15}~\mbox{GeV}.
\eea
However, the numerical value of this maximum temperature turns out to be of the order of same as the limiting perturbative value for $\bar{\varGamma_\phi}=1$ as shown in the figure Fig.\ref{plotexact}.
Therefore, above temperature can be naturally identified as maximum possible reheating temperature. This also corresponds to the maximum possible value of scalar spectra index $n_s^{max}$. Identifying associated temperature of the produced radiation in eq.\ref{eqnreexact} with eq.\ref{eqtre}, we arrive at the following exact expression for $(N_{re},T_{re})$,
\begin{eqnarray} \label{nretre}
&& T_{re} = {\cal G}_k\left(1-\frac{1}{P}\frac{\pi^2g_{re} {\cal G}_k^4}{3^2.5V_{end}}\right)^{\frac {1}{4-3(1+w)-\bar{\varGamma_\phi}}} . \\
&& N_{re} = \frac {1}{4-3(1+w)-\bar{\varGamma_\phi}}\ln \left[1-\frac {1}{P}\frac{\pi^2g_{re} {\cal G}_k^4}{3^2.5V_{end} }\right]
\end{eqnarray}
In the fig.\ref{plotexact}, we have considered three possible values of $\bar{\varGamma_\phi}$ for quadratic inflaton potential. The special value is $\bar{\varGamma_\phi}=1$, for which the equilibrium condition between the inflaton and the radiation can be achieved at the maximum temperature shown as a {\bf black dot}.
The maximum value of scalar spectral index turned out to be $n_s^{max}\simeq 0.9654$. 
This analysis motivates us to subsequently analyze more general case, and we will show that this conclusion still holds.  

\subsection{Example II: Standard pertrubative case} 
In this section we will consider the standard perturbatively decaying inflaton parameterizing by decay constant $\varGamma_{\phi}$ as follows,
\bea
&&{\dot \rho}_{\phi} + 3 H (1+w)\rho_{\phi} = -\varGamma_{\phi} \rho_{\phi}(1+w), \nno\\ 
\implies && \rho_{\phi}(t) =  \rho^i_{\phi} e^{-3(1+w)n_{re}} e^{- \varGamma_{\phi} (t-t_i)(1+w)} ,
\eea 
Where $\varGamma_{\phi}$ is effective time independent inflaton decay constant. It is the phenomenological term which acts as a damping force during the oscillating inflaton. This term can
be related to the total decay rate of inflaton to radiation.  However, we believe our conclusion will remain same for time dependent $\varGamma_{\phi}$, which we will study later. 
Before doing any numerical analysis, let us examine the approximate solution
which has already been discussed in the literature \cite{tmax}.
During the early stage of evolution, approximating
$\rho_{\phi}^i e^{-\varGamma_{\phi} t}\simeq \rho_{\phi}^i$, the radiation density can be calculated as
\bea
 \frac{\rho^f_{rad}}{\rho^i_{\phi}} \simeq
 \frac {2\varGamma_{\phi} e^{-4 N_{re}}}{(5-3w)\dot{n}_{re}(t_i)}\left(e^{\frac{5-3w}{2}N_{re}} - 1 \right) .
 \eea
Similar to the exactly solvable case in eq.\ref{tmaxexact}, the above equation also leads to a maximum radiation temperature \cite{tmax} for a given $\varGamma_{\phi}$,
\bea \label{tmax}
T_{rad}^{max} \simeq \left(\frac {39^2 M_p^2  (\dot{n}^i_{re})^2}{ \pi^2 g_{re}} \right)^{\frac 1 8} \sqrt{T_{re}} .
\eea
Where, the relation $T_{re}= 0.45 \left({200}/{g_{re}}\right)^{1/4}\sqrt{\varGamma_{\phi} M_p}$ has been used. In the same way as our earlier exactly solvable case, maximum possible reheating temperature could be obtained, if one identifies a special point where two temperature meets, $T_{rad}^{max} = T_{re}$. 
Our numerical analysis also shows the maximum reheating temperature at the aforementioned special point, 
\bea \label{tmaxre}
T_{re}^{max} \simeq \left(\frac {39^2  \rho^i_{\phi}}{3 \pi^2 g_{re}} \right)^{\frac 1 4}.
\eea
Interestingly, the maximum reheating temperature $T_{re}^{max}$ can also be computed for another exactly solvable case with $w=1/3$. Corresponding result is as follows,   
\bea \label{tmaxw13}
T_{rad}^{max}|_{w=\frac 1 3} &\simeq&  \left(\frac {30 \rho_{\phi}^i  }{ \pi^2 g_{re}} \frac {\varGamma_{\phi}}{4 \dot{n}_i + \varGamma_{\phi}} \right)^{\frac 1 4} \\\nno
T_{re}^{max}|_{w=\frac 1 3} & =& \lim_{\varGamma_{\phi}\gg 4 \dot{n}_i}T_{rad}^{max}|_{w=\frac 1 3} = \left(\frac {30 \rho_{\phi}^i  }{ \pi^2 g_{re}} \right)^{\frac 1 4} .
\eea
This expression is exactly the same as previously discussed. For this special value of $w=1/3$, we also have exact expression for all the reheating parameters $(T_{re}, N_{re})$ as follows,
\begin{eqnarray} \label{nretrerad}
&& T_{re} = {\cal G}_k\left(1-\sqrt{\frac{4\rho^i_{\phi}}{3 M_p^2 \varGamma_{\phi}^2}} \ln\left[1 - \frac{\pi^2g_{re} {\cal G}_k^4}{3^2.5V_{end} }\right]
 \right)^{-\frac 1 2 }. \\
&& N_{re} = \frac {1}{2}\ln \left[1-\sqrt{\frac{4\rho^i_{\phi}}{3 M_p^2 \varGamma_{\phi}^2}} \ln\left[1 - \frac{\pi^2g_{re} {\cal G}_k^4}{3^2.5V_{end} }\right]\right]
\end{eqnarray}

At this point let us again emphasize the fact that as long as we are in the perturbative regime, the relation among the scalar spectral index $n_s$ and the reheating temperature $T_{re}$ can be understood from our detail analysis above. However, existence of maximum reheating temperature will come if we extrapolate all our formulas for large $\varGamma_{\phi} > \sqrt{2{\rho}_i/(3 M_p^2)}$. For low scale inflation, 
$\varGamma_{\phi}$ could always be in the perturbative regime.
For large scale inflation, this could lead 
to non-perturbative regime, which will be discussed in our subsequent section. We will discuss about possible limits on the value of $\varGamma_{\phi}$ below which our analysis will be valid. To this end, it is important to point out that in the effective reheating equation of state description \cite{martin}, the maximum temperature can be explained in the limit of zero reheating e-folding number $N_{re}$. Therefore, large $\varGamma_{\phi}$ limit in our analysis can be thought of as equivalent to the zero $N_{re}$ limit of the previously studied reheating constraint analysis. However, it is important to remember that those two facts are certainly not identical. Corresponding to our maximum temperature, we have a minimum reheating efolding number. Our prediction of maximum reheating temperature $\sim 10^{15}$ GeV, and its model independence could be robust and they are intimately connected with the observed CMB scale. 

Never the less the main point of our study is to understand the effect of decaying inflaton into the reheating constraint analysis. We think this is the appropriate procedure to understand the relation among $(T_{re}, n_s)$. Another advantage of our procedure is that we can easily generalize our analysis to include any other decay products during reheating such as dark matter which is observed to be dominant matter component of our universe\cite{rubin}-\cite{Bertone:2004pz}, and that can shape the observed pattern in the CMB. Before, this, our main motivation would be to incorporate the non-perturbative aspects of reheating into our formalism.  
\section{Regime of validity of perturbative reheating}\label{valid}
In this section we will try to mention the possible limits on the inflaton decay constant assuming some specific form of the interactions among the inflaton and the reheating field.  
As emphasized throughout the present work, we have assumed that the inflation decay to other components( for the present work the radiation component) is effectively described by a phenomenological decay term $\varGamma_{\phi}$. In fact, this was the first attempt to reheat the universe\cite{purturbative}. However, it was soon realized that once the particle production initiates, the inflaton decay is subject to various non-perturbative resonance production and feedback mechanisms. Those processes can change the reheating scenario dramatically. Though it has been argued in \cite{Drewes:2015coa} that all such feedback mechanisms will have no effect on the CMB. Depending upon the coupling the parametric resonance can be very efficient which may complete the reheating era within a few efolding and in such cases the CMB will have a very little to tell about the reheating phase. Despite that the situation may not be such helpless as noted in \cite{GarciaBellido:2008ab} that the interactions amongst the produced particles can delay the parametric resonance extending the efolding number of reheating. This will eventually improve the situation of CMB constrain on reheating phase. It must also be noted that we can always choose the coupling constant small enough to evade the parametric resonance. Below, we will briefly mention the  space of parameter region in which the perturbative treatment of reheating will be valid over the parametric resonance.
\subsection{Inflaton decaying into scalar particle}
\subsubsection{Scalar $\phi \chi^2$ interaction}
 
 First let us consider the case when inflaton decays into another scalar particle $\phi \to \chi\chi$ with the following interaction term
 $ 	\mathcal{L} = -g\phi \chi^2
 $.
 In this case the vacuum decay width for the decay process $\phi \to \chi\chi$ is given by\cite{Peskin:1995ev} 
 \begin{eqnarray}\label{decay2}
 \varGamma_{\phi \to \chi\chi} = \frac{g^2}{8\pi m_{\phi}}\sqrt{1 - \left(\frac{2m_{\chi}}{m_{\phi}}\right)^2} \simeq \frac{g^2}{8\pi m_{\phi}},
 \end{eqnarray}
 where, $m_{\phi}, m_{\chi}$ are the mass of the inflaton and produced particle respectively, and $g$ is the coupling constant. The mode function $\chi_k$ of the decay product can be  cast into the following Mathieu equation,
\begin{eqnarray}
\ddot{\chi_k} + (A_k - 2q \cos(2z))\chi_k = 0.
\end{eqnarray}
Where, $z=(m_{\phi}t - 2z-\pi/2$),  $A_k=4k^2/m_{\phi}^2$, $q=4g\Phi/m_{\phi}^2$. $\Phi$ is the initial amplitude of the inflaton during oscillations. The Mathieu equation is known to show resonance solutions of the form $\chi_k \propto exp(\mu_k z)$. The condition for the resonance to be efficient is formulated as
\begin{equation}
	q^2 m \gtrsim H
\end{equation} 
This can be transformed into the following condition\footnote{In deriving this condition the initial amplitude $\Phi$ has been replaced by $\phi_{end}$, which implies that this is essentially a lower bound on the decay width as in the case of preheating $\Phi < \phi_{end}$} on the dimensionless coupling constant $\tilde{g} = g/m_{\phi}$ indicating the regime of perturbative validity \cite{Drewes:2017fmn}
\begin{equation}\label{scalar2}
\tilde{g} \leq \frac{V_{end}^{\frac{1}{4}}}{\phi_{end}}\left(\frac{m_{\phi}}{24M_p}\right)^{\frac{1}{2}} ,
\end{equation}
which can further expressed in terms of  decay constant as 
\bea \label{gamma2}
	\Gamma_{\phi} \leq \frac{V_{end}^{\frac{1}{2}}}{\phi_{end}^2}\left(\frac{m_{\phi}^2}{192\pi M_p}\right)  \implies \Gamma_{\phi}^{cri}(\mbox{model}) = \frac{V_{end}^{\frac{1}{2}}}{\phi_{end}^2}\left(\frac{m_{\phi}^2}{192\pi M_p}\right)
\eea
Therefore, we see that if the decay width satisfies aforementioned condition, the perturbative reheating will be the only mechanism and our perturbative analysis will be at work. Given a model $\Gamma_{\phi}^{cri}(\mbox{model})$ is the point which qualitatively separates the perturbative and non-perturbative effect of inflaton decay. 
\subsubsection{Scalar $\phi \chi^3$ interaction} 
In this case, inflaton couples to another light scalar via interaction
\bea
\mathcal{L} = -y\phi \chi^3~~,
\eea
where y is the coupling constant. The vacuum decay rate of the inflaton field into three bodies $\phi \to \chi\chi\chi$ can be determined by Dalitz plot \cite{Patrignani:2016xqp} as,
\bea\label{decay3}
\varGamma_{\phi\to\chi\chi\chi}=\frac{y^2 m_\phi}{3 ! 64\left(2\pi\right)^3}~~.
\eea
At the tree level, the mode function $\chi_k$ following the same Mathieu equation and the consideration from the previous scalar $\phi\chi^2$ interaction can be correlated if one replaces $\tilde{g}m_\phi\Phi\rightarrow h^2 \Phi^2$. Therefore, the condition to treat the dynamics of reheating perturbatively is roughly 
\bea\label{scalar3}
q\backsim \frac{y^2\Phi^2}{m_\phi^2}\leq 1~~.
\eea
To estimate the lower bound on the coupling for the resonance, we make a substitution $\Phi\rightarrow \phi_{end}$. The above condition for the effectiveness of perturbative reheating can be written in terms of decay rate as,
\bea\label{gamma}
\varGamma_{\phi}\leq\frac{m_\phi^3}{3!64\left(2\pi\right)^3 \phi_{end}^2}~~\implies \Gamma_{\phi}^{cri}(\mbox{model}) = \frac{m_\phi^3}{3!64\left(2\pi\right)^3 \phi_{end}^2}
\eea
Thereafter, in our analysis, we want to examine whether this above condition consistent with our analysis or not for the different inflationary models. 
\subsection{Inflaton decaying into a pair of fermions}
Let us now consider the case when the inflaton decays into a pair of massless fermions with the following Yukawa interaction
\bea
\mathcal{L}_{int} = -h\phi\bar{\psi}\psi~~,
 \eea
where $h$ is the dimensionless coupling constant. Now the vacuum decay rate is given by
\begin{equation}\label{decay4}
\varGamma_{\phi \to \bar{\psi}\psi} = \frac{h^2 m_{\phi}}{8\pi}~~.
\end{equation}
The condition for the validity of perturbative reheating in this case, as shown in\cite{Greene:1998nh}, can be written as,
\begin{equation}\label{fermion}
	q = \frac{h^2 \Phi^2}{m_{\phi}^2} \leq 1~~.
\end{equation}
Hence, in connection with decay rate, the equation (\ref{fermion}) is rewritten as,
\bea\label{gamma1}
\varGamma_{\phi}\leq\frac{m_\phi^3}{\phi_{end}^2\left(8\pi\right)}~~\implies \Gamma_{\phi}^{cri}(\mbox{model}) = \frac{m_\phi^3}{\phi_{end}^2\left(8\pi\right)}
\eea
In our proposed effective two phase dynamical scenario we will observe the existence of similar critical inflaton decay constant associated with the reheating e-folding number. We will see how aforementioned three different interacting model dependent critical decay constants restricts in initial parameter space of the reheating dynamics.  
In the following sections our attempt will be to build up a formalism which can effectively incorporate the non-perturbative dynamics at the initial stage of the reheating. 
 \section{regime of effective non perturbative and perturbative reheating }\label{section4}
 \begin{figure}
 	\begin{center}
 \includegraphics[width=18.0cm,
 height=09.0cm]{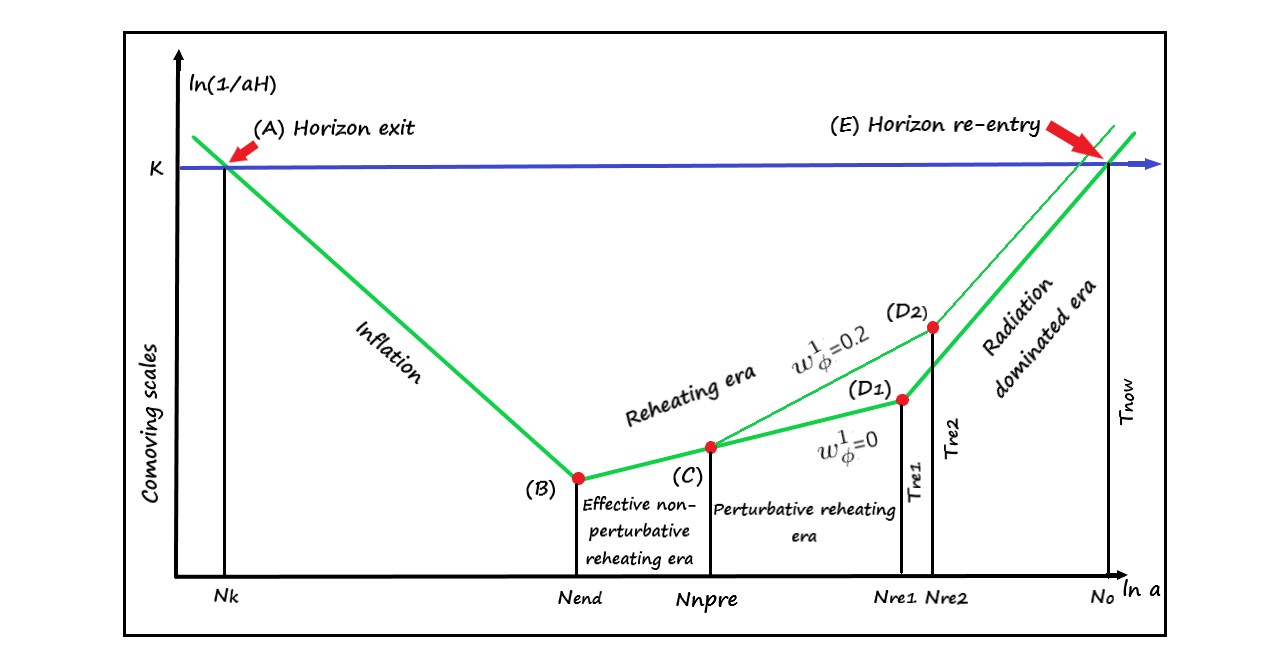}
 		\caption{\scriptsize  	The evolution of the comoving Hubble scale ($\frac{1}{aH}$) connects the inflationary phase with the CMB. The end of the inflation denoted by point B, and the ending of the radiation dominated era denoted by the points $D_1$ and $D_2$. The inflationary phase and radiation dominated era connect through the reheating phase, which contains two different regions, the effective non-perturbative reheating era and the perturbative reheating era. $ C$ denotes the ending point of the non-perturbative reheating era. The points $D_1$ and $D_2$ are the ending point of the perturbative reheating era for two different inflaton equations of state during perturbative reheating $\omega_\phi^1=(0,0.2)$ respectively. For the perturbative reheating era with the inflaton equation of state   $\omega_\phi^1=(0,0.2)$, the e-folding number, basically, the duration of the perturbative process are different. For a particular value of the spectral index (lower values of $n_s$, towards $n_s^{min}$), the decay width calculating by considering $\omega_\phi^1=0.2$ is quite lower in comparison with $\omega_\phi^1=0$. That's why for that particular values of $n_s$, the duration of the perturbative era is quite wider for $\omega_\phi^1=0.2$ in connection with $\omega_\phi^1=0$ ($N_{re2}>N_{re1}$).  }
 		\label{picture}
 	\end{center}
 \end{figure}

The standard and well-studied mechanism to consider the non-perturbative effect during reheating is called preheating. This stage is essentially the combination of a highly non-linear process of parametric resonance and subsequent thermalization. The well-known fact that generically non-perturbative preheating mechanism does not completely decay inflaton into the radiation field. Therefore, subsequent perturbative decay will be necessary to complete the reheating process. To the best of our knowledge, the reference \cite{Maity:2018dgy} has considered this issue for the first time and studied perturbative reheating, followed by the preheating considering a specific model of chaotic type inflation. However, generically the preheating mechanism is model dependent. Hence, combining the end of preheating and subsequent model-independent perturbative reheating is somewhat irreconcilable. Therefore, our objective in the following sections would be to make these two phases reconcilable. 
  
Instead of dwelling into explicit non-perturbative computation during the preheating stage, we will adopt an effective model-independent approach following the reference \cite{martin}. 
The basic idea is to assume the dynamics of preheating to be solely governed by an effective equation of state $\omega_{eff}$ supplemented with the total energy conservation law in terms of its constituents. As already emphasized in the introduction, the information about the actual non-perturbative dynamic will be encoded though considering its universal features into our effective dynamics.    

As has been pointed out already,  during the non-perturbative dynamics, inflaton decay is not complete, and typically it is around $50\%$ of its total comoving energy, which is being transferred into the daughter fields. Furthermore,  inflation models with quadratic potential near the minimum, the non-perturbative reheating does not lead to the equation of state, $\omega=\frac{1}{3}$, which is expected at the end point of reheating \cite{Podolsky:2005bw,Maity:2018qhi}.  Our essential idea would be to correctly utilize those results as the end point conditions of our proposed effective dynamics in place of preheating. After the end of this, the usual Boltzmann perturbative reheating process will follow. The second phase completes the reheating process by leading to the correct state equation with relativistic degrees of freedom as the dominant components collectively called radiation. The Fig.\ref{picture} illustrates our methodology of calculation. Throughout this paper, we call this as {\bf two-phase} reheating process. \\
{\bf phase-I}:({\em Effective non-perturbative phase}) During the early stage of reheating, the phase will be described by total energy density $\rho_T = \rho_R + \rho_{\phi}$ and the constant effective equation state $w_{eff}$. Hence the evolution will be described by,
\bea \label{density}
\rho_{T} =  \rho_{Te} \left(\frac{a_{end}}{a}\right)^{3(1+w_{eff})}~,
\eea
where $\rho_{Te}$ is the total energy density at the end of the inflation. $a_{end}$ is the scale factor at the end of the inflation. In this section we will build up our formalism considering two matter component with $\rho_{\phi}$, $\rho_{R}$ as the inflaton energy density and radiation energy density respectively at any instant of time. In the subsequent section we will add dark matter as a third component as an extension. Nonetheless, from the total energy expression one can write down the following equation follows from Eq.\ref{density},
\bea\label{nonper}
\dot{\rho_\phi} + \dot{\rho_R} + 3 H (1+w_{eff}) (\rho_\phi + \rho_R )= 0~.
\eea
To reduce the number of unknown parameters, to this end we will also utilize total energy conservation relation considering individual equation of state of the inflaton $(\omega_{\phi})$ and the radiation field  $(\omega_R = 1/3)$  described as
\bea \label{nonperturbative}
\dot{\rho_\phi} + 3 H (1+w_{\phi})\rho_\phi + \dot{\rho_R} + 4 H \rho_R = 0~.
\eea
Given the aforementioned constraint relation one obtains the possible restricted value of the effective equation of state, $\omega_{eff}$. To find those restrictions we combine above two eqs.(\ref{nonper}) and (\ref{nonperturbative}), and obtain the following {\it consistency relation},
\bea \label{state}
\frac{\rho_{R}}{\rho_{\phi} + \rho_{R}}= \frac{3 \left(w_\phi-w_{eff} \right)}{3 w_\phi -1}~.
\eea
Right at the end of inflation or the beginning of the preheating phase, the energy density of the radiation part will naturally be closed to zero  $\rho_{R} \simeq 0$. As time evolves, $\rho_{R}$ increases due to decaying inflaton. This initial condition automatically restricts the possible values of $w_{eff}$ to be very closed to that of the inflaton equation of state $w_\phi$. {\em Of course more appropriate approach would be to assume the inflaton equation of state evolving from the value very closed to $\omega_{\phi}$ to the value required in the next phase.} We will comment on this issue at the appropriate place during our discussion. However, from the usual numerical lattice simulation, the preheating phase's significant duration is dominated by the inflaton. Hence, the equation of state will naturally be closed to that of the inflaton field. Thus, our effective dynamics approach towards preheating truly captures all these necessary properties of the non-perturbative dynamics. As $w_{eff}$ turned out to be no longer a free parameter and constrained by the above consistency relation (\ref{state}), our following analysis will be based on this important result. Throughout our study we consider models for which the inflaton equation of state during phase-I $\omega_{\phi} = 0$, and consequently following Eq.\ref{state} we choose two values of effective equation of state $\omega_{eff} =(10^{-6},10^{-3}$). This choice will automatically fixes the initial radiation densities during phase-I as  $\rho_{R}/(\rho_{\phi}+\rho_{R}) = (3 \times 10^{-6}, 3\times 10^{-3})$. We will see the maximum reheating temperature crucially depends upon these initial conditions.

{\bf Phase-II}: ({\em Perturbative phase}) Once the preheating dynamics ends the usual Boltzmann perturbative reheating follows. During this period various components of the total energy density satisfy the following standard Boltzmann equations \cite{Riotto},
\begin{equation}\label{per1}
\dot\rho_{\phi}+ 3 H(1+w_{\phi}^1)\rho_{\phi}+\Gamma_{\phi}\rho_{\phi} (1+w_{\phi}^1) = 0~,
\end{equation}
\begin{equation}\label{per2}
 \dot\rho_{R}+ 4 H \rho_R - \Gamma_{\phi}\rho_{\phi} (1+w_\phi^1)= 0~,
\end{equation}
where the inflaton field $\phi$ decays into radiation with the decay rate $\Gamma_\phi$. $\omega_\phi^1$ represents inflaton equation of state during perturbative reheating. Important to note that inflaton equation of state during phase-I, $\omega_{\phi}$ is taken to be different than that of the phase-II, $\omega_{\phi}$. {\em This is where we will again consider lattice simulation results as another important input.}     

Now that we have identified the full reheating phase in terms of two distinct stages, we will numerically solve all those equations self consistently. With the appropriate dimensionless rescaled variables for the inflaton and radiation energy densities,
\begin{equation}\label{scale}
\varPhi= \frac{\rho_{\phi} a^{3(1+w_\phi)}}{m_\phi^{(1-3 w_\phi)}}  ~ ;  ~ R(t)= \rho_R a^4~~,
 \end{equation}
governing equations for the effective dynamical preheating phase turn into the following form, 
\bea \label{nonperturbative 1}
\left.\begin{array}{c}
\frac{\varPhi'}{A^{3 w_\phi}}+ \frac{R'}{A}= 0~, \\
\frac{\varPhi'}{A^{2+3 w_\phi}}+ \frac{R}{A^4} [3 (1+w_{eff})-4]+ \frac{R'}{A^3}+ \frac{3 \varPhi (w_{eff}-w_\phi)}{A^{3(1+w_\phi)}}=0~
\end{array}\right\},~~ \mbox{\bf Phase I}
\eea
and the associated governing equations for the subsequent perturbative phase will reduce into
 \bea \label{perturbative 1}
\left.\begin{array}{c}
 \varPhi^{'} + C_1 (1+w_\phi^1) \frac{A^{1/2}\varPhi}{X}= 0~,\\
 \label{perturbative 2}
 R^{'}- C_1 (1+w_\phi^1) \frac{A^\frac{3 (1-2 w_\phi^1)}{2}\varPhi}{X}= 0~
\end{array}\right\}.~~~~~~~\mbox{\bf Phase-II}
 \eea
The rescaled scale factor is defined as $A= \frac{a}{a_{end}}$. "Prime"($'$) represents derivative with respect to A. The constant $C_1$ and redefined variables are,
 \begin{equation}
  X= \frac{\varPhi}{A^{3 w_\phi^1}}+ \frac{R}{A}~;~ C_1= \frac{\sqrt{3} M_p \varGamma_\phi}{m_{\phi}^2}~.
 \end{equation}
$m_\phi$ is the mass of the inflaton. In next section we will describe the methodology for solving the above set of equations numerically. 
 \section{Procedure for numerical analysis and boundary conditions}\label{sec5} 
 Let us describe the strategy of our numerical study. We first identify the inflation model-dependent input parameters as $N_k, H_k, V_{end}$ for a particular CMB scale $k$. For a given a canonical inflaton potential $V(\phi)$, the inflationary e-folding number $N_k$  and Hubble constant $H_k$ can be expressed as  
\bea \label{nk}
N_k  = \log\left(\frac{a_{end}}{a_k}\right)=\int_{\phi_{end}}^{\phi_k} \frac 3 2 \frac{V(\phi)}{V'(\phi)} d\phi= \int_{\phi_{k}}^{\phi_{end}} \frac{|d\phi|}{\sqrt{2\epsilon_v} M_p} ~~,~~H_k = \frac{1}{3 M_p^2} V(\phi_k)= \frac{\pi M_p\sqrt{r_k A_s}}{\sqrt{2}}~,
\eea
where, the field values at a particular scale $k$, $(\phi_{end},\phi_k)$ are computed form the condition of end of inflation,
\bea \label{boundary}
\epsilon(\phi_{end}) = \frac{1}{2 M_p^2} \left(\frac{V'(\phi_{end})}{V(\phi_{end})}\right)^2 =1, 
\eea
and equating a particular value of scalar spectral index with $n_s(\phi_k)$.
Therefore, we will get explicit relations between $(N_k,n_s^k)$ and $(H_k, n_s^k)$. The well known inflationary input parameters can be found out from the following equations 
\begin{equation}
 n_{s}^{k}= 1- 6 \epsilon(\phi_k)+ 2 \eta(\phi_k)~,~r_k=16\epsilon(\phi_k)~~,
\end{equation}
which are expressed in terms of   slow-roll parameters 
\begin{equation}
 \epsilon_v= \frac{M_{p}^2}{2}\left(\frac{V'}{V}\right)^2~~;~~|\eta_v|  = M_{p}^2 \frac{|V''|}{V}~.
\end{equation}
The reheating parameters $N_{re},T_{rad}$ will implicitly depend upon the scalar spectral index $n_s^k$ for a given scale.
The above expression can be inverted to find $\phi_k$ in terms of the scalar spectral index.
After identifying all required parameters from inflation, we will set the initial conditions for subsequent reheating dynamics. Using all these relations among those parameters, one can establish the connection between CMB anisotropy and reheating through inflation.

{\bf Phase-I initial condition}: The initial conditions for phase-I of the reheating dynamics (effective non-perturbative era) are set by the end of inflation at $A=1$ and the equation \ref{state}. Those are as follows,  
\begin{equation}\label{boundary1}
 \varPhi(A=1)= \frac{3}{2} \frac{V_{end}(\phi)}{m_{\phi}^4}~~;~~R(A=1)= \frac{3\left(\omega_{eff}-\omega_\phi\right)}{1-3\omega_{eff}}~\varPhi(A=1)~.
\end{equation}
Where $V_{end}(\phi)$ which is defined at the end of inflation, fixed by $\phi_{end}$. The initial Hubble expansion rate is expressed as $H_{I}$= ${\rho_{\phi}^{end}}/{3 M_{p }^2}$.

Subsequent perturbative dynamics will now crucially depend on the end point of the first phase of reheating namely the phase-I. On this issue we rely on the actual non-perturbative lattice simulation results \cite{Podolsky:2005bw,Figueroa:2016wxr,Maity:2018qhi} considering specific model of reheating where the inflaton field is assumed to couple with the reheating field. This system has been studied quite extensively \cite{Allahverdi:2010xz,Amin:2014eta} in the literature by using the publicly available numerical package LATTICEEASY\cite{Felder:2000hq} and its parallelized version CLUSTEREASY\cite{Felder:2007nz}. The non-perturbative analysis for different inflationary models has been proved to yield some universal results which will be our important input for the numerical analysis. Extensive works on non-perturbative reheating analysis yields an important fact that only the $50\%$ of the total comoving inflaton energy density is getting transferred into the daughter field. Additionally the inflaton equation of state tends to achieve a steady state value depending upon the power law form of the inflaton potential near its minimum.  
For example if one assumes the  inflaton potential to be of power law form $V \sim \phi^n$, for chaotic type model namely $n=2$, non-perturbative phase ends with steady value of the equation of state $\sim 0.2$. However, for other value of $n\geq4$, the equation of state approaches $\omega=\frac{1}{3}$ at the end point of the non-perturbative reheating. These are the crucial quantitative results from non-perturbative preheating dynamics we will be utilizing in our analysis for the phase-II dynamcis.   


{\bf Phase-II initial condition:} After the phase-I dynamics, the pertubative dynamics will automatically follow. However, important point would be to identify the appropriate boundary conditions. The starting moment of phase-II will be set by the normalized scale factor  $A_{npre}={a_{npre}}/{a_{end}}$ which is the ratio between the scale factor at the end of the effective non-perturbative epoch namely phase-I $a_{npre}$, and the end of the inflation. The initial conditions for the dimensionless comoving densities are
\bea
\varPhi = \varPhi(A_{npre})~~;~~ \frac{R(A_{npre})}{R(A_{npre}) +\varPhi(A_{npre})} \simeq \frac 12 .
\eea
It is important to realize that the initial condition is determined by $50\%$ decay of the total comoving energy density $\rho_T$. Further, we have numerically checked that our results do not seem to depend both qualitatively as well as quantitatively much on the amount of decay within $40\% - 60\%$ of the total energy at the end of phase-I.  For the analysis, we further assume the inflaton equation of state  $\omega_{\phi}^1 \simeq 0.2$ irrespective of the models under consideration. {\em This approximate value is again another important input from the Lattice simulation.} For comparison, we also consider the cases where either phase-I or phase-II evolution completely governs the reheating dynamics.  

{\bf Determining the reheating parameters:}
Once we numerically solve the reheating dynamics, we define one of the important parameters called reheating temperature $T_{re}$, which is generically identified as the radiation temperature $T_{rad}$ when the condition $H(t)= \Gamma_\phi$ is satisfied,
\bea \label{reheating 1}
 H(A_{re})^2= \left(\frac{\dot A_{re}}{A_{re}}\right)^2= \frac{\rho_\phi(\Gamma_\phi,A_{re},n_{s}^k)+ \rho_{R}(\Gamma_\phi,A_{re},n_{s}^k))}{3 M_p^2}=\Gamma_{\phi}^2~,
\eea
where $A_{re}$ is the normalized scale factor at the end of the reheating. Accordingly, the reheating temperature in terms of radiation temperature ($T_{rad}$) is expressed as,
\bea \label{reheating 2}
T_{re}= T_{rad}^{end}= \left(\frac{30}{\pi^2 g_*(T)}\right)^{1/4}\rho_{R}(\Gamma_\phi,A_{re},n_{s}^k)^{1/4}~.
\eea
Furthermore, the e-folding number during reheating $N_{re}$ consists of two contributions born out of two distinct phases as 
\begin{equation}\label{nre3}
 N_{re}= \log \left(\frac{a_{re}}{a_{end}}\right)= \log\left(\frac{a_{re}}{a_{npre}} \frac{a_{npre}}{a_{end}}\right)= N_{pre}+ N_{npre}~,
\end{equation}
\bea \label{nre1}
 N_{pre}= \log \left(\frac{a_{re}}{a_{npre}}\right)~~;~~N_{npre}= \log \left(\frac{a_{npre}}{a_{end}}\right)~,
\eea
where $N_{pre}$ and $N_{npre}$ are the e-folding number during perturbative and effective nonperturbative region  respectively. Combining equations (\ref{eqtre}) and (\ref{nre3}), we obtain the most important modification of Eq.\ref{eqtre} relating the reheating and inflationary parameters,
\bea \label{reheating 3}
T_{re}= \left (\frac{43}{11 g_{re}}\right)^{1/3} \left(\frac{a_0T_0}{k}\right) H_k e^{-N_k} e^{-N_{npre}} e^{-N_{pre}}~.
\eea
Now connecting equations (\ref{reheating 1}), (\ref{reheating 2}) and (\ref{reheating 3}), we can establish one to one correspondence between $T_{re}$ and $\Gamma_\phi$. 

As described before, we will consider three possible cases and compare the results
\bea
\begin{array}{ccc}
\mbox{Case-I} & N_{npre} \neq 0~,~N_{pre} \neq 0 &~~\mbox{Phase-I + Phase-II}\\
\mbox{Case-II} & N_{npre} \neq 0~~;~~N_{pre} = 0 &~~\mbox{Kamionkawski et al 2014 \cite{martin}} \\ 
\mbox{Case-III} & N_{npre} = 0~~;~~N_{pre} \neq 0 &~~\mbox{Phase-II, Discussed in the previous section}
\end{array}
\eea
To this end let us specifically mention about the case-II, when perturbative dynamics ceases to exist. 
This particular procedure proposed in \cite{martin}, has been studied quite extensively in the literature \cite{reheating}. In this particular phase, dynamics is solely governed by the effective equation of state $\omega_{eff}$. The explicit decay of inflaton does not appear in the computation. However, information about the decay constant $\Gamma_{\phi}$ is extracted from the equilibrium condition  $\Gamma_\phi = H$, where the reheating temperature ($T_{re}$) is defined as $T_{re}=0.2 \left(\frac{200}{g_*}\right)^{1/4}\left(\Gamma_\phi M_{pl}\right)^{1/2}$, with $g_*$ being the effective number of relativistic degrees of freedom. However, not to ignore an important difference between the phase-I described before and the approach devised in \cite{martin} or case-II for the present study is the additional conservation equation \ref{state}. 
This essentially differentiates the regime of applicability of these two approaches. Phase-I dynamics is assumed to be applicable in the early non-perturbative regime. Whereas, since condition \ref{state} does not exist, the original Kamionkowski et al\cite{martin} approach is effectively applicable throughout the full period of reheating without any microscopic details. Further, the value of $\omega_{eff}$ is no longer constrained to be very closed to the inflaton equation of state during phase-I. This relation essentially helps us to compare the results for various scenarios we consider. {\bf To avoid symbol confusion whenever we study case-II, we use the symbol $\omega_{eff}^K$ instead $\omega_{eff}$ which we reserve for two-phase reheating dynamics.}  
 
 \section{Maximum radiation temperature and reheating temperature: analytic study}\label{analytic}
Before moving on to a particular model, let us analytically estimate the maximum reheating temperature and its dependence upon the initial condition following the same line as before. Considering the standard definition of the radiation temperature $
T_{rad}=\left(\frac{30}{\pi^2 g_*}\rho_R\right)^{1/4}
$, and computing the radiation energy density during phase II following the Eqs.(\ref{per1}), (\ref{per2}), (\ref{nonper}) and (\ref{nonperturbative}) the approximate radiation temperature assumes the following form (see appendix \ref{calculation} for details calculation)
 \bea\label{reheat}
  T_{rad}=\left(\frac{\rho_\phi^{in}\Gamma_\phi (1+\omega_\phi^1)}{\beta x^4 H_{in}}\left[\frac{2}{5-c}\left(x^{\frac{5-c}{2}}-1\right)+\frac{\rho_R^{in}}{\rho_\phi^{in}}\left(\frac{1-x^{\frac{c+3}{2}}}{c+3}+\frac{H_{in}}{\Gamma_\phi(1+\omega_\phi^1)}\right)\right]\right)^{1/4}~~,
 \eea
 where $x$, $\beta$, $c$ and $H_{in}$ express as
\bea
x=\frac{a}{a_{npre}}~,~\beta=\frac{\pi^2 g_*(T)}{30}~,~c=3\omega_\phi^1~,~H_{in}=\frac{\sqrt{\rho_\phi^{in}}}{\sqrt{3}M_P}~~.
\eea
In the above expression $\rho_\phi^{in}$, $\rho_R^{in}$ represent inflaton and radiation energy density respectively at the end of phase-I or the beginning of  phase-II
\bea
\rho_\phi^{in}=\rho_\phi(a=a_{npre})~,~\rho_R^{in}=\rho_R(a=a_{npre})~~.
\eea
The maximum radiation temperature defined at the point  $x_{max}=a_{max}/a_{npre}$, where  $\frac{dT_{rad}}{dx} = 0 $, which gives us the maximum radiation temperature for two phase reheating expressed in terms of dimensionless comoving densities, 
\bea
T_{rad}^{max}\simeq D^{1/4}\left[1+\frac{(3+c)R(A_{npre})}{8\Phi(A_{npre})A_{npre}^{1-c}}\left(\frac{1-x_{max,p}^{\frac{c+3}{2}}}{c+3}+\frac{\sqrt{\Phi(A_{npre})A_{npre}^{-3(1+\omega_\phi^1)}m_\phi^4}}{\sqrt{3}M_p\Gamma_\phi(1+\omega_\phi^1)}\right)\right]~~
\eea
\bea
D=\left(\frac{2\Gamma_\phi\sqrt{3M_p^2\Phi(A_{npre})A_{npre}^{-3(1+\omega_\phi^1)}m_\phi^4}}{\left(3+c\right)\beta x_ {max,p}^4}\right)^{1/4}~~,~x_{max,p}=\left(\frac{8}{3+c}\right)^{\frac{2}{5-c}}~~.
\eea
One particularly notices the correction term in the maximum radiation temperature due to initial comoving radiation density $R(A_{npre})$ at the beginning of phase-II. It boils down to well know expression $T_{re}^{max}= D^{1/4}$ in the $R(A_{npre}) = 0$ limit same as Eq.\ref{tmax}. In this above expression, we ignored the contribution of dark matter. However, generically during the reheating period, dark matter is not the dominant component, therefore, the numerical value of the reheating temperature will not be affected. The analytic expression for the dimensionless comoving density during phase II related to the density at the end of inflation will be
\bea
\Phi(A_{npre})=\left(1-3\omega_{eff}\right)\Phi(A=1)A_{npre}^{-3\omega_{eff}}~~.
\eea
Where $A_{npre}$ is the normalized scale factor at the end of the effective dynamics (phase I)
\bea\label{initial41}
A_{npre}=\frac{1-3\omega_{eff}}{3\omega_{eff}}~~.
\eea
One of our important results from the above expression for the maximum reheating temperature is the highest radiation temperature, which is defined at
$T_{rad}^{max} = T_{re}^{max}$ corresponding to a given $n_s^{max}$. As we change the value of $\omega_{eff} = (10^{-3} \rightarrow 10^{-6}$), the maximum reheating temperature changes as $T_{re}^{max} = (10^{13}\to 10^{10})$ GeV. Once we set $R(A_{npre}) = 0$, the maximum reheating temperature becomes $T_{re}^{max} \sim 10^{15}$ GeV as expected (see Eqs.\ref{tmax},\ref{tmaxre}).
Proceeding further, we can also obtain the approximation expression for the reheating temperature itself. Utilizing the expression of Hubble constant at the equilibrium point  ($H_{re}=\Gamma_\phi$), and subsequent  entropy conservation, one arrives at the following expression 
\bea\label{fixreheating}
T_{re}^4\simeq \frac{x_{re}^4\rho_\phi^{in}}{3\beta M_p^2}\left[\frac{G^4\beta}{\rho_\phi^{in}}+\frac{5-c}{2\left(c+3\right)}\frac{\rho_R^{in}}{\rho_\phi^{in}}\left(\frac{G^4\beta}{\rho_\phi^{in}}-\frac{\rho_R^{in}}{\rho_\phi^{in}}\right)x_{re}^{c-1}\right]~~.
\eea
Where, $x_{re}=a_{re}/a_{npre}$ can be recognize as
\bea\label{fixxre}
x_{re}=\left(\frac{\alpha}{\eta}\right)^{\frac{1}{c-1}}~~,
\eea
Here 
\bea\label{fixxre1}
\alpha=\frac{G^4\beta}{\rho_\phi^{in}}~,~\eta=\frac{5-c}{2}\left(\frac{G^4\beta}{\rho_\phi^{in}}-\frac{\rho_R^{in}}{\rho_\phi^{in}}\right)\left[\frac{\rho_R^{in}}{\left(c+3\right)\rho_\phi^{in}}+\frac{5-c}{2}\frac{3M_p^2H_{in}^2}{\rho_\phi^{in}\left(1+\omega_\phi^{1}\right)^2}\left(\frac{G^4\beta}{\rho_\phi^{in}}-\frac{\rho_R^{in}}{\rho_\phi^{in}}\right)\right]~.
\eea
The detailed derivation of all the aforementioned equations for the reheating temperature in the appendix  \ref{calculation1}. Now we will consider a class of inflationary models of inflation and analyze our proposal of two-phase reheating scenario.  
\section{Inflation models and numerical resutls}\label{models}
 
Based on our methodology discussed above, we will now consider a class of inflationary models for which the inflaton potentials assume quadratic form. We will also point out the regime of validity of the effective non-perturbative and perturbative era for the different inflationary models. After the inflation, the inflaton field generically oscillates around the minimum of its potential $V(\phi)$. 
Reheating fields coupled with the oscillating inflaton is generically prone to non-perturbative particle production. Our objective is to replace this non-perturbative dynamics by an effective dynamical equation, which is solely governed by the effective equation of state, $\omega_{eff}$ supplemented with the additional constraint relation eq.\ref{state}. We have already observed that during phase-I, $\omega_{eff}$ is closed to that of the inflaton equation of state, $\omega_{\phi}$.  
Near the minimum of the potential if the form is taken to be power law as
$\propto\phi^n$, over multiple oscillations, the average inflaton equation of state is expressed as \cite{mukhanov} 
\bea
\omega_{\phi}=\frac{P_\phi}{\rho_\phi}\approx \frac{\langle\phi V'(\phi)-2V(\phi)\rangle}{\langle\phi V'(\phi)+2V(\phi)\rangle}=\frac{n-2}{n+2}~~.
\eea
For $n=2$ model, $\omega_\phi$ assumes dust like equation of state $(\omega_\phi=0)$. Throughout the subsequent study, we consider those inflationary models which have quadratic potential near their minimum. Therefore, during phase-I of reheating, we set $\omega_\phi=0$. To this end, let us emphasize again that during phase-II, when the reheating dynamics enter into the perturbative phase, we assume the inflaton equation of state  $\omega_\phi^1 \simeq 0.2$, which is one of the important lattice simulation results mentioned earlier. Further, we analyze phase-I dynamics considering two specific choices of the effective equation of state  $\omega_{eff}=(\omega_{\phi}+10^{-3},
\omega_{\phi}+10^{-6})$ which are closed to $\omega_{\phi}$. 
\begin{figure}
 	\begin{center}
 \includegraphics[width=11.3cm,
 height=08.2cm]{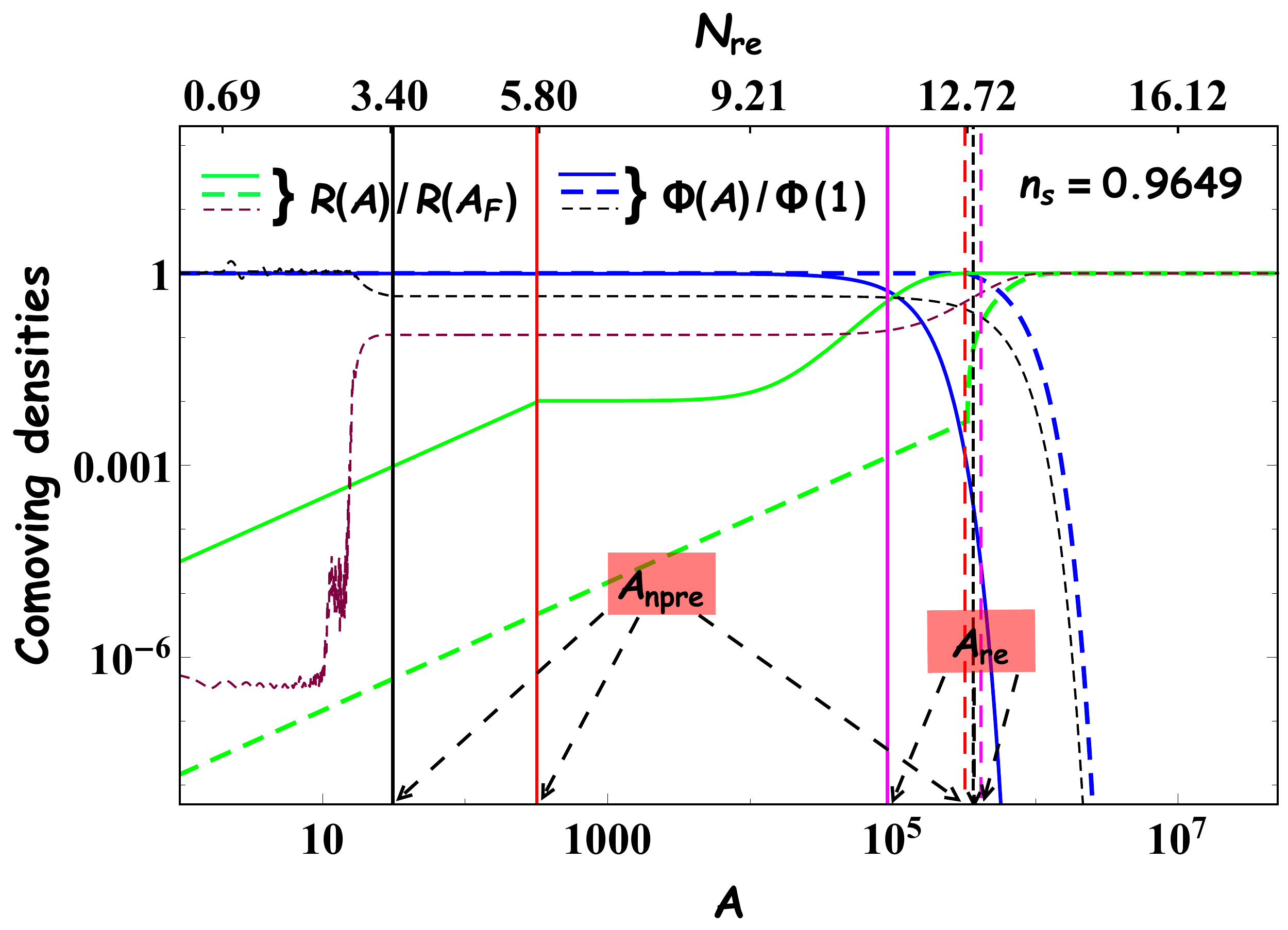}
 		\caption{\scriptsize  	We plot the evolution of the different energy components (inflaton and radiation) with the normalized scale factor for chaotic inflation model with $n=2$. The \textcolor{blue}{\bf{blue}} and \textcolor{green}{\bf{green}} curve indicates the variation of comoving densities, the inflaton, and radiation density, respectively, for our proposed two-phase dynamics (case-I). The \textcolor{red}{\bf{red}} and \textcolor{Rhodamine}{\bf{pink}} line represents the normalized scale factor at the ending of phase I and  II accordingly. Furthermore, the solid and dashed curves correspond to the two different values of the effective equation of state $\omega_{eff}=(10^{-3},10^{-6})$. Whereas, the result for considering standard non-perturbative lattice simulation, during phase I, shown by the dashed \textcolor{black}{\bf{black}} and \textcolor{Brown}{\bf{brown}} line.  }
 		\label{compare1}
 	\end{center}
 \end{figure}
 
{\bf Plots and important model independent observations} : Before we go into detailed discussion on various inflationary models, let us fist illustrate different plots and  important model independent observations. For each model we have drawn two different plots: one in $(\Gamma_{\phi}~vs~N)$ space where it shows the variation of reheating e-folding number $N$ depending upon the inflaton decay constant. As noted earlier, we have considered different scenarios. For our proposed two-phase reheating scenario (case-I), we have studied two possible values of phase-I effective equation of state, $\omega_{eff} = 10^{-3}$ corresponding to {\bf {\textcolor{green}{solid green}}} and {\bf solid black} curves, and $\omega_{eff}=10^{-6}$ corresponding to {\bf {\textcolor{green}{dotted green}}} and {\bf {\textcolor{black}{dotted black}}} curves. For all cases, $\omega_{\phi} = 0$. One of the most important outcomes of our analysis is the emergence of a critical inflaton decay constant $\Gamma_{\phi} = \Gamma^{cri}_{\phi}$ denoted by {\bf {\textcolor{red}{red dots}}} associated with each particular $\omega_{eff}$. This indicates the fact that for $\Gamma_{\phi} > \Gamma^{cri}_{\phi}$, the reheating period will be dominated by phase-I, effective non-perturbative dynamics, otherwise it is perturbative  dominated. The critical value of inflaton decay constant increases with the decreasing $\omega_{eff}$. This can be understood from several interconnecting physical effects. First of all most important  Eq.\ref{state},
\bea
R(A=1)= \frac{\left(\omega_{eff}-\omega_\phi\right)}{1-3\omega_{eff}}~\varPhi(A=1),
\eea
 which not only fixes the approximate value of $\omega_{eff}$ but also sets the initial condition for phase-I dynamics.
Further, larger the value of $\omega_{eff}$, higher will be initial radiation density $R(A=1)$ which automatically leads to smaller value of phase-I e-folding number $N_{npre}$. Therefore, a particular $\Gamma_{\phi}$ will naturally lead to larger $N_{pre}$, as associated with each
 $\Gamma_{\phi}$ there exits a reheating temperature which follows from $T_{re} \propto e^{-N_{re}} = e^{-(N_{pre}+N_{npre})}$. On the other hand critical $\Gamma_{\phi}^{cri}$ is a point where $N_{pre} = N_{npre}$ in $N~vs~\Gamma_{\phi}$ space. From these two conditions one can argue that transition from perturbative to non-perturbative reheating phase would occur for larger critical value $\Gamma_{\phi}^{cri}$ for larger $\omega_{eff}$ value. Given a reheating model 
with specific inflaton-daughter field interaction, we have also discussed about the existence of critical inflation decay constant $\Gamma_{\phi}^{cri}(model)$ which were shown by vertical {\bf {\textcolor{red}{red lines}}} in the plots. From the theoretical values of the critical inflaton decay constant,  $(\Gamma_{\phi}~vs~N_k)$ plots indicates that the value of $\omega_{eff}$ must lie within $(10^{-3}, 10^{-6})$ irrespective of the inflationary models considered. At this point let us understand the physical meaning of non-vanishing initial radiation density $\rho_R(A=1) \simeq (10^{-3},10^{-6}) \rho^{in}_{\phi}$ considering $\omega_{eff}=(10^{-3},10^{-6})$. We replace the full non-perturbative dynamics by an effective dynamics, which naturally does not capture the complete picture. Typically non-perturbative phase contains three distinct phases: parametric resonance phase, thermalization phase and steady state phase. And this is the initial parametric resonance phase, where explosive particle production can naturally give raise to required initial radiation density $\rho_R(A=1) \simeq 10^{-6} \sim 10^{-3}$ in unit of total density $\rho_{\phi}$ almost instantly.  

To see whether our proposed phase I dynamics is justified or not, we compare our result with actual non-perturbative results. In order to do that, we use non-perturbative lattice simulation during the preheating, considering a specific inflaton-reheating field interaction $\frac{1}{2}g^2\phi^2\chi^2$. In all the lattice simulation results, the initial radiation density typically assumes $\rho_R(A=1)\simeq 10^{-4}$ in units of initial inflaton energy density, which essentially lies within what we have considered. Once the preheating phase reaches the steady-state condition, we again solve perturbative dynamics, and found that the reheating ends at around the same value of $A_{re}$ (shown by the \textcolor{black}{\bf{dashed black line}}) where our two-phase reheating ends for $\omega_{eff}=10^{-3}$ (\textcolor{Rhodamine}{\bf{solid pink line}}) and for $\omega_{eff}=10^{-6}$ (\textcolor{Rhodamine}{\bf{dashed pink line}}) accordingly. Therefore, our effective two-phase reheating approach seems to capture the essential properties of non-perturbative lattice results, except the non-perturbative e-folding number, which will be taken up in the future. 

Nevertheless, for comparison, in the same plot, we also have drawn total reheating e-folding number for other two cases:{\bf{ \textcolor{Rhodamine}{dotted pink lines}}} for case-II and {\bf {\bf \textcolor{blue}{solid blue lines}}} for case-III mentioned before. It turns out that total number of reheating e-folding number for case-II, case-III, and the case-I, $N_{re} = (N_{npre} + n_{pre})$ are almost the same for all different values of the equation of state. 

In an another class of plots in $(n_s~vs~T_{re})$ space, we describe the variation of reheating temperature $T_{re}$ with respect to the scalar spectral index $n_s$. From these plots, we can read that two-phase reheating process (case-I) is crucially dependent upon the value of $\omega_{eff}$. Furthermore, case-I results are qualitatively similar to that of the case-II $\omega_{eff}^K = 0.212$ (equation of state at the starting point of phase-II in two-phase analysis). On the other hand, perturbative reheating (case-III) results are qualitatively similar to that of the case-II for $\omega_{eff}^K = 0$. 
For usual perturbative reheating scenario (case-III) the semi-analytic approach discussed before reveals the existence of maximum possible reheating temperature $\sim 10^{15}$ GeV. Our numerical computation also indicates the same thorough {\bf {\bf \textcolor{blue}{solid blue lines}}}.  Further, case-II scenario also has the same prediction of model independent maximum reheating temperature $T_{re}^{max}$ irrespective of the value of its effective equation of state $\omega_{eff}^K = (0, 0.212)$ shown though {\bf {\bf \textcolor{Rhodamine}{solid pink lines}}} and {\bf {\bf \textcolor{Rhodamine}{dotted pink lines}}} respectively. For conventional reheating dynamics (case-II $\&$ case-III) the maximum reheating temperature directly corresponds to instantaneous reheating with total e-folding number $N_{re}\to 0$. This can also be straightforwardly connected with the maximum possible scalar spectral index $n_s^{max}$. The proposed  two-stage reheating dynamics (case-I) instead predicts very different results in this regard.  First of all instantaneous reheating ceases to exit in this scenario because of its underlying assumptions. As $N_{pre}\to 0$, $N_{re}\to N_{npre}$, which automatically leads to different values of $(T_{re}^{max},n_s^{max})$ followed from the condition $N_{re} = N_{npre}$, which naturally assumes model independent values such as 
$N_{npre} \sim 6$ for $\omega_{eff} =10^{-3}$, and $N_{npre} \sim 12$ for $\omega_{eff} =10^{-6}$. Smaller the effective equation of state during phase-I, larger will be its duration $N_{npre}$ and consequently $T_{re}^{max}$ will be reduced.
As expected from our earlier analytical  calculation the important results are the values of maximum reheating temperature $T_{re}^{max} \sim (10^{13}, 10^{10})$ GeV for $\omega_{eff} = (10^{-3}, 10^{-6})$ respectively. Physical origin of this two different limiting temperature is clear from the fact that increase of $T_{re}$ is directly connected with the increase of $\Gamma_{\phi}$. Hence with the increasing temperature reheating dynamics undergoes a transition from perturbative to non-perturbative regime at particular critical temperature $T_{re}^{cri}$ associated with $\Gamma_{\phi}^{cri}$, leading to a distinct value of $N_{npre}$  which is different for different $\omega_{eff}$ value. This leads to different $T_{re}^{max}$. Therefore, an important conclusion we can arrive at  
is that given the approximate estimates
of model specific critical decay width $\Gamma_{\phi}^{cri}(model)$, the maximum reheating temperature $T_{re}^{max} $ should be within $(10^{10} - 10^{13})$ GeV, irrespective of the dynamics of the second phase-II and inflationary model under consideration.
 However, we must note that the associated maximum values of the $n_s^{max}$ are model dependent, which will be discussed for each model.

\subsection{Chaotic inflation  \cite{chaotic}}
\begin{figure}[t!]
 	\begin{center}
 		\includegraphics[width=008.1cm,height=05.6cm]{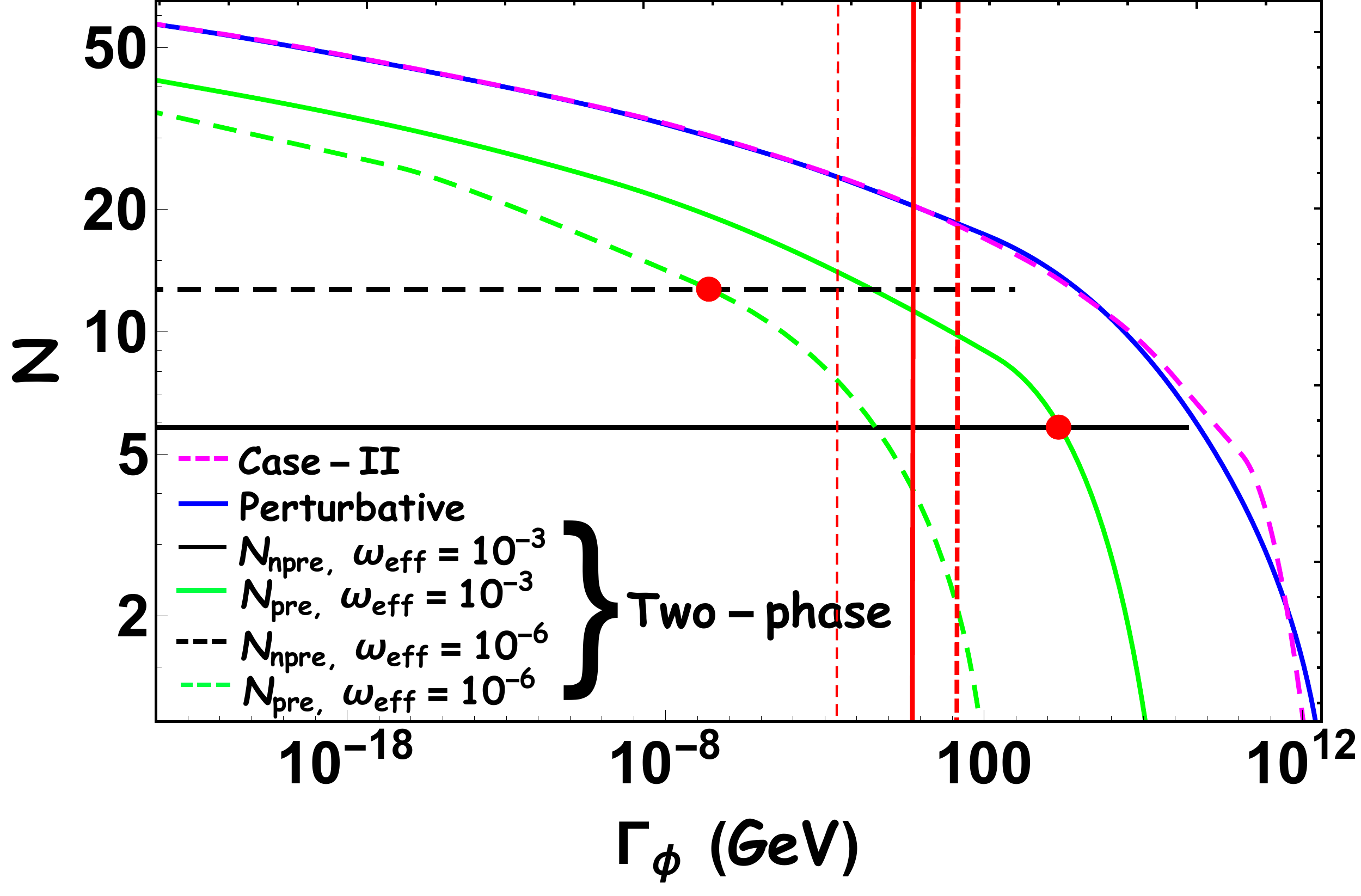}
 		\includegraphics[width=008.1cm,height=05.6cm]{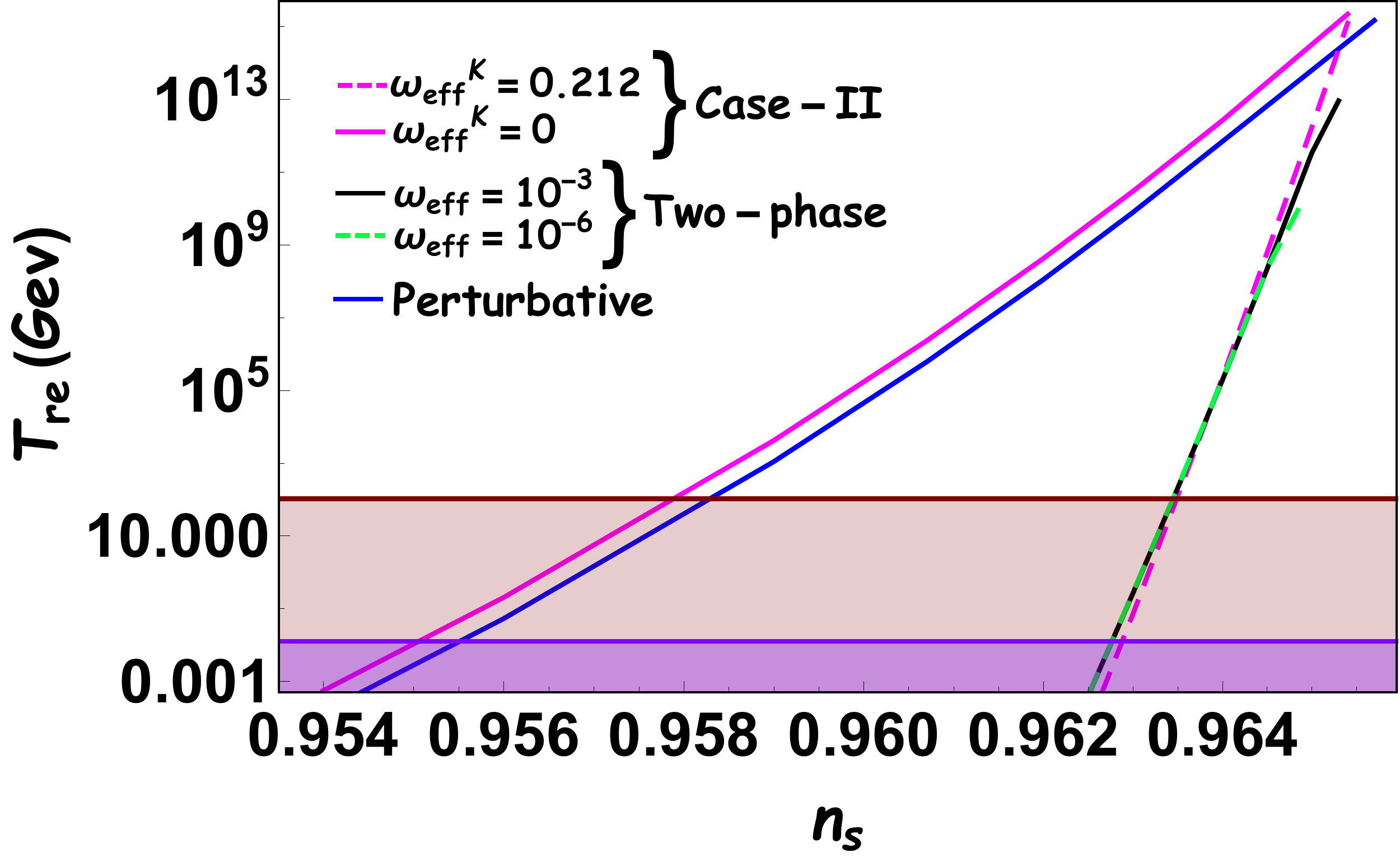}

 		\caption{\scriptsize We plot on the left side, the variation of the e-folding number as a function of inflation decay width ($\Gamma_\phi$) and on the right side, variation of reheating temperature ($T_{re}$) as a function of $n_s$ for chaotic inflation model  with $n=2$. The plot on the left side, variation of $N_{npre}$ ( e-folding number during first phase of reheating), $N_{pre}$ (e-folding number during perturbative reheating) are shown by {\bf {\bf \textcolor{black}{black}}},  {\bf {\bf \textcolor{green}{green}}} lines (solid and dashed) for two different values of $\omega_{eff}=(10^{-3},10^{-6})$. The intersection points of $N_{npre}$ and $N_{pre}$ for different values of $\omega_{eff}$ are showen by {\bf {\bf \textcolor{red}{red circle}}}. The {\bf {\bf \textcolor{blue}{ blue}}} and {\bf {\bf \textcolor{Rhodamine}{dashed pink}}} lines indicates the variation of e-folding number during reheating for purely perturbative, and the analysis is given by Kaminkowski et al. 2014 \cite{martin} with $\omega_{eff}^K=0$ respectively. The thick dashed, thin dashed, and solid {\bf {\bf \textcolor{red}{red line}}} corresponds to the three different values of the decay constant at the transition point of non-perturbative to the perturbative era from the theoretical point of view provided by equations (\ref{gamma1}), (\ref{gamma2}) and (\ref{gamma}). All plots are drawn within $2\sigma$ range of $n_s$ \cite{Akrami:2018odb}. The {\bf {\bf \textcolor{brown}{light brown}}} region is below the electro weak scale $T_{ew}\backsim 100 ~GeV$ and the {\bf {\bf \textcolor{violet}{violet}}} region below $10^{-2} ~GeV$ would ruin the predictions of big bang nucleosynthesis (BBN).}. 
 		\label{chaotic1}
 	\end{center}
 \end{figure}
Even though chaotic inflation is observationally disfavored, we consider this potential for its simple nature. For usual chaotic inflation the potential looks like, 
\bea \label{chao}
V(\phi)= \frac 1 2 m^{4-n} \phi^n.
\eea
Where $n = 2,4,6 \dots$. If we consider only the absolute value of the field, $n=3,5,\dots$ can also be included. $m$ is parameter of mass dimension. For the purpose of our study, we only consider $n=2$ mainly because $\omega_{\phi} = 0$.

{\bf Initial conditions for phase-I:} The initial densities to solve dynamical equation during phase-I can be calculated as 
\begin{equation}
\varPhi(A=1)= \frac{3}{4} \frac{m^{4-n}}{m_{\phi}^4} \left(\frac{n M_p}{\sqrt{2}}\right)^n~,~R(A=1)= \frac{3\left(\omega_{eff}-\omega_\phi\right)}{1-3\omega_{\phi}}~\varPhi(A=1)~,
\end{equation}
where 
\bea
 m= M_p \left(3 \pi^2 r_k A_{\delta\phi}\right)^{\frac{1}{4-n}} \left(\frac{1-n_s^k}{n(n+2)}\right)^{\frac{n}{2(4-n)}}~.
 \eea
$A_{\delta\phi}\sim 10^{-9}$ is the amplitude of the inflaton fluctuation which is measured from CMB observation. $m_{\phi}$ is defined as second derivative of the inflaton potential. To establish the connection among inflationary and reheating parameters the inflationary e-folding number, $N_k$ and tensor to scalar ratio, $r_k$ are similarly calculated as,
 \bea
 ~N_k = \frac{n+2}{2(1-n_s^k)} - \frac{n}{4} ~, ~ r_k= \frac{8n}{n+2} \left(1-n_s^k \right)~.
\eea

{\bf Initial conditions for phase-II:}
 Additionally the initial conditions for the phase-II will be set at the normalized scale factor $A_{npre}$ where phase-I ends. The conditions are
 \bea \label{phase2cond}
 \varPhi = \varPhi(A_{npre})~~;~~ \frac{R(A_{npre})}{R(A_{npre}) +\varPhi(A_{npre})} \simeq \frac 12 .
 \eea
 To establish the relation between reheating temperature ($T_{re}$) and inflationary index ($n_s^k$), we follow the methodology explained in the previous section.\\
{\bf Observations:} Important results for chaotic inflation are depicted in Fig.(\ref{chaotic1}). As stated at length, the initial effective equation of state $\omega_{eff}$ plays a crucial role in driving the whole reheating dynamics. For our purpose we took two sample values $(10^{-3}, 10^{-6})$. According to these two values, the critical values of the inflaton decay constants are found to be $\Gamma_\phi^{cri}=(2.46\times10^3, 2.73\times10^{-7})$ GeV.  
Similarly, we can address critical values (transition from perturbative to non-perturbative reheating) in terms of reheating temperature. For this model, the critical values of the reheating temperature set to be $T_{re}^{cri}\simeq (2.7\times 10^{10},3\times 10^5)$ GeV with for the equation of state $\omega_{eff}=(10^{-3},10^{-6})$. This entails the fact that If $\Gamma_{\phi} > \Gamma_{\phi}^{cri} ~(T_{re}>T_{re}^{cri})$, the reheating phase will be dominated by non-perturbative process.

For concreteness, let us bring specific reheating models into consideration. We have discussed three different interaction models with associated non-perturbative constraints equations (\ref{gamma}), (\ref{gamma2}) and (\ref{gamma1}). Associated with those we have theoretical values of the critical inflaton decay constants $\Gamma_\phi^{cri} (model)=(0.003,0.5,11.8)$ GeV respectively. The first two values correspond to inflaton decaying into the scalar particle, and the third one corresponds to decaying into a pair of fermionic particles. Interestingly, comparing those numerical and theoretical values of $\Gamma_{\phi}^{cri}$, one can observe that the initial effective equation of state $\omega_{eff}$ during phase-I must lie within $(10^{-3},10^{-6})$. This essentially suggests that all the three models of inflaton interaction will lead to initial radiation density within the value $(10^{-3},10^{-6})$ instantaneously, which we can immediately read off from the Fig.\ref{compare1}.      

In all the reheating scenarios discussed and proposed so far, there exists a model-independent maximum reheating temperature. However, the associated maximum value of the spectral index $n_s^{max}$ turned out to be model dependent. In the conventional perturbative reheating  discussed before, and also the constraints from reheating (case-II) scenario, $N_{re} \rightarrow 0$ provides the condition for $n_s^{max}$. 
For two phase reheating scenario (case-I)
the phase-I effective dynamics is inevitable, which leads to different condition $N_{re}\approx N_{npre}$ for the maximum possible $n_s^{max}$ compatible with CMB observation. Furthermore, for each model one can define minimum spectral index $n_s^{min}$ which can be associated with minimum possible reheating temperature set by BBN constraints \cite{Kawasaki:1999na}-\cite{Fields:2014uja}, which is $T_{re}^{min}=10^{-2}$ GeV. Taking into account both the possibilities, for case-I we obtain the possible bound on the spectral index $0.9628\leq n_s\leq 0.9653$ and  $0.9628\leq n_s\leq0.9649$ for $\omega_{eff} = (10^{-3},10^{-6})$ respectively. For case-II \cite{martin}, the bound is $0.955\leq n_s\leq0.9654$, $0.9629\leq n_s\leq0.9654$ for $\omega_{eff}^K=(0,0.212)$ respectively.  Additionally, for purely perturbative dynamics case-III, one obtains $0.9555\leq n_s\leq0.9657$. Important to remind at this point, all these bounds are consistent with CMB within $2\sigma$ error of $n_s$.  
From the maximum $n^{max}_s$, the maximum value of the inflationary e-folding number ($N_k^{max}$) can be obtained. For example for case-I scenario we have  $N_k^{max}\simeq(57,56)$ with effective equation of state $\omega_{eff}=(10^{-3},10^{-6})$ respectively. Whereas, for case-II, $N_k^{max}\simeq 57$ and for case-III, $N_k^{max}\simeq 58$.

The variation of the reheating temperature as a function of the spectral index for the different reheating mechanism is shown in the fig.\ref{chaotic1}. The behavior of reheating temperature with respect to $n_s$ appears to be model-independent.
 
\subsection{Axion inflation \cite{axion,Freese:2014nla}}

The potential for the axion/natural inflation is 
\bea \label{naturalinflation}
V(\phi)=  \Lambda^4 \left[1 - \cos\left(\frac{\phi}{f}\right) \right] .
\eea
where, $(\Lambda,f)$ are the scale of inflation and axion decay constant of this present model. By tuning the value of the decay constant, this model marginally consistent with the recent observation \cite{Gerbino:2016sgw}. To be consistent with CMB data, we consider two sample super-Planckian values of the axion decay constant, $f=(10,50)M_p$. The scale of this inflation, $\Lambda$ fixes by the CMB normalization. 

{\bf Initial conditions for phase-I:} The initial conditions to solve the differential equations for effective non-perturbaive era are set at the end of inflation to be,
\begin{equation}
 \varPhi(A=1)=\frac{3}{2} \frac{2 \Lambda^4 M_p^2}{(2 f^2 +M_p^2) m_\phi^4}~,~R(A=1)=\frac{3\left(\omega_{eff}-\omega_\phi\right)}{1-3\omega_{eff}}~\varPhi(A=1)~,
\end{equation}
where
\begin{equation}
 \Lambda=\left(\frac{3\pi^2 M_p^2 A_s (f^4(1-n_s)^2 -M_p^4)}{2 f^2}\right)^{\frac{1}{4}}~,~m_\phi=\frac{\Lambda^2}{f}~.
\end{equation}
In addition the inflationary e-folding number, $N_k$ and tensor to scalar ratio, $r_k$ for natural inflation model are expressed in terms of scalar spectral index and model parameters as
\begin{equation}
 N_k=\frac{f^2}{M_p^2} \ln \left(\frac{2f^2 (f^2 (1-n_s)+M_p^2)}{(2f^2 +M_p^2) (f^2 (1-n_s)- M_p^2)}\right)~,~r_k= 4\left(\frac{f^2 (1-n_s)- M_p^2}{f^2}\right)~.
\end{equation}
\begin{figure}[t!]
 	\begin{center}
 		\includegraphics[width=008.1cm,height=05.6cm]{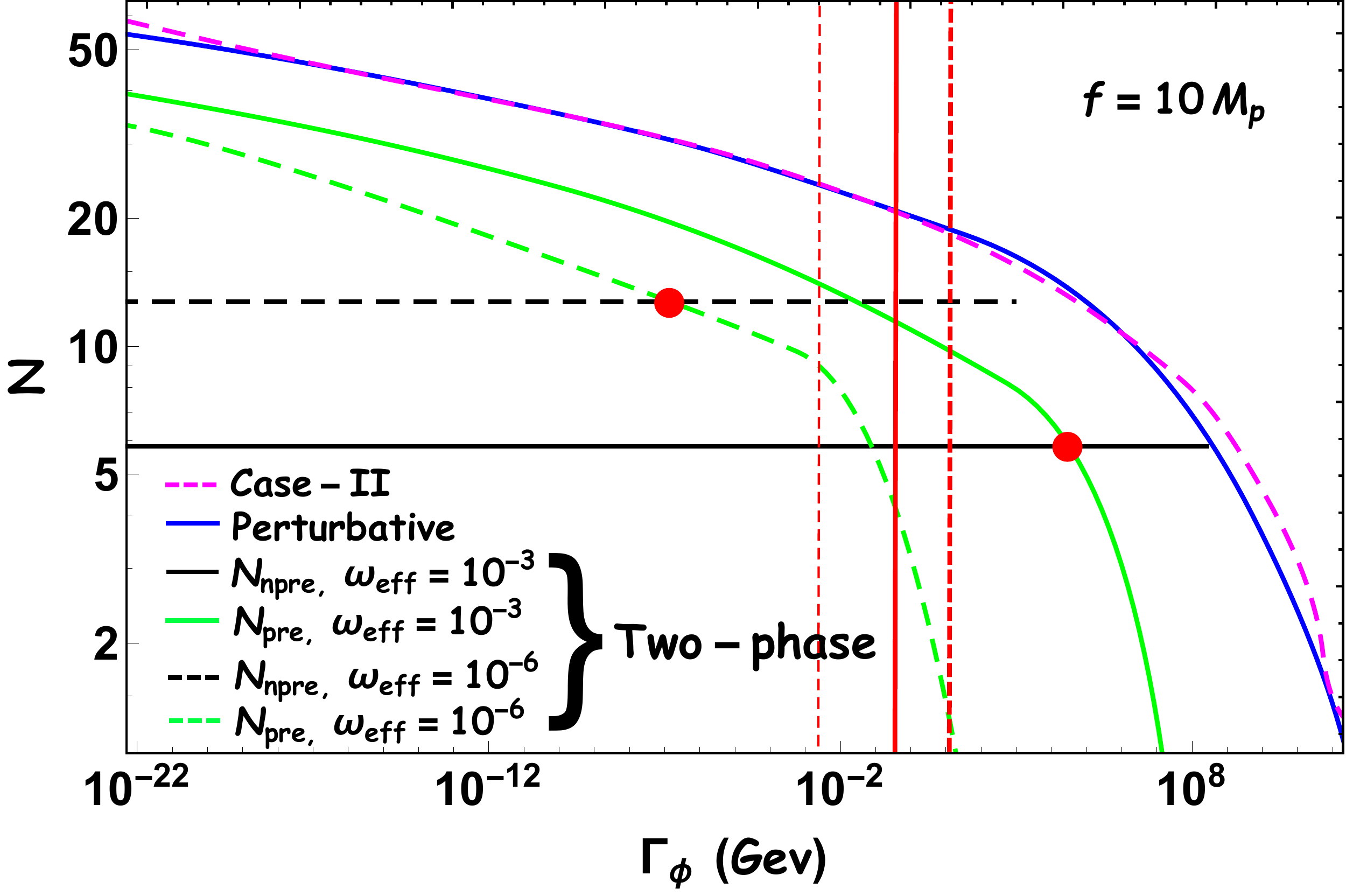}
 		\includegraphics[width=008.1cm,height=05.6cm]{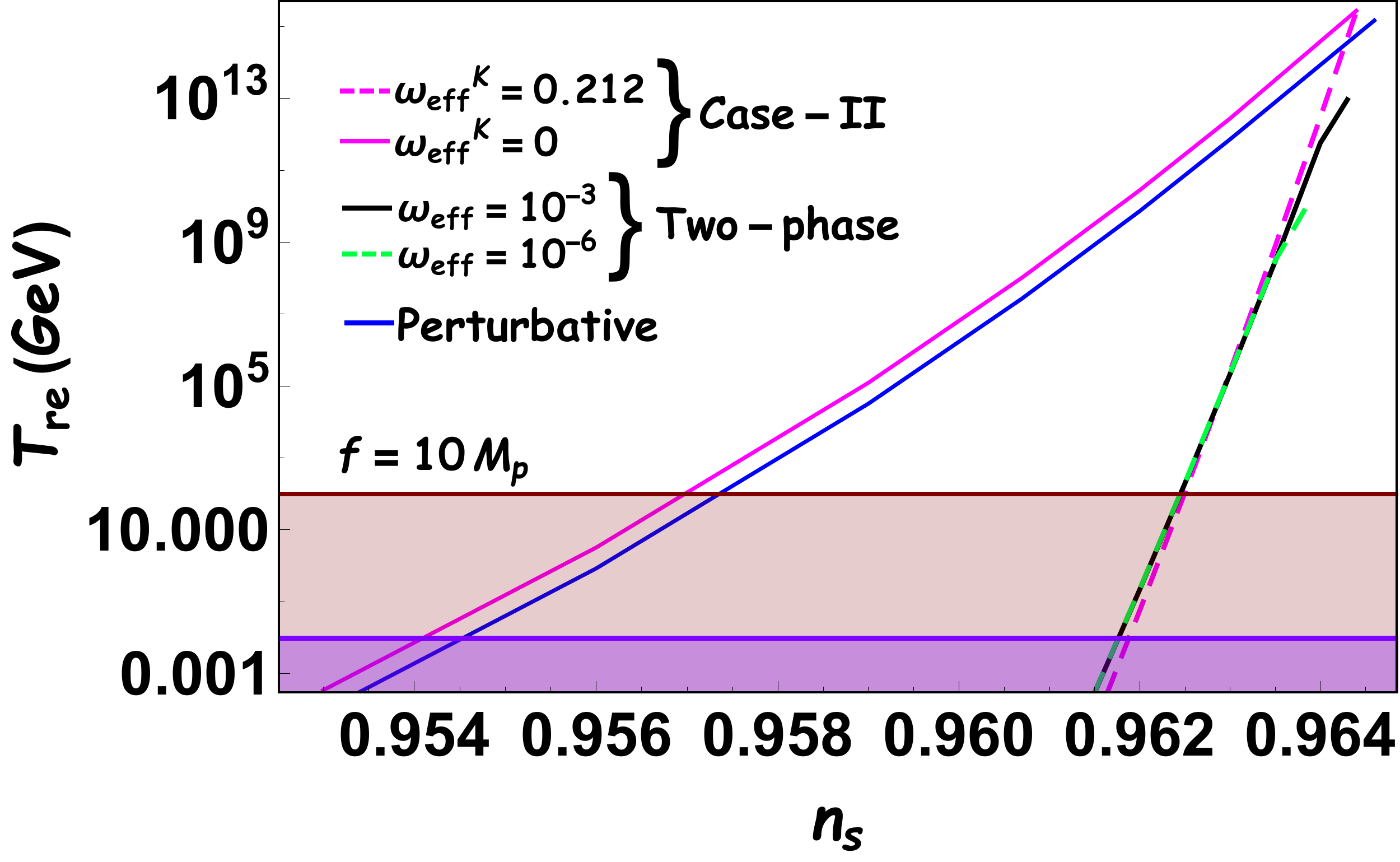}
 		\includegraphics[width=008.1cm,height=05.6cm]{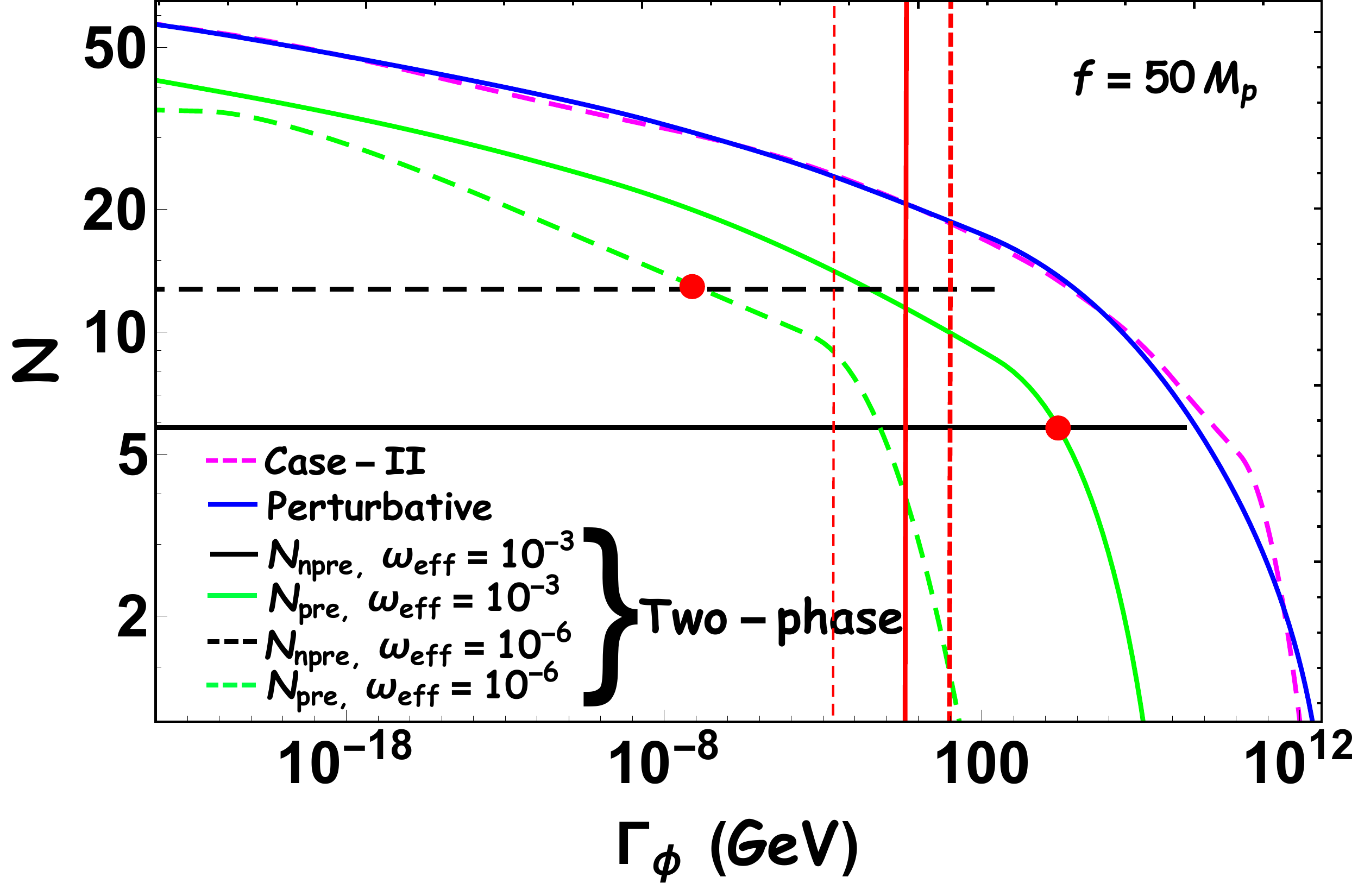}
 		\includegraphics[width=008.1cm,height=05.6cm]{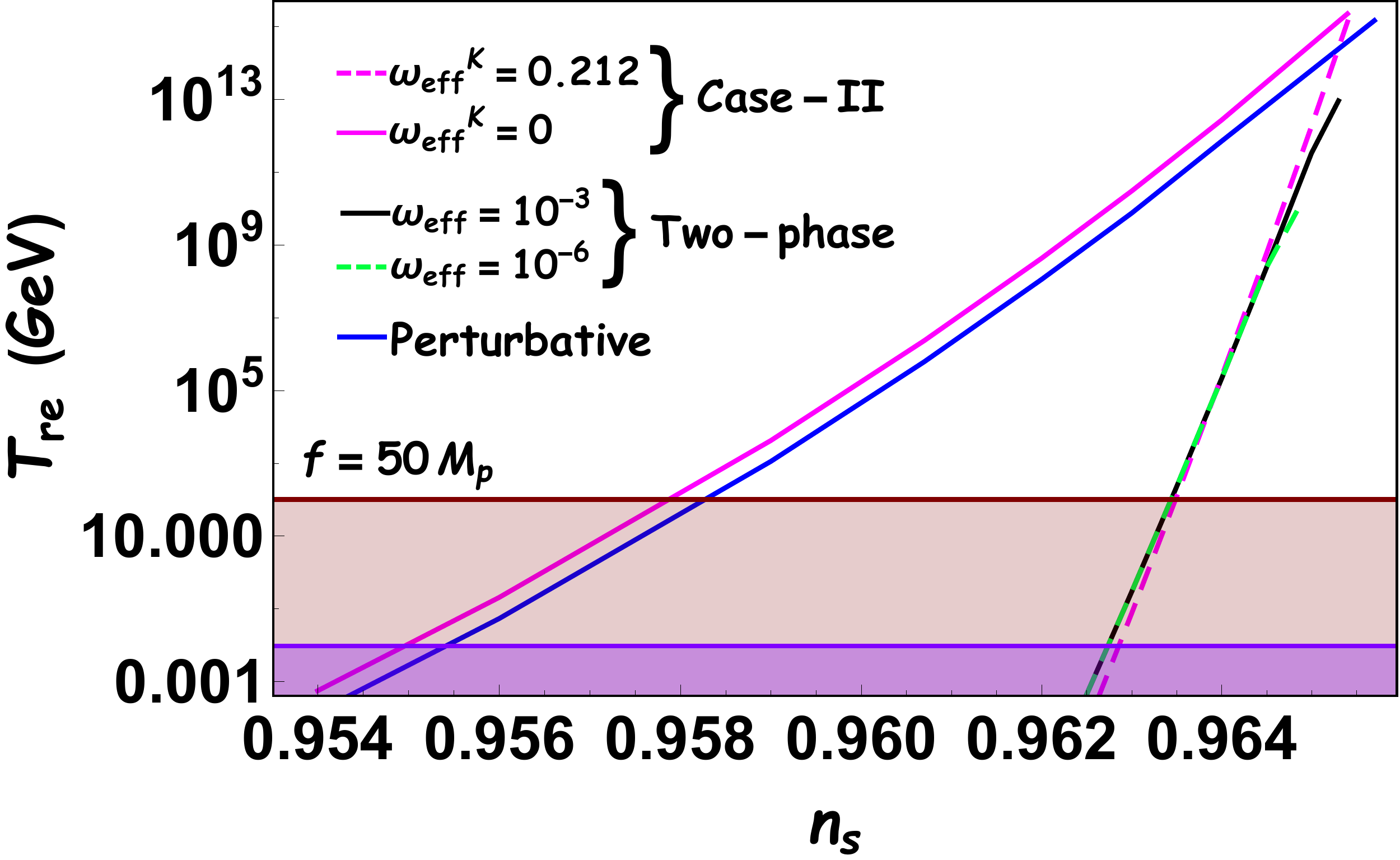}
 		\caption{\scriptsize  All plots are same as in the previous Fig.\ref{chaotic1}. The main difference is that, here we have plotted for natural inflation
model for $f = (10, 50) M_p$.}.
 		\label{naturalplot}
 	\end{center}
 \end{figure}

The initial condition for the phase-II dynamics will be the same as chaotic inflation has given in Eq.\ref{phase2cond}. 

{\bf Observations:} 
Main results of axion inflationary model are depicted in Fig. (\ref{naturalplot}). As has been mentioned earlier we have considered two sample values of the axion decay constant $f=(10,50) M_p$. For a fixed value of axion decay constant, $f=10 M_p$, the critical values of inflaton decay constant assume $\Gamma_\phi^{cri}\simeq (3.7 \times10^4,1.1 \times10^{-7})$ GeV, and that of the reheating temperatures are $T_{re}^{cri}\simeq (4\times 10^{10},1.8\times 10^5)$ GeV. Similarly  for $f=50M_p$, $\Gamma_\phi^{cri} \simeq (2.7 \times10^{4},1.1 \times10^{-7})$ GeV and $T_{re}^{cri}\simeq (9\times 10^{10},1.9\times 10^5)$ GeV. For both the cases the effective equation of states are taken to be $\omega_{eff} =(10^{-3},10^{-6})$.  
One the other hand the theoretical value of the critical inflaton decay constants for three different interacting reheating models are calculated to  be, $\Gamma_{\phi}^{cri}(model)=(2.8\times10^{-3},0.45,10.8)$ GeV for $f =10 M_P$, and $\Gamma_{\phi}^{cri}(model) = (3.2 \times 10^{-3},0.51,12.3)$ GeV for $f=50 M_p$. Those values of decay constants are determined from equations (\ref{gamma}), (\ref{gamma2}) and (\ref{gamma1}) accordingly.
Let us point out again that the first two values correspond to inflaton decaying into the scalar particle, and the third one corresponds to decaying into a pair of fermionic particles. Here again, from the left panel of Fig.\ref{naturalplot}, one concludes that if the universe undergoes two-phase reheating, considering the specific interaction during reheating the initial $\omega_{eff}$ during phase-I must lie within $(10^{-3},10^{-6})$.   
\begin{table}[t!]
	\caption{Reheating models and their associated bound on inflationary parameters (Axion inflation)}
	$f=10 M_p$
	\begin{tabular}{|p{2cm}|p{2.5cm}|p{2.5cm}|p{2.5cm}|p{2.5cm}|p{4cm}| }
		\hline
		
		Inflationary parameter&\multicolumn{2}{c|}{Case-I (Two-phase)}&\multicolumn{2}{c|}{ Case-II}& ~~Case-III (Perturbative)\\
		\cline{2-6}
		& \quad$\omega_{eff}=10^{-3}$ &\quad$\omega_{eff}=10^{-6}$&\quad $\omega_{eff}^K=0$&\quad $\omega_{eff}^K=0.212$ &\qquad \qquad $\omega_{\phi}=0$\\
		\hline
		$n_s^{min}$&\quad $0.9618$&\quad $0.9618$&\quad $0.9541$&\quad $0.9619$&\qquad \qquad $0.9545$\\
		$n_s^{max}$&\quad $0.9643$&\quad $0.9639$&\quad $0.9644$&\quad $0.9644$&\qquad \qquad $0.9646$\\
		$N_k^{max}$&\quad $57.06$&\quad $56.39$&\quad $57.23$&\quad $57.23$&\qquad \qquad $57.58$\\
		
		\hline
		
	\end{tabular}
	$f=50 M_p$
	\begin{tabular}{|p{2cm}|p{2.5cm}|p{2.5cm}|p{2.5cm}|p{2.5cm}|p{4cm}| }
		\hline
		
		Inflationary parameter&\multicolumn{2}{c|}{Case-I (Two-phase)}&\multicolumn{2}{c|}{ Case-II}& ~~Case-III (Perturbative)\\
		\cline{2-6}
		& \quad$\omega_{eff}=10^{-3}$ &\quad$\omega_{eff}=10^{-6}$&\quad $\omega_{eff}^K=0$&\quad $\omega_{eff}^K=0.212$ &\qquad \qquad $\omega_{\phi}=0$\\
		\hline
		$n_s^{min}$&\quad $0.96275$&\quad $0.96275$&\quad $0.9549$&\quad $0.9629$&\qquad \qquad $0.9554$\\
		$n_s^{max}$&\quad $0.9653$&\quad $0.9649$&\quad $0.9654$&\quad $0.9654$&\qquad \qquad $0.9657$\\
		$N_k^{max}$&\quad $57.14$&\quad $56.48$&\quad $57.31$&\quad $57.31$&\qquad \qquad $57.81$\\
		
		\hline
	\end{tabular}
	\label{tableaxion}
\end{table}

The lower limit of $n_s$ has been set by the minimum possible reheating temperature due to big-bang nucleosynthesis (BBN) constraint. With increasing spectral index from its minimum value $n_s^{min}$ along with decay width, the perturbative e-folding number $N_{pre}$ depreciates towards zero and the total e-folding number, $N_{re}$ approaches towards $N_{npre}$ which is identified as the point of $T_{re}^{max}$ and $n_s^{max}$. Following the discussion of chaotic inflation model, in the table \ref{tableaxion}, we provide possible limiting value the inflationary parameters $(n_s^{min},n_s^{max},N_k^{max})$ parameters for three different reheating scenarios. These limiting values, in turn, will restrict the possible values of reheating parameters. Therefore, the more we decrease the error of the inflationary parameter more precisely, we will be able to fix the reheating parameters. 

\subsection{\bf $\alpha-$attractor model\cite{alpha}}\label{alpha11}
\begin{figure}[t!]
 	\begin{center}
		\includegraphics[width=008.1cm,height=5.6cm]{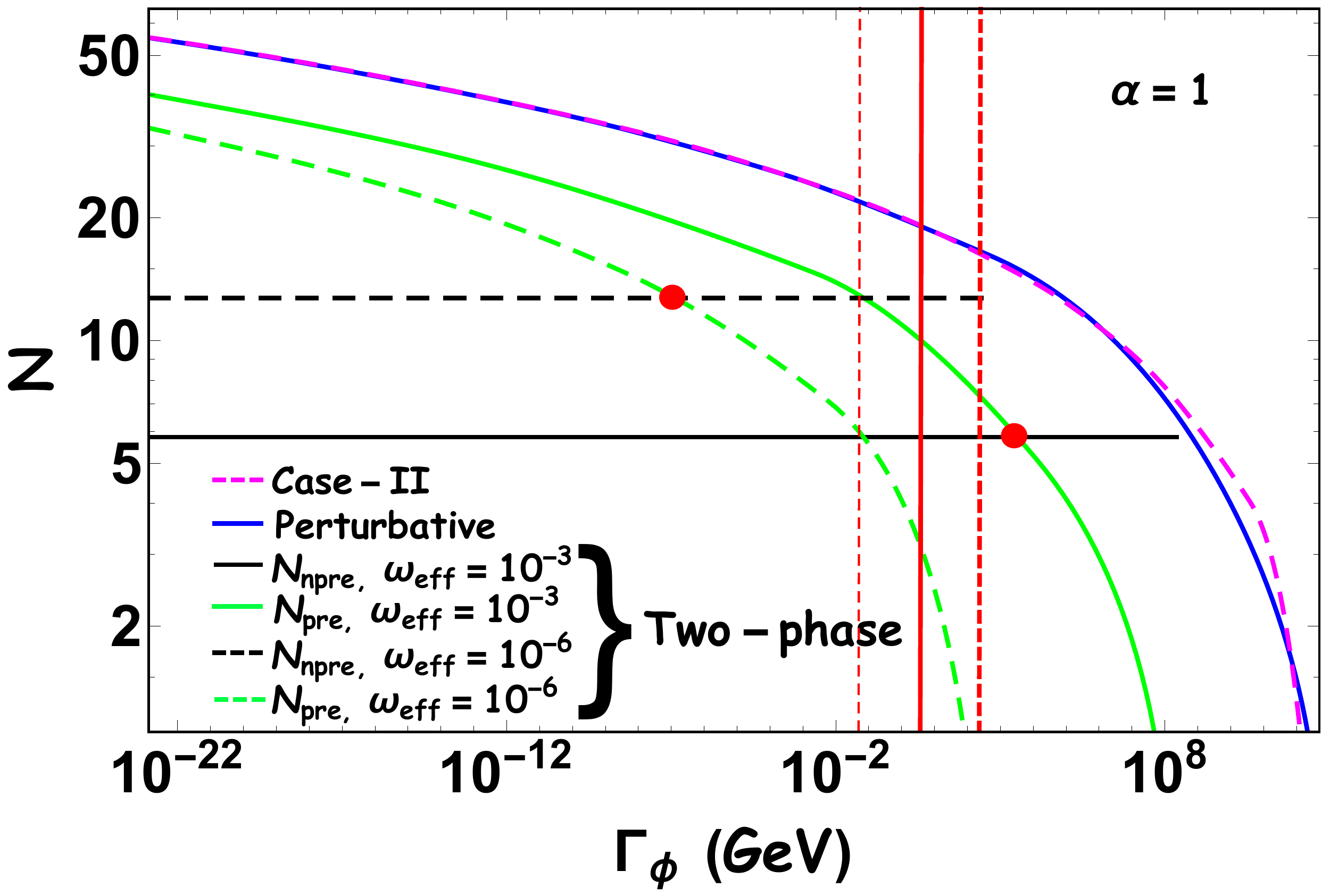}
 		\includegraphics[width=008.1cm,height=5.6cm]{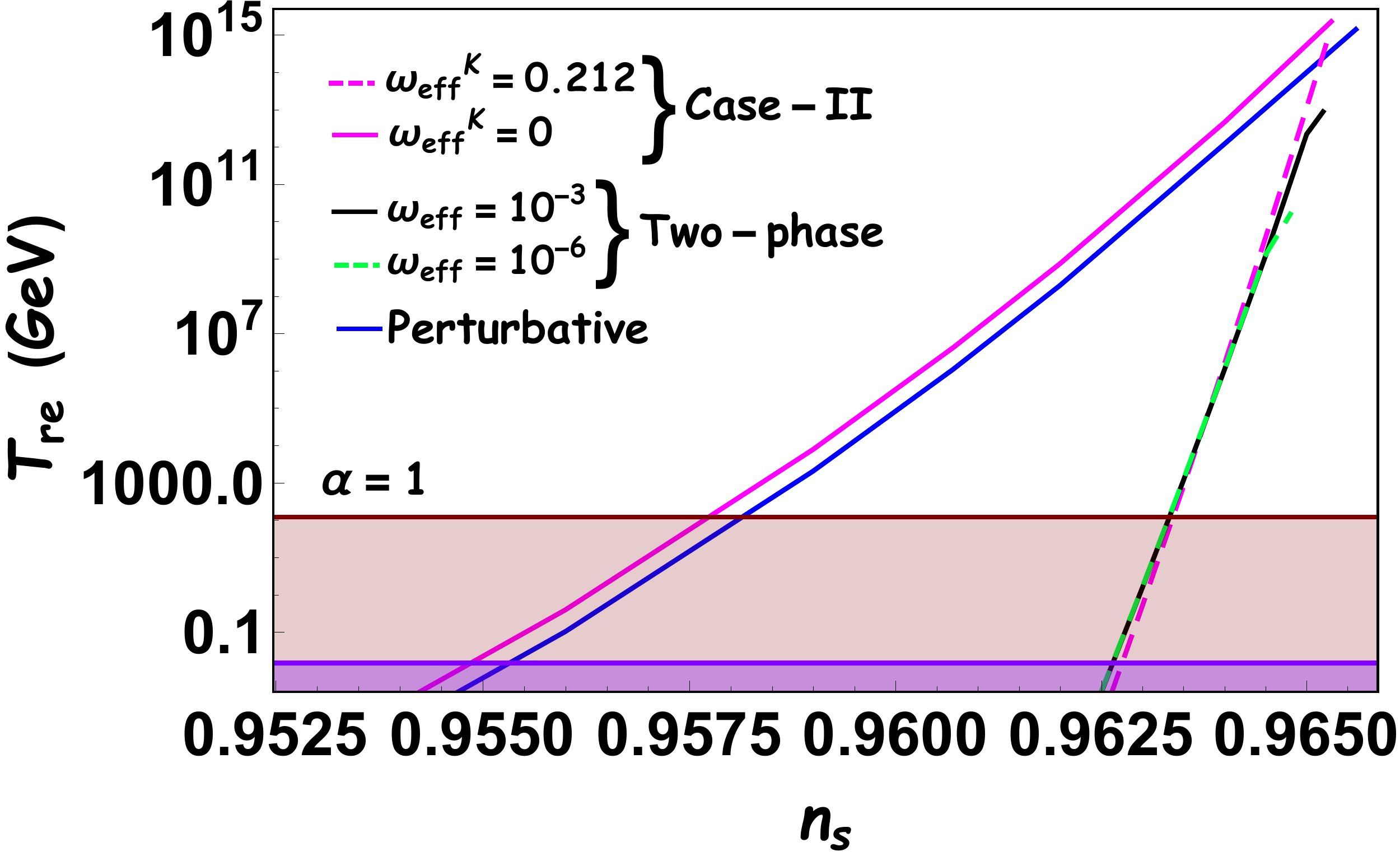}
 		\includegraphics[width=008.1cm,height=5.6cm]{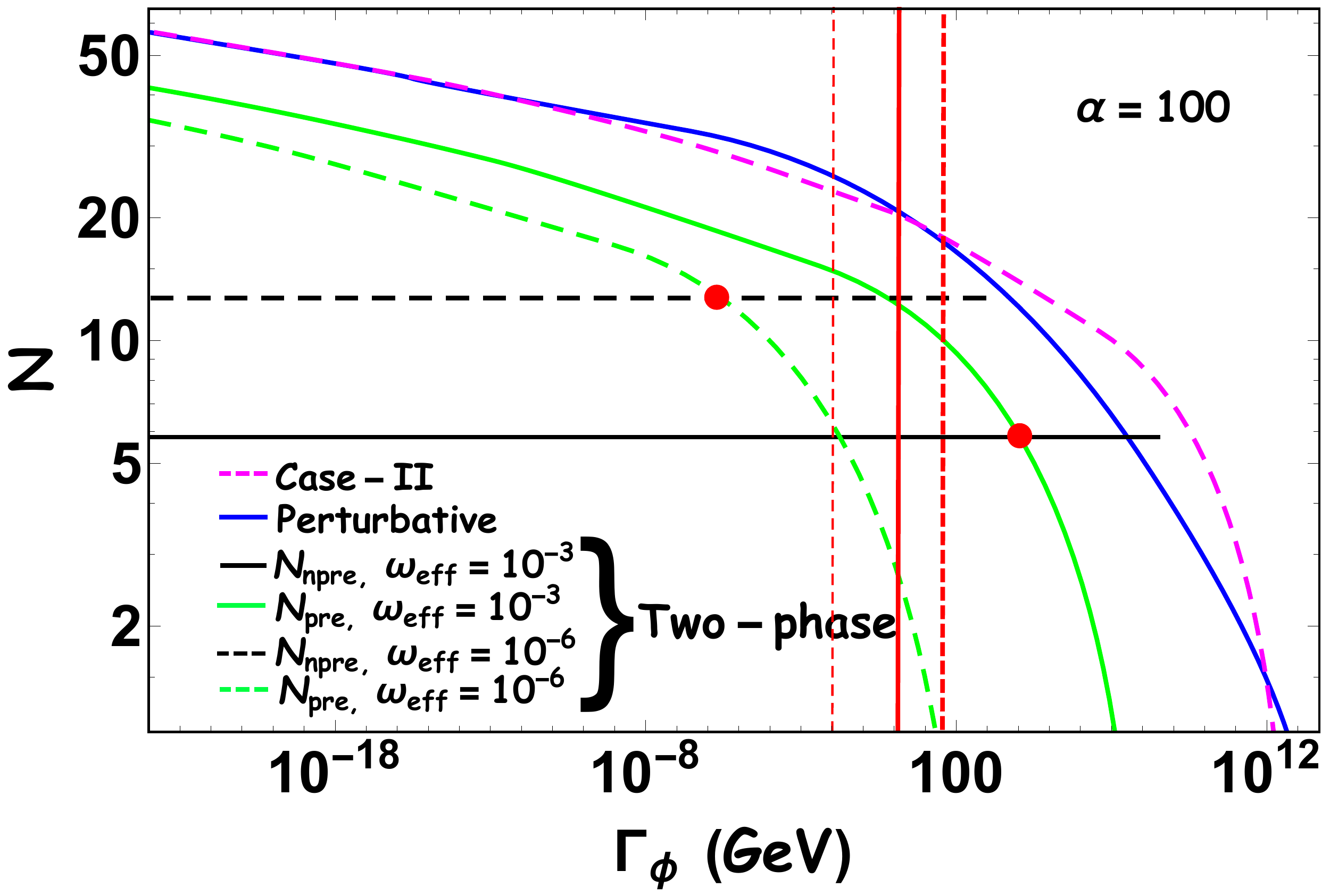}
 		\includegraphics[width=008.1cm,height=5.6cm]{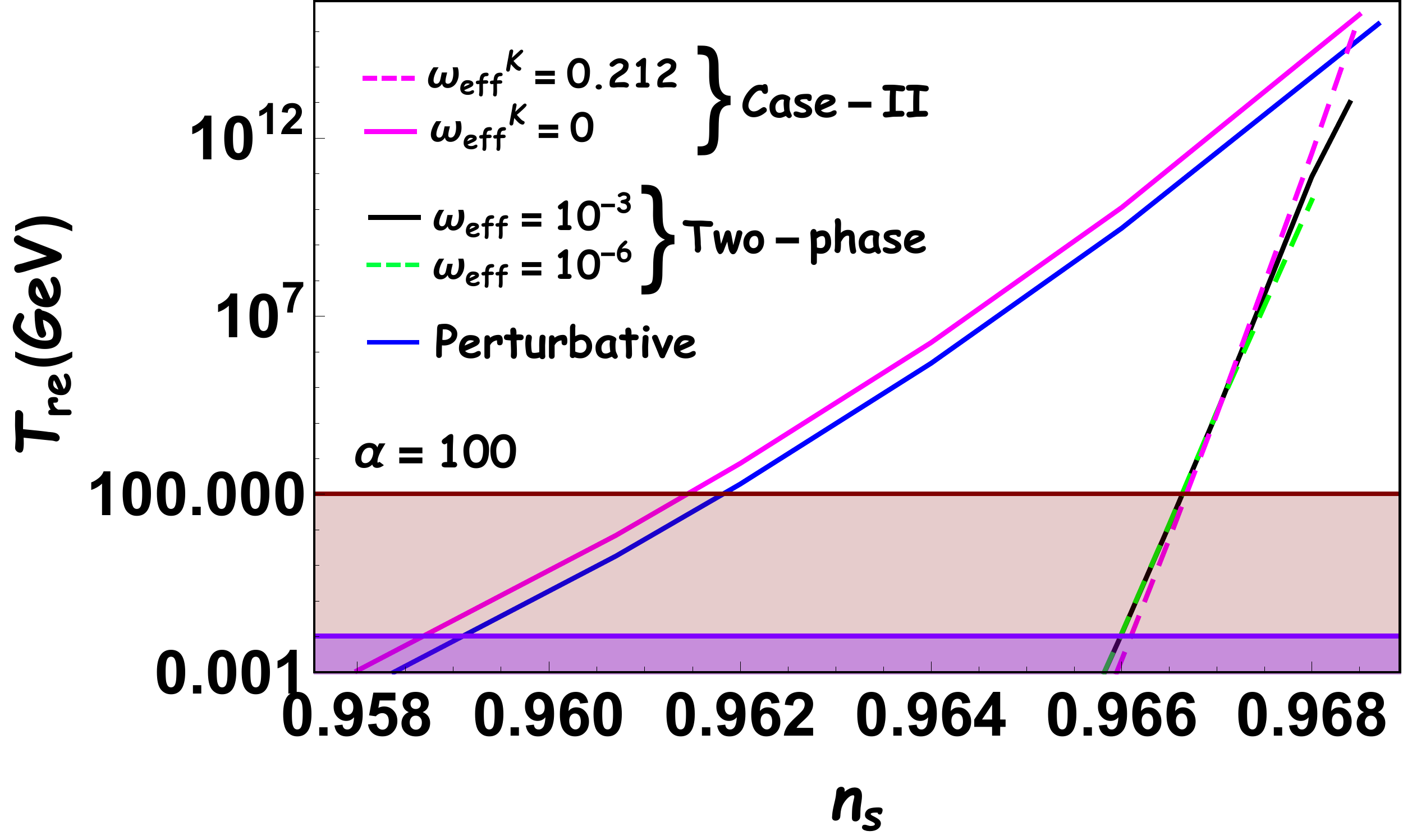}
 		\caption{\scriptsize  All plots are same as in the previous Fig.\ref{chaotic1}. The main difference is that, here we have plotted for $\alpha$-attractor
model for $\alpha = (1,100)$ with $n=1$. However, the plot for $\alpha=1$ and $n=1$ is for Higgs-Starbinsky model.}.
 		\label{alphaplot}
 	\end{center}
 \end{figure}  
This is a new class of models that unifies many of the existing inflationary models in a single framework and was first proposed in \cite{alpha}. This is currently the most favored model from the observational point of view. A class of $\alpha-$ attractor potential, known as the $E-$model, is given as
\begin{equation} \label{a}
V(\phi) = \Lambda^4 \left[  1 - e^{ -     \sqrt{\frac{2}{3\alpha} }    \frac{\phi}{M_p}     } \right]^{2n}.
\end{equation}
Where the mass scale $\Lambda$ is fixed from the CMB power spectrum. An important feature of this class of potential is a large plateau region for the large field value. It also predicts a very low value of the scalar-to-tensor ratio for different $n$ and $\alpha$. However, it is worth noting that for $n=1$, $\alpha=1$, this model reduces to the Higgs-Starobinsky model. So the form of the potential for the Higgs-Starobinsky model is as follows,
\bea \label{sat}
V(\phi)=\beta \left(1 - e^{-\sqrt{\frac{2}{3}} \frac {\phi}{M_p}} \right)^2 ,
\eea
where the dimension full parameter $\beta$ takes the following forms,
\bea
\beta_{S} = \frac {1} {4 \alpha}~~~;~~~
\beta_{H} = \frac  {\lambda M_p^4}{\xi^2} .
\eea
Prefixes, $S, H$ stand for Starobinsky and Higgs model, respectively.
The aforementioned coupling parameters appear in the non-canonical Lagrangian are as follows,  
\bea
&&{\cal L}_{S} = \frac{M_p^2}{2} R_J(1 + \alpha R_J) + \dots \\
&&{\cal L}_{H} =\frac{M_p^2}{2} R_J + 
\frac{2\xi R_J}{M_p^2}h^2 -\frac 1 2 \pr_{\mu}h \pr^{\mu}h - \frac {\lambda}{4} h^4+ \dots , \nno
\eea 
where, $R_J$ is the Ricci scalar in the Jordan frame.
For the Higgs inflation model one assumes $(\xi > 1, h/M_p > 1)$ during inflation. The inflaton degree of freedom $\phi$ in the  eq.\ref{sat}, are expressed as,
\bea
\phi_{S} = \sqrt{\frac{2}{3}} \ln \left(1+ 2 \alpha R_J \right) ~;~
\phi_{H} = \sqrt{\frac{2 }{3}} \ln \left(1+ \frac{\xi h^2} {M_p^2} \right), \nno
\eea 
in unit of $M_p$. For our purpose we have taken two values of $\alpha$(1,100) with $n=1$ and compare their outcomes. 
 \begin{table}[t!]
	\caption{Reheating models and their associated bound on inflationary parameters ($\alpha$-attractor model)}
	Higgs-Starobinsky model ($\alpha=1$)
\begin{tabular}{|p{2cm}|p{2.5cm}|p{2.5cm}|p{2.5cm}|p{2.5cm}|p{4cm}| }
 \hline
 
 Inflationary parameter&\multicolumn{2}{c|}{Case-I (Two-phase)}&\multicolumn{2}{c|}{ Case-II}& ~~Case-III (Perturbative)\\
 \cline{2-6}
 & \quad$\omega_{eff}=10^{-3}$ &\quad$\omega_{eff}=10^{-6}$&\quad $\omega_{eff}^K=0$&\quad $\omega_{eff}^K=0.212$ &\qquad \qquad $\omega_{\phi}=0$\\
 \hline
 $n_s^{min}$&\quad $0.9626$&\quad $0.9626$&\quad $0.9548$&\quad $0.9628$&\qquad \qquad $0.9552$\\
  $n_s^{max}$&\quad $0.9652$&\quad $0.9648$&\quad $0.9653$&\quad $0.9653$&\qquad \qquad $0.9656$\\
  $N_k^{max}$&\quad $55.36$&\quad $54.72$&\quad $55.52$&\quad $55.52$&\qquad \qquad $56.02$\\
 
 \hline
 
 \end{tabular}
$\alpha=100$
\begin{tabular}{|p{2cm}|p{2.5cm}|p{2.5cm}|p{2.5cm}|p{2.5cm}|p{4cm}| }
 \hline
 
 Inflationary parameter&\multicolumn{2}{c|}{Case-I (Two-phase)}&\multicolumn{2}{c|}{ Case-II}& ~~Case-III (Perturbative)\\
 \cline{2-6}
 & \quad$\omega_{eff}=10^{-3}$ &\quad$\omega_{eff}=10^{-6}$&\quad $\omega_{eff}^K=0$&\quad $\omega_{eff}^K=0.212$ &\qquad \qquad $\omega_{\phi}=0$\\
 \hline
 $n_s^{min}$&\quad $0.966$&\quad $0.966$&\quad $0.9587$&\quad $0.9661$&\qquad \qquad $0.959$\\
  $n_s^{max}$&\quad $0.9684$&\quad $0.968$&\quad $0.9685$&\quad $0.9685$&\qquad \qquad $0.9657$\\
  $N_k^{max}$&\quad $56.73$&\quad $56.03$&\quad $56.91$&\quad $56.91$&\qquad \qquad $57.27$\\
 
 \hline
 \end{tabular}
  \label{tablealpha}
\end{table}	

{\bf Initial conditions for phase-I:} Initial coditions to solve the differential equations for the effective non-perturbative era  in the context of present model can be  expressed as,
 \bea
 \varPhi(A=1)= \frac{3}{2} \Lambda^4 \left(\frac{2n}{2n+\sqrt{3 \alpha}}\right)^{2n}~~,~~R(A=1)=\frac{3\left(\omega_{eff}-\omega_\phi\right)}{1-3\omega_{eff}}~\varPhi(A=1)~,
 \eea
 where 
 \bea
 \Lambda= M_p \left(\frac{3\pi^2 r A_s}{2}\right) \left[\frac{2n(1+2n)+\sqrt{4n^2+6\alpha(1+n)(1-n_s)}}{4n(1+n)}\right]^{\frac{n}{2}}~.
 \eea
  
 The inflationary e-folding number, $N_k$ and tensor to scalar ratio, $r_k$ can be written interms of inflationary spectral index($n_s$) as,
  \bea
  N_k=\frac{3\alpha}{4n} \left[e^{\sqrt{\frac{2}{3\alpha}}\frac{\varPhi_k}{M_p}}-e^{\sqrt{\frac{2}{3\alpha}}\frac{\varPhi_{end}}{M_p}}-\sqrt{\frac{2}{3\alpha}}\frac{(\varPhi_k-\varPhi_{end})}{M_p}\right]~,~r_k=\frac{64 n^2}{3\alpha \left(e^{\sqrt{\frac{2}{3\alpha}}\frac{\varPhi_k}{M_p}}-1\right)^2}~~.
  \eea

Furthermore, the initial conditions to solve the Boltzmann equations for different energy components during the perturbative epoch are determined by the spectral index at the ending point of the effective dynamics, $A_{npre}$. All the initial conditions for phase-II will be the same as before, provided in Eq.\ref{phase2cond}.
 
 {\bf Observaions:} We have chosen two sample values of $\alpha = (1, 100)$. With these two values the model dependent critical values of the inflaton decay constant assume $\Gamma_\phi^{cri}(model) =(0.069,5.03,260.3)$ GeV and $(0.01,1.6,42.2)$ GeV for three different kinds of the decay processes. Where as our numerical analysis predicts the critical decay constant to be  $\Gamma_\phi^{cri}=(3.44\times10^{3},1.37\times10^{-7})$ GeV for $\alpha=1$ and $\Gamma_\phi^{cri}=(1.27\times10^{4},3.90\times10^{-6})$ GeV for $\alpha=100$ with $\omega_{eff}=(10^{-3},10^{-6})$ respectively.   Within these values all the model dependent critical decay constant must lie. In addition to that, the reheating temperature connected with the critical value of the inflaton decay constant turns out to be $T_{re}^{cri}\simeq (2.3\times 10^5,3.5\times10^{10})$ $GeV$ for $\alpha=1$ and for $\alpha=100$, $T_{re}^{cri}\simeq(1.2\times10^6,7.2\times10^{10})$ $GeV$ with $\omega_{eff}=(10^{-6},10^{-3})$ accordingly. Similar to the other inflation model discussed above, in the table \ref{tablealpha} the possible constraints on the inflationary parameters can be obtained. 

\subsection{Minimal plateau inflation model\cite{Maity:2019ltu}}\label{minimal}
  \begin{figure}[t!]
 	\begin{center}
		\includegraphics[width=8.1cm,height=5.6cm]{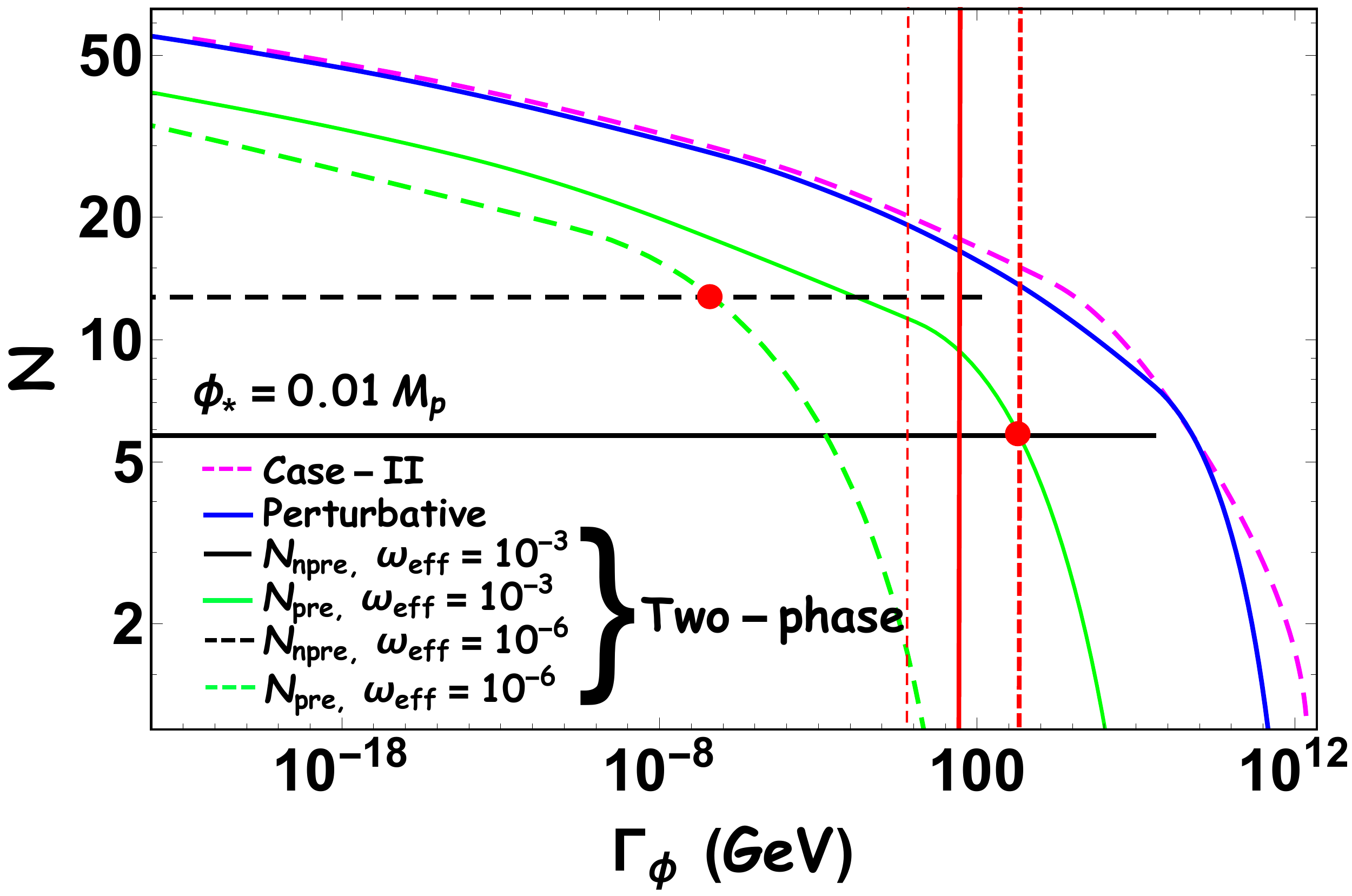}
 		\includegraphics[width=8.1cm,height=5.6cm]{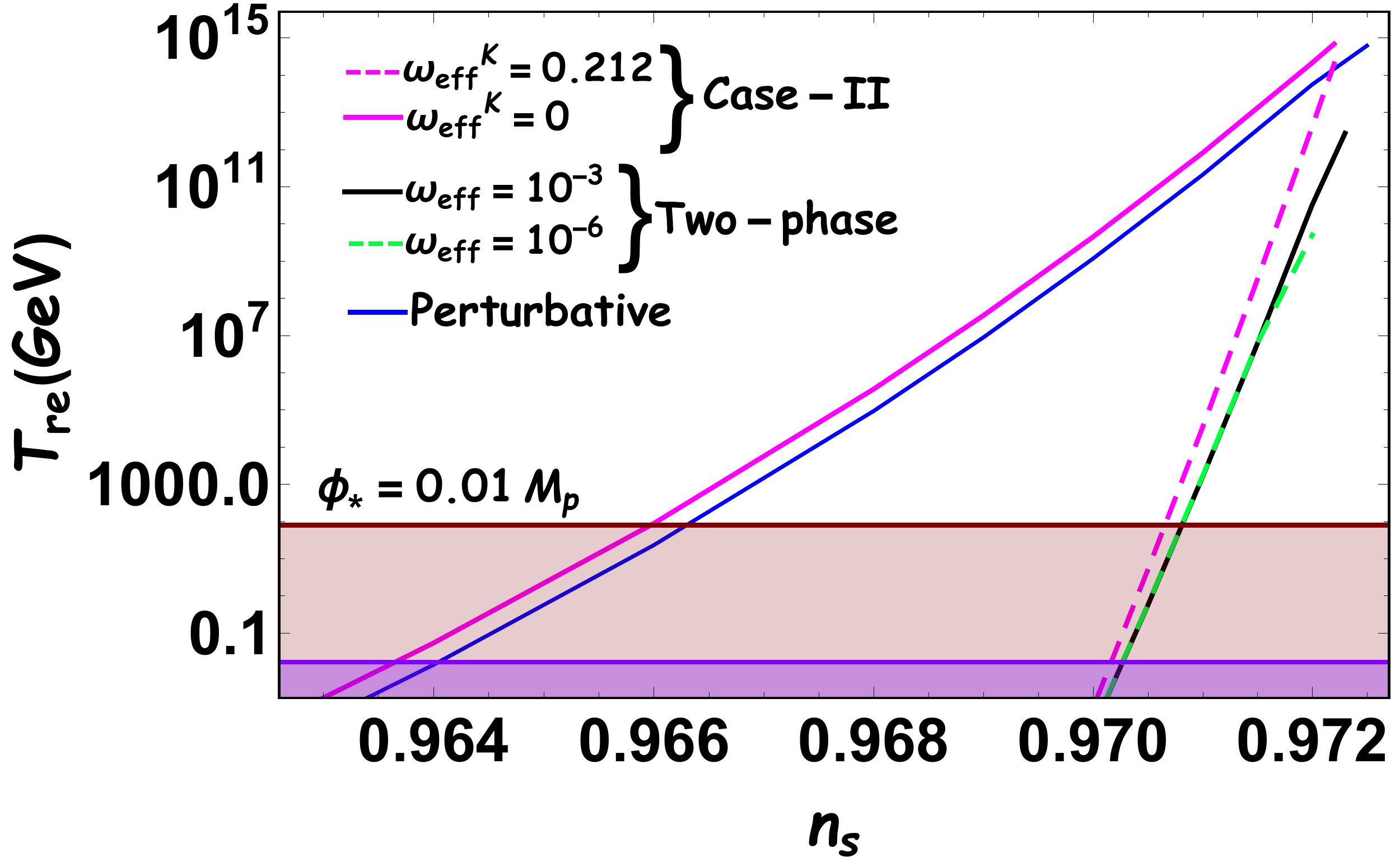}
 		\includegraphics[width=8.1cm,height=5.6cm]{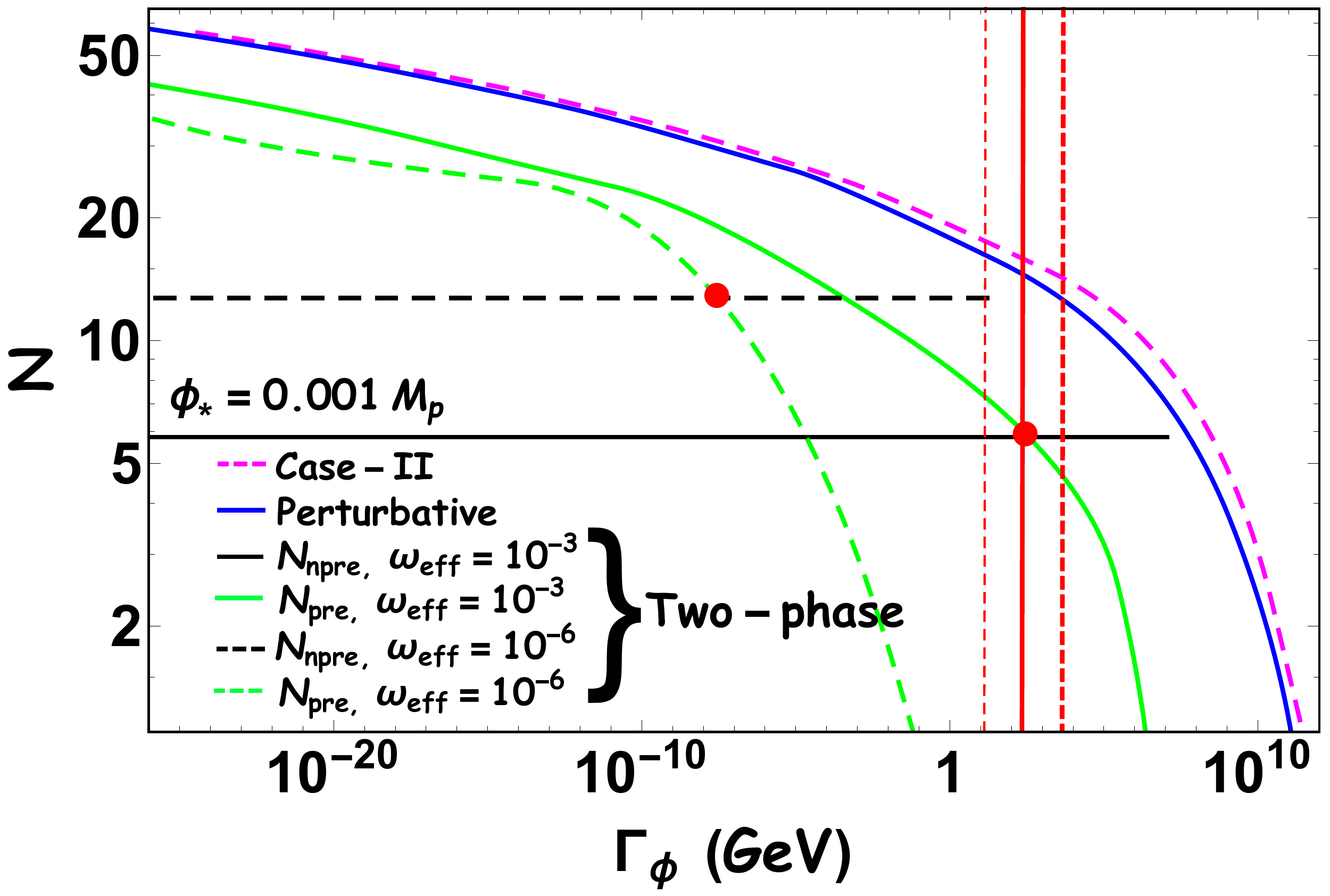}
 		\includegraphics[width=8.1cm,height=5.6cm]{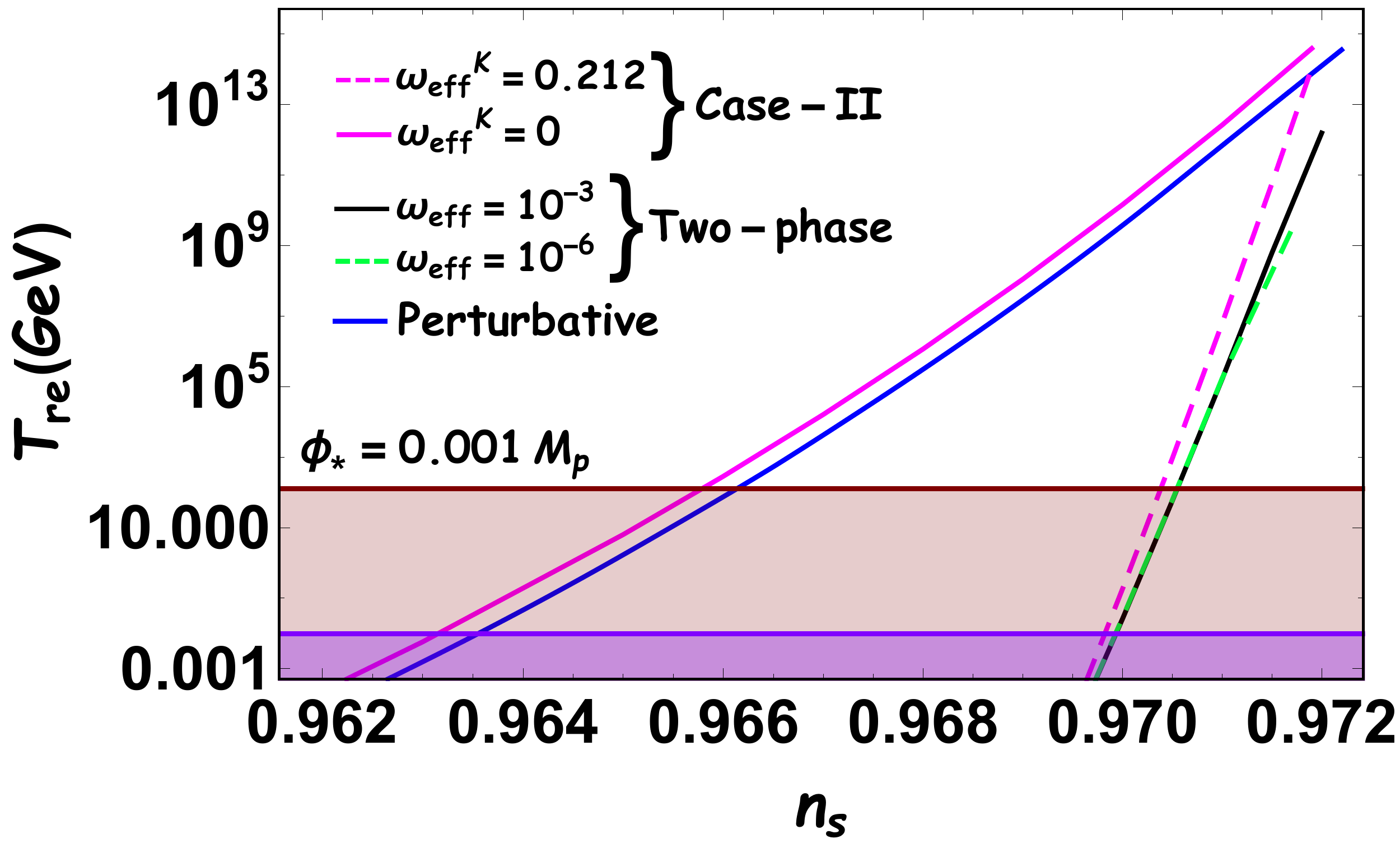}
 		\caption{\scriptsize  All plots are same as in the previous Fig.\ref{chaotic1}. The main difference is that, here we have plotted for minimal inflation model with $\phi_*=(0.01, 0.001)M_p, n=2$.}
 		\label{minimalplot}
 	\end{center}
 \end{figure} 
The minimal plateau inflationary model is a non-polynomial modification of the power-law chaotic potential. The potential for this inflation is given by,
\bea
V_{min}=\Lambda \frac{m^{4-n}\phi^n}{1+\left(\frac{\phi}{\phi_*}\right)^n}~~,
\eea
here $n$, $\Lambda$ and $m$ has the same role as in the power-law chaotic inflation model, and their values are fixed from WMAP normalization \cite{Komatsu:2010fb}. Only even values of $n$ are taken, as in the case of the chaotic inflation model. The new scale of $\phi_*$ controls the shape of the potential. For a wide range of $\phi_*$, this model predicts lower values of scalar-to-tensor ratio for different values of $n$, satisfies the latest PLANCK data \cite{Akrami:2018odb}. For numerical purpose, we consider $\phi_*=(0.001,0.1) M_p$ with $n=2$.
\begin{table}[t!]
	\caption{Reheating models and their associated bound on inflationary parameters (Minimal plateau model)}
	$\phi_*=0.01 M_p$
	\begin{tabular}{|p{2cm}|p{2.5cm}|p{2.5cm}|p{2.5cm}|p{2.5cm}|p{4cm}| }
		\hline
		
		Inflationary parameter&\multicolumn{2}{c|}{Case-I (Two-phase)}&\multicolumn{2}{c|}{ Case-II}& ~~Case-III (Perturbative)\\
		\cline{2-6}
		& \quad$\omega_{eff}=10^{-3}$ &\quad$\omega_{eff}=10^{-6}$&\quad $\omega_{eff}^K=0$&\quad $\omega_{eff}^K=0.212$ &\qquad \qquad $\omega_{\phi}=0$\\
		\hline
		$n_s^{min}$&\quad $0.9703$&\quad $0.9703$&\quad $0.9637$&\quad $0.9702$&\qquad \qquad $0.9640$\\
		$n_s^{max}$&\quad $0.9723$&\quad $0.972$&\quad $0.9722$&\quad $0.9722$&\qquad \qquad $0.9725$\\
		$N_k^{max}$&\quad $54.16$&\quad $53.58$&\quad $53.96$&\quad $53.96$&\qquad \qquad $54.55$\\
		
		\hline
		
	\end{tabular}
	$\phi_*=0.001 M_p$
	\begin{tabular}{|p{2cm}|p{2.5cm}|p{2.5cm}|p{2.5cm}|p{2.5cm}|p{4cm}| }
		\hline
		
		Inflationary parameter&\multicolumn{2}{c|}{Case-I (Two-phase)}&\multicolumn{2}{c|}{ Case-II}& ~~Case-III (Perturbative)\\
		\cline{2-6}
		& \quad$\omega_{eff}=10^{-3}$ &\quad$\omega_{eff}=10^{-6}$&\quad $\omega_{eff}^K=0$&\quad $\omega_{eff}^K=0.212$ &\qquad \qquad $\omega_{\phi}=0$\\
		\hline
		$n_s^{min}$&\quad $0.9700$&\quad $0.9700$&\quad $0.9632$&\quad $0.9698$&\qquad \qquad $0.9636$\\
		$n_s^{max}$&\quad $0.9720$&\quad $0.9717$&\quad $0.9719$&\quad $0.9719$&\qquad \qquad $0.9722$\\
		$N_k^{max}$&\quad $53.57$&\quad $53.00$&\quad $53.38$&\quad $53.38$&\qquad \qquad $53.96$\\
		
		\hline
	\end{tabular}
	\label{tableminimal}
\end{table}

{\bf Initial conditions for phase-I:} The initial conditions are set as,
\bea
\Phi(A=1)=\frac{3}{2}\frac{V_{end}}{m_{\phi}^4}~~,~~R(A=1)=\frac{3\left(\omega_{eff}-\omega_\phi\right)}{1-3\omega_{eff}}~\varPhi(A=1)~,
\eea
where
\bea
V_{end}=\frac{m^{4-n}\phi_{end}^n}{1+\left(\frac{\phi_{end}}{\phi_*}\right)^n}~~,~~m=\left(\frac{3\pi^2M_p^4 r_k A_s}{2\Lambda \phi_k^n}\left(1+\left(\frac{\phi_k}{\phi_*}\right)^n\right)\right)^{\frac{1}{4-n}}~~.
\eea
We set $\Lambda=1$, except for $n=4$. Constraining the parameter $\Lambda$ for $n=4$  has been studied in the context of minimal Higgs inflation in \cite{Maity:2016zeu}. The inflationary parameters $N_k$ and $r_k$ can be written as,
\bea
r_k=\frac{8M_p^2n^2}{\phi^2\left(1+\left(\frac{\phi}{\phi_*}\right)^n\right)^2}~~,~~N_k=\int\limits_{\phi_k}^{\phi_{end}}-\frac{\phi\left(\phi_*^n+\phi^n\right)}{n M_p^2 \phi_*^n }d\phi~~.
\eea  

Similar to the other inflationary models, the initial conditions the second phase boundary condition is set at the normalized scale factor at $A_{npre}$ thought the equation Eq.
\ref{phase2cond}. As we mentioned earlier, our main intention is to see the modification in reheating parameters ($T_{re}, N_{re}$) in comparison with the usual analysis.

{\bf Observations:} In this model the value of $\Gamma_\phi^{cri}(model)$ for three different decay process assume $\Gamma_\phi^{theo}=(0.7,34.2,2749.2)$~GeV for $\phi_*= 0.01$, and $(15.3,231.7,5.8\times10^4)$~GeV for $\phi_*= 0.001 M_p$. As usual those values are obtained from Eqs.(\ref{gamma},\ref{gamma2},\ref{gamma1})
 with $\phi_*=(0.01,0.001) M_p$ accordingly. On the other hand our numerical analysis estimates the value of $\Gamma_\phi^{cri}=(2.3\times10^3,4.8\times10^{-7})$ GeV for $\phi_*=0.01 M_p$, and $\Gamma_\phi^{cri}= (394.7,2.7\times10^{-8})$ GeV for $\phi_*=0.001 M_p$. As discussed for other inflationary scenarios, for each model parameter value of $\phi_*$ two bracketed values of $\Gamma_{\phi}^{cri}$ are calculated for $\omega_{eff}=(10^{-3},10^{-6})$ respectively.   The reheating temperature linked with the decay width $\Gamma_\phi^{cri}$, can be found to be $T_{re}^{cri}\simeq(2.80\times10^5, 2.25\times10^{10})$ GeV  considering $\phi_*=0.01M_p$, and for $\phi_*=0.001M_p$, $T_{re}^{cri}\simeq(5.80\times10^4,8.76\times10^{9})$ GeV with $\omega_{eff}=(10^{-6},10^{-3})$ accordingly. Interestingly, for this minimal inflation scenario a specific choice of $\phi_*=0.01 M_p$, $ \omega_{eff}=10^{-3}$, $\Gamma_\phi^{cri}$ approximately matches with that of $\Gamma_{\phi}^{cri}(model)$ for a specific reheating scenario when inflaton decaying into a pair of fermionic particles with the interaction $\phi {\bar \psi}{\psi}$. Similarly for $\phi_*=0.001 M_p$, $\omega_{eff}=10^{-3}$, we found  $\Gamma_\phi^{cri} \simeq \Gamma_\phi^{cri}(model)$ when reheating dynamics is governed by the inflaton decaying into pair of scalar particles with the interaction $\phi \chi^2$. Associated with the reheating temperature, the bound on the inflationary parameters are given in the table \ref{tableminimal}.
 	

  \section{Constraining the inflaton coupling parameters}\label{bound}
  \begin{figure}[t!]
  	\begin{center}
  		\includegraphics[width=008.1cm,height=6.1cm]{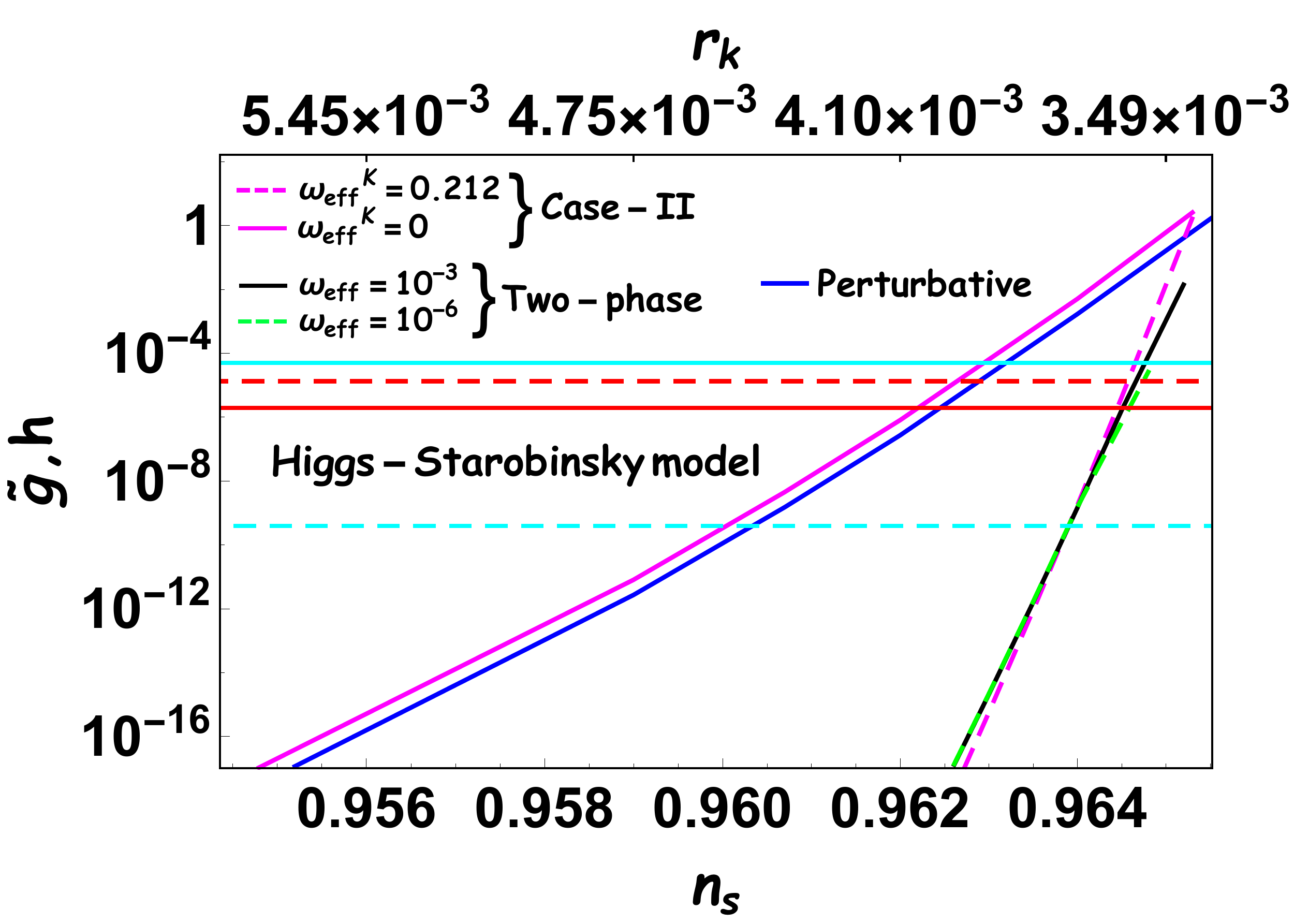}
  		\includegraphics[width=008.1cm,height=6.1cm]{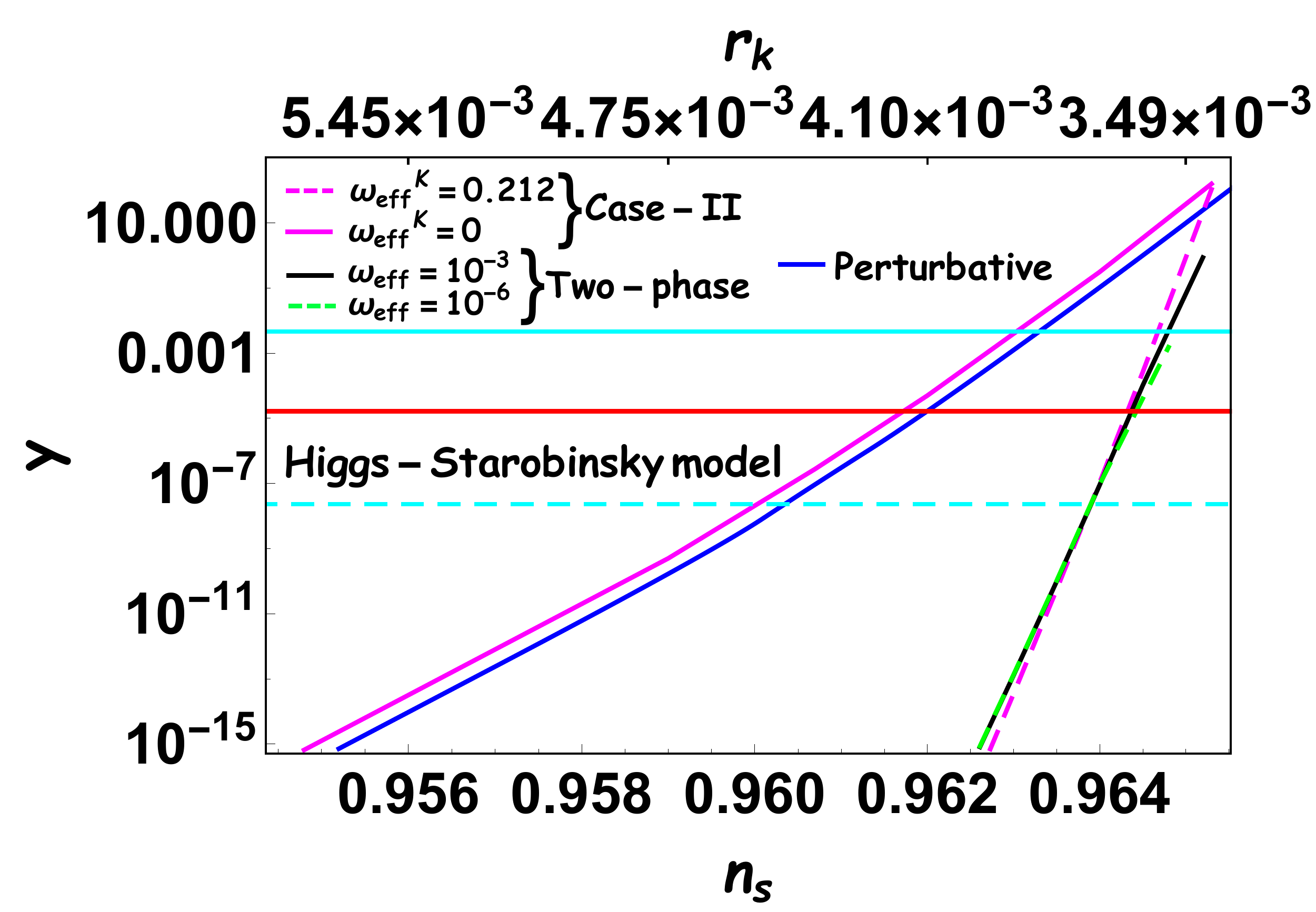}
  		\includegraphics[width=008.1cm,height=6.1cm]{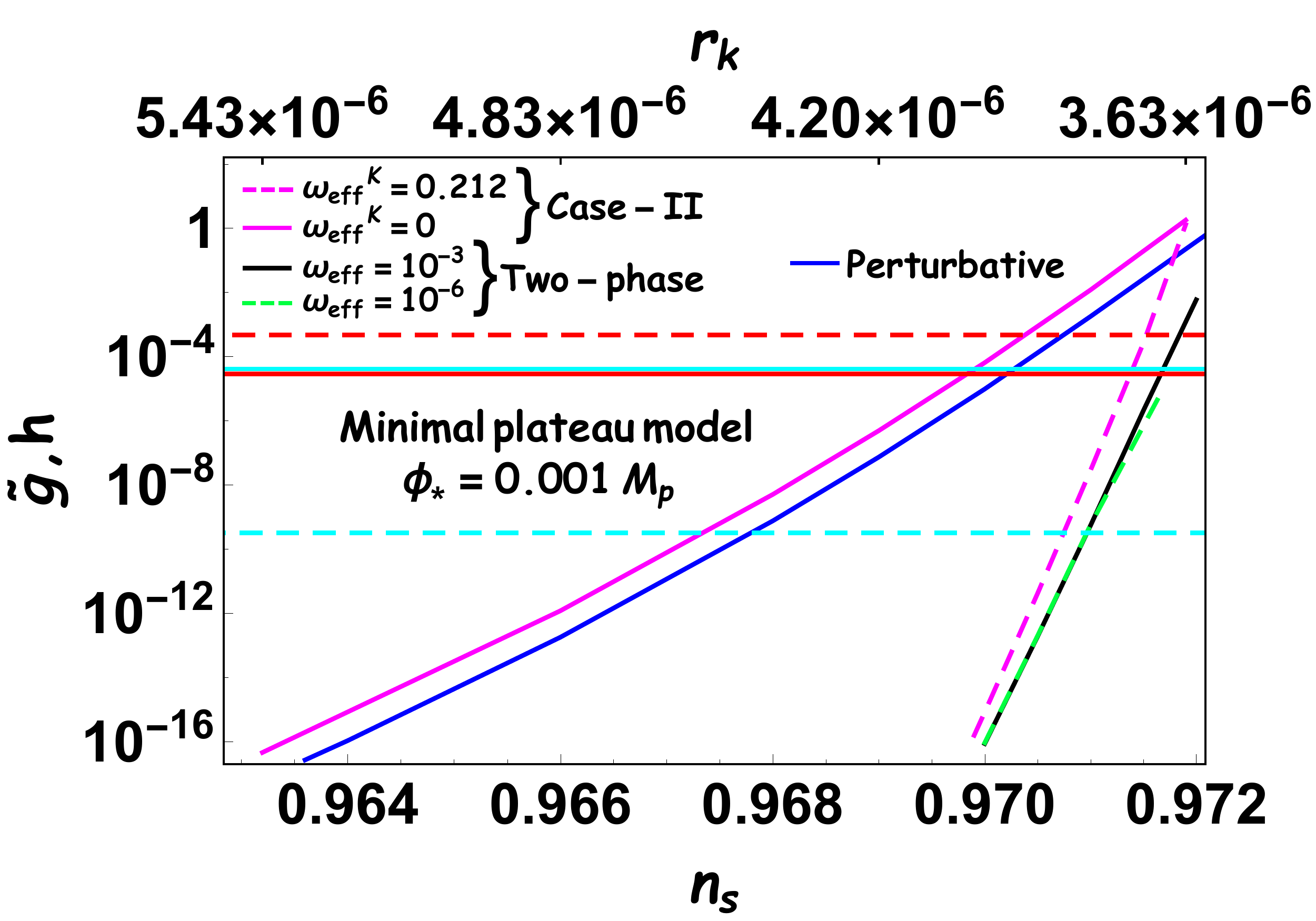}
  		\includegraphics[width=008.1cm,height=6.1cm]{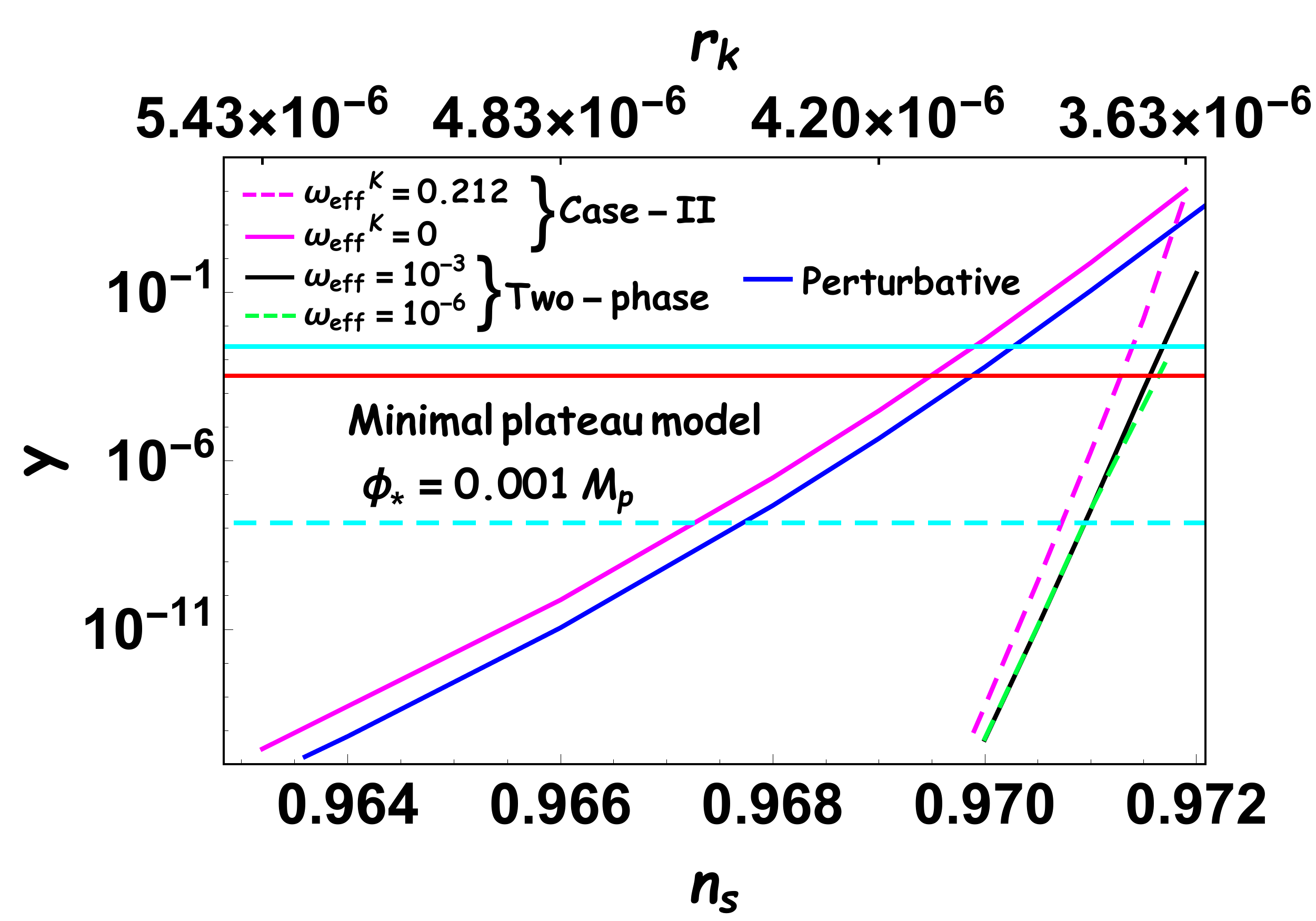}
  		\caption{\scriptsize We have plotted the spectral index dependence of the dimensionless coupling constant $\tilde{g}=\frac{g}{m_\phi}$ with $g\phi\chi^2$ interaction, $y$ with three bodies $y\phi\chi^3$ interaction  and Yukawa coupling with $y\phi\psi\bar{\psi}$ interaction. The upper two plots are for the Higgs-Starobinsky inflation model, and the lower two plots are for the minimal plateau inflation model with $\phi_*=0.001 M_p$, $n=2$. The solid and dashed \textcolor{Rhodamine}{\bf pink line}  corresponds to the usual reheating dynamics given by Kaminkowski et al. \cite{martin} for $\omega_{eff}^K = (0, 0.212)$ respectively. The \textcolor{blue}{\bf solid blue line} indicate the results for perturbative analysis. The results for our developed two-phase reheating mechanism represented by the \textcolor{black}{\bf solid black line} and \textcolor{green}{\bf dashed green line} for $\omega_{eff}=(10^{-3},10^{-6})$ accordingly.  In the first and third plot, the solid and dashed \textcolor{red}{\bf red line} implies the transition point from parametric resonance to perturbative dynamics for two different kinds of interaction, $g\phi\chi^2$ and $y\phi\psi\bar{\psi}$, measured from theoretical constraints provided by the equation (\ref{scalar2}), (\ref{fermion}).  Similarly, in the second and fourth plot, the solid \textcolor{red}{\bf red line} corresponds to the three bodies $y\phi\chi^3$ interaction. Additionally, the solid and dashed \textcolor{ProcessBlue}{\bf sky blue line} indicates the coupling constant at the intersection points of the e-folding numbers, $N_{npre}$ and $N_{pre}$, above which value the effective dynamics start dominating over perturbative dynamics for $\omega_{eff}=(10^{-3}, 10^{-6})$ respectively. All the plots are drawn within the minimum and maximum values of the spectral index. The minimum values of the spectral index $(n_s^{min})$ corresponds to $T_{re}\approx 10^{-2}$ $GeV$ and for maximum values of spectral index $(n_s^{max})$, $N_{re}\approx N_{npre} $ in our analysis and $N_{re}\to 0$ in conventional reheating dynamics.  }
  		\label{coupling}
  	\end{center}
  \end{figure} 
  So far, we have discussed mainly understanding the reheating parameters and their constraints from reheating. In this section we qualitatively translate those results into constraints on coupling parameters $(\tilde{g}= g/m_{\phi}, y, h)$ corresponding to specific inflaton-scalar interactions $\tilde{g} m_{\phi} \phi \chi^2, y \phi \chi^3$, and inflaton-fermion interaction $h\phi \bar{\psi}\psi$ respectively. So far, our analysis was independent of the specific inflaton interaction model. Therefore, the inflaton decay width was a free parameter with one-to-one correspondence with the reheating temperature. Constraining reheating models is very challenging from the perspective of its observational limitations. Therefore, indirect constraints on the inflaton coupling parameters through reheating dynamics would be significant from the model building point of view. Reheating temperature directly estimates the allowed ranges of dimensionless coupling parameter via the inflaton decay constant $\Gamma_\phi$. In this section for illustration only considers two observationally viable inflationary models: Higgs-Starobinsky and minimal plateau models with $n=2$, which are consistent with the current observational bound on $r<0.064$ \cite{Akrami:2018odb}.
  \begin{table}[t!]
	\caption{Reheating models and their associated bound on coupling parameters}
	Higgs-Starobinsky model
\begin{tabular}{|p{2cm}|p{2.5cm}|p{2.5cm}|p{2.5cm}|p{2.5cm}|p{4cm}| }
 \hline
 
 Coupling ~parameter&\multicolumn{2}{c|}{Case-I (Two-phase)}&\multicolumn{2}{c|}{ Case-II}& ~~Case-III (Perturbative)\\
 \cline{2-6}
 & \quad$\omega_{eff}=10^{-3}$ &\quad$\omega_{eff}=10^{-6}$&\quad $\omega_{eff}^K=0$&\quad $\omega_{eff}^K=0.212$ &\qquad \qquad $\omega_{\phi}=0$\\
 \hline
 $\tilde{g}_{min},h_{min}$&\quad $1.31\times 10^{-17}$&\quad $1.38\times 10^{-17}$&\quad $1.06\times 10^{-17}$&\quad $1.13\times 10^{-17}$&\qquad \qquad $1.15\times 10^{-17}$\\
  $\tilde{g}_{max},h_{max}$&\quad $0.01$&\quad $2.55\times10^{-5}$&\quad $2.52$&\quad $1.71$&\qquad \qquad $2.48$\\
  $y_{min}$&\quad $8.07\times 10^{-16}$&\quad $8.47\times 10^{-16}$&\quad $6.51\times 10^{-16}$&\quad $6.98\times 10^{-16}$&\qquad \qquad $7.10\times 10^{-16}$\\
 $y_{max}$&\quad $0.85$&\quad $1.50\times 10^{-3}$&\quad $155.10$&\quad $105.31$&\qquad \qquad $152.76$\\
 \hline
 
 \end{tabular}\\[.2cm]
Minimal plateau model ($\phi_*=0.001 M_p$)\\[.1cm]
\begin{tabular}{|p{2cm}|p{2.5cm}|p{2.5cm}|p{2.5cm}|p{2.5cm}|p{4cm}| }
 \hline
 
 Coupling ~parameter&\multicolumn{2}{c|}{Case-I (Two-phase)}&\multicolumn{2}{c|}{ Case-II}& ~~Case-III (Perturbative)\\
 \cline{2-6}
 & \quad$\omega_{eff}=10^{-3}$ &\quad$\omega_{eff}=10^{-6}$&\quad $\omega_{eff}^K=0$&\quad $\omega_{eff}^K=0.212$ &\qquad \qquad $\omega_{\phi}=0$\\
 \hline
 $\tilde{g}_{min},h_{min}$&\quad $8.74\times 10^{-17}$&\quad $9.27\times 10^{-17}$&\quad $4.62\times 10^{-17}$&\quad $1.58\times 10^{-16}$&\qquad \qquad $2.70\times 10^{-17}$\\
  $\tilde{g}_{max},h_{max}$&\quad $5.90\times 10^{-3}$&\quad $1.19\times10^{-5}$&\quad $1.79$&\quad $1.30$&\qquad \qquad $1.12$\\
  $y_{min}$&\quad $5.38\times 10^{-15}$&\quad $5.71\times 10^{-15}$&\quad $2.84\times 10^{-15}$&\quad $9.73\times 10^{-15}$&\qquad \qquad $1.66\times 10^{-15}$\\
 $y_{max}$&\quad $0.36$&\quad $7.37\times 10^{-4}$&\quad $110.26$&\quad $80.31$&\qquad \qquad $68.93$\\
 \hline
 
 \end{tabular}
  \label{tablecoupling}
\end{table}	
  
  {\bf Bounds on couplings:} The constraints on  different coupling constants are shown in figure \ref{coupling}. Plots show how the dimensionless coupling parameter $\tilde{g}$, $h$, and $y$ are intimately linked with CMB anisotropy via the inflationary observables such as $(n_s, r_k)$ for different types of reheating dynamics. The mapping $T_{re}\rightarrow \Gamma_{\phi} \rightarrow  (\tilde{g}, y, h)$ are directly followed from Eqs.(\ref{decay2}, \ref{decay3}, \ref{decay4}, \ref{reheating 2}). From these equations, we obtain the constraints on the coupling parameters with respect to the inflationary parameters. 
  Any realistic scenario of reheating should include all possible inflaton coupling based on underlying symmetry. Therefore, the assumption of a specific inflaton coupling's contribution to be the dominant one throughout the entire reheating period may not be relevant. Hence, a more pragmatic approach would be to construct particle physics motivated models which we left for our future study. However, as a toy model analysis, the present study may guide us in building scenarios that include all the standard model fields.
  Nevertheless based on our reheating discussions so far, we compare the constraints for all the cases. To this end let us point out that in terms of mathematical expression, the decay width $\Gamma_{\phi}$ associated with the coupling parameters $\tilde{g}$ and $h$ are same. Therefore, for each model under consideration we have two different figures in the $(\tilde{g}/h, n_s)$ and $(y, n_s)$ space. Given the observation from CMB temperature anisotropy, the coupling parameters for the Higgs-inflation model, which are assumed to be responsible for entire reheating process, are found to be constrained  within $1.31\times10^{-17}\leq(\tilde{g},h)\leq0.01$ ({\bf solid black curve}) and $1.38\times10^{-17}\leq (\tilde{g}, h)\leq2.55\times10^{-5}$ (\textcolor{green}{\bf dotted green curve}) for two different values of effective equation state $\omega_{eff}=(10^{-3},10^{-6})$ respectively. Where as for the same values of the effective equation of state the coupling constant $y$ for three body  interaction $(y\phi\chi^3)$ lies within $8.07\times10^{-16}\leq y \leq0.85$ ({\bf solid black curve}), and  $8.47\times10^{-16}\leq y \leq1.50\times10^{-3}$ (\textcolor{green}{\bf dotted greed curve}). Important but straightforward to note that the largest values of the coupling constant  $\tilde{g}_{max}/h_{max} =  (0.01, 2.55\times10^{-5}) $, and $y_{max} = (0.85,1.50\times10^{-3})$ correspond to maximum reheating temperature $T_{re}^{max} = (10^{10}, 10^{13})$ GeV respectively.
  Reemphasizing the fact that two different limiting values of coupling constants are realizable only in the high-temperature limit for two different $\omega_{eff}$. All the above estimates are for the two-phase reheating process (case-I). For the other two scenarios, the bounds on the coupling constant can be read from the table
  \ref{tablecoupling}. 
  
 The interesting interplay among the inflationary theory parameters and the emergent reheating parameters governed by the CMB anisotropy gives important constraints on the theory itself. Apart from having the maximum possible values of the coupling constants, compatible with CMB observations, there exists a critical value of the same born out of $\Gamma_{\phi}^{cri}$, which entails whether the reheating is perturbative or non-perturbative phase dominated. For minimal plateau model, we found $(\tilde{g}_{cri}\approx3.46\times10^{-5})$ for $\omega_{eff}=10^{-3}$, which closely matches with the associated perturbative constraints $\tilde{g}_{cri}(model)\approx2.38\times10^{-5}$. 
 For the Higgs-Starobinsky inflation model, $h_{cri}\simeq5.3\times10^{-5}$ for $\omega_{eff}=10^{-3}$  and the associated perturbative constraints for Yukawa interaction is $h_{cri}(model)\simeq1.33\times10^{-5}$. 
 Therefore, we can infer from this observation that our two-phase reheating scenario essentially captures the necessary features of the non-perturbative phase.

So far, we have discussed the reheating dynamics considering inflaton and radiation as the two dynamical components. However, as we all know, dark matter is another important constituent of our present universe. One of this component's important properties is that it's coupling with the standard model fields must be very weak. Apart from this, not much is known about its other fundamental properties, such as charge, mass, and coupling. Experimental searches of this particle are going on across the globe without much success till now. 
The searches include both directly as well as indirectly observing the properties of this object and, finally, jointly constrain the parameter region. This paper will study the dark matter phenomenology based CMB parameter space following our previous work \cite{Maity:2018exj}. We essentially generalize our two-phase reheating formalism and include the dark matter as the third dynamical matter component.  

\section{Unifying the dark sector}\label{dark}
In the previous section, we discussed the two-phase reheating process, where inflaton decays only into radiation. In the present discussion, we add additional dark matter components and discuss the impact on dark matter phenomenology. The assumption is that inflaton decays into radiation and then radiation to dark matter. The methodology of the analysis will be the same as before, except the new additional dynamical equations for dark matter. 

{\bf phase-I}:({\em Effective non-perturbative phase}) dynamics is governed by  
\bea \label{density2}
\rho_t =\rho_{\phi} +\rho_{R}+\rho_{X}=  \rho_{end} \left(\frac{a_{end}}{a}\right)^{3(1+w_{eff})}~,
\eea
where the new component $\rho_{X}$ is the energy density of the dark matter particle with mass $M_X$ and energy of the dark matter is expressed as $\left\langle E_X \right\rangle=\sqrt{M_X^2+9T^2}$ \cite{Giudice:2000ex}. $T$ is the temperature. The above equation can be written in differential form as,
\bea\label{nonper2}
\dot{\rho_\phi} + \dot{\rho_R}+\dot{\rho_X} + 3 H (1+w_{eff}) (\rho_\phi + \rho_R +\rho_X)= 0~.
\eea
Besides the above equation, we consider additional conservation equation characterizing the dynamics of every individual energy components during this phase as,
\bea\label{nonper4}
\dot{\rho_\phi} +3 H (1+w_{\phi}) \rho_\phi + \dot{\rho_R}+4H\rho_R+\dot{\rho_X} +3H\rho_X= 0~.
\eea
 To solve the above equations (\ref{nonper4}) and (\ref{nonper2}), we need one more condition. We define the ratio of the dark matter and the radiation energy density as $\gamma = \frac{\rho_X}{\rho_R}$. After combining the above two equations one finds,
\bea \label{weff}
\frac{\rho_R}{\rho_\phi+\rho_R+\rho_X}=\frac{\rho_R}{\rho_\phi+\rho_R+\gamma\rho_R}=\frac{3\left(\omega_\phi-\omega_{eff}\right)}{3~\omega_\phi\left(1+\gamma\right)-1}~~.
\eea
At the initial stage of the reheating, radiation energy density must be very small $\rho_R \simeq 0$. Hence, as discussed extensively for the two component reheating, here also  $\omega_{eff}$ must assume the value very closed to the inflaton equation of state $\omega_\phi$, at least near the beginning. In terms of dimensionless variable this phase can be written as
\bea\label{nonper1}
 \frac{\varPhi'}{A^{3 w_\phi}}+ \frac{R'}{A}+\frac{\langle E_X\rangle X'}{m_\phi}= 0~,
 \eea
 \bea\label{nonper3}
  \frac{\varPhi'}{A^{2+3 w_\phi}}+ \frac{R}{A^4} [3 (1+w_{eff})-4]+\frac{3\langle E_X \rangle X}{m_\phi A^3}\omega_{eff}+ \frac{R'}{A^3}+ \frac{3 \varPhi (w_{eff}-w_\phi)}{A^{3(1+w_\phi)}}+\frac{X'\langle E_X \rangle}{m_\phi A^2}=0~,
 \eea
 here the dimensionless dark matter density $X=\frac{\rho_X}{\langle E_X\rangle} a^3$.

{\bf phase-II} ({\em perturbative phase}) The subsequent perturbative phase will now be governed by two more parameters related to the dark matter component. Apart from the inflaton equation of state $\omega_\phi^1$ and the inflaton decay constant $\Gamma_{\phi}$, we have a thermal average of dark matter annihilation cross-section $\langle \sigma v\rangle$, and the dark matter mass $M_X$. The corresponding dimensionless comoving energy densities' dynamics will be governed by the Boltzmann equation  \cite{Maity:2018dgy}.
    \bea
   {\Phi'} &=& -c_1 \frac{A^{1/2}\Phi}{\sqrt{\frac{\Phi}{A^{3\omega_\phi^1}}+ \frac{R}{A}+ \frac{X \langle E_X\rangle}{m_\phi}}}~~,\\
     {R'} &=& c_1 \frac{A^{\frac{3(1-2\omega_\phi^1)}{2}}\Phi}{\sqrt{\frac{\Phi}{A^{3\omega_\phi^1}}
   		+ \frac{R}{A}+ \frac{X \langle E_X\rangle}{m_\phi}}} + c_2 \frac{A^{-3/2}\langle \sigma v\rangle 2\langle E_X\rangle M_{pl} }{\sqrt{\frac{\Phi}{A^{3\omega_\phi^1}}+ \frac{R}{A}+\frac{ X \langle E_X\rangle}{m_\phi}}}\left(X^2-X_{eq}^2\right)~~,\\ 
      {X'}&=& - c_2 \frac{A^{-5/2}\langle \sigma v\rangle M_{pl} m_\phi }{\sqrt{\frac{\Phi}{A^{3\omega_\phi^1}}+ \frac{R}{A}+ \frac{X \langle E_X\rangle}{m_\phi}}}\left(X^2-X_{eq}^2\right)~~.
   \eea
 The equilibrium number density of the dark matter particle can be described in terms of the modified Bessel function of the second kind \cite{Giudice:2000ex}
\bea
n_X^{eq}= \frac{g T^3}{2 \pi^2} \left(\frac{M_X}{T}\right)^2 K_2 \left(\frac{M_X}{T}\right)~,
\eea
 and the constants $c_1$ and $c_2$ are delineate as,
 \begin{equation}
  c_1= \frac{\sqrt{\frac{3}{8\pi}} M_{pl} \varGamma_\phi}{m_{\phi}^2}~,~c_2=\sqrt{\frac{3}{8\pi}}~~.
  \end{equation}
 We consider fermionic type dark matter particles with internal degrees of freedom $g$. 
 
{\bf Initial conditions:} The general form of the initial conditions during the first phase of reheating (phase-I) are,
\bea
 \Phi(1)=\frac{3}{2}\frac{V(\phi_{end})}{m_\phi^4}~,~R(1)=\frac{3\left(\omega_{eff}-\omega_{\phi}\right)}{1-3~\omega_{eff}\left(1+\gamma\right)}\Phi(1)~,~X(1)=\frac{\gamma~ m_\phi}{\langle E_X\rangle}R(1)~~.
 \eea
 The initial values of the energy densities for the phase-II will be 
 set at the normalized scale factor $A_{npre}$ as 
 \bea
 \varPhi = \varPhi(A_{npre})~~;~~ \frac{R(A_{npre})}{R(A_{npre}) +\varPhi(A_{npre}) + X(A_{npre})} \simeq \frac 12~~;~~X(A_{npre}) = \frac{\gamma~ m_\phi}{\langle E_X\rangle}R(A_{npre})~~. 
 \eea
As described in detail in section \ref{sec5}, radiation energy density is again assumed to be $50\% $ of the total comoving energy density right after the completion of phase-I. Therefore, the dark matter number density will automatically be fixed for a given $\gamma$ value. All the required equation of states for two different phases are assumed to take the same approximate values  $\omega_\phi^1 \simeq 0.2$ and $\omega_\phi \simeq 0$. The methodology of solving the dynamics will be the same as before except some additional constraints in the dark sector after the end of reheating. 

{\bf Boundary condition from observations :} The condition for ending the reheating dynamics is set by the following equation, 
 \bea
 H^2= \frac{ \rho_\phi(\varGamma_\phi,N_{re},n_{s}^k)+ \rho_{R}(\varGamma_\phi,N_{re},n_{s}^k))+\rho_{X}(\varGamma_\phi,N_{re},n_{s}^k)}{3M_p^2}=\varGamma_{\phi}^2~~~~.
 \eea 
 supplemented with the observational constraint namely the relation between reheating temperature follows from the above equation and present CMB temperature $T_0 = 2.7 K \simeq 2.35\times10^{-13}~GeV$ though the relation Eq.\ref{reheating 3}. Further additional observational constraint is the observed value of the dark matter abundance defined as $\Omega_X$ \cite{Jarosik:2010iu,Aghanim:2018eyx}
 \bea
 \Omega_X h^2= \langle E_X\rangle \frac{X(T_F)~ T_F ~A_F}{R(T_F) ~T_{0} ~m_\phi} \Omega_R h^2~ = 0.1199\pm 0.0022,
 \eea
which is expressed in terms of radiation  abundance $\Omega_R$ ($\Omega_R h^2=4.3\times10^{-5}$). $T_F$ is the temperature at very late time when both dark-matter and radiation energy components become stationary. While solving the Boltzmann equations during perturbative reheating (phase-II), these condition will constrain the dark matter parameter $\langle\sigma v\rangle$ (thermal average of the cross-section times velocity) for a fixed value of the dark matter mass, $M_X$, and the inflaton decay constant in terms reheating temperature. The detailed analysis only on phase-II has already been done in \cite{Maity:2018dgy} including the dark matter phenomenology. Nevertheless 
we only consider dark matter production  via freeze-in mechanism. This mechanism indicates that the dark matter will never reach equilibrium with the thermal bath. This kind of dark matter is known as FIMP (feebly interacting dark matter) \cite{Tenkanen:2016twd}-
\cite{Hall:2009bx}.  We can illustrate the production of dark matter via Freeze-in mechanism   through the heavy mediator during reheating is sensitive  to the early history of the universe before the UV dominated era \cite{Mambrini:2013iaa,Garcia:2018wtq,Dutra:2018gmv,Allahverdi:2018iod,Arias:2019uol,Bernal:2019uqr,Bernal:2019mhf,Heurtier:2019eou,Miller:2019pss,Ahmed:2020fhc}.

{\bf Physical constraints:} Further constraints on the dark matter parameter space will be inherited if one considers various theoretical limits on the scattering cross-section. Cross-section can not be arbitrarily large. Perturbative  unitarity usually limits the cross-section $\langle\sigma v\rangle$ in term of mass, $\langle \sigma v\rangle_{max}=\frac{8\pi}{M_X^2}$ \cite{Griest:1989wd}, which are shown by \textcolor{Rhodamine}{pink solid lines} in Figs.(\ref{alphadarkmatterplot}, \ref{minimaldarkmatterplot}). On the other hand, we will also have another bound on the cross-sections coming from the fact that during reheating dark matter production peaks around the temperature of $T_*=\frac{M_X}{4}$ \cite{Giudice:2000ex}. This provides a natural condition on the dark matter number density $n_X(T)<n_X^{eq}(T_*)$ as for $T<T_*$ the dark matter production would be frozen, and it must be diluted subsequently due to the expansion of the universe. Aforementioned condition on the dark matter number density sets an upper bound on the cross-section $\langle \sigma v\rangle \approx \langle \sigma v\rangle_{T=T_*}$ \cite{Giudice:2000ex} \cite{Fornengo:2002db}
\bea\label{upperbound}
\langle\sigma v\rangle_*\leq 7\times 10^{-14}\left(\frac{2}{g}\right)\left(\frac{g_*(T_*)}{10}\right)\left(\frac{10}{g_*(T_{re})}\right)^{\frac{1}{2}}\left(\frac{M_X}{10GeV}\right)\left(\frac{100 MeV}{T_{re}}\right)^2GeV^{-2}~~.
\eea
We call it as {\bf reheating bound} in the plot. This condition is depicted by {\bf black  solid lines} and {\bf black dotted lines} for perturbative and two-phase reheating scenarios respectively in Figs.(\ref{alphadarkmatterplot}, \ref{minimaldarkmatterplot}). We have shown both of these bounds in the subsequent plots for different inflation model once we fixed the dark matter mass. {\em Interesting observation that can be made from this theoretical constraint is that given a dark matter mass, the perturbative unitarity bound and the dynamical condition Eq.\ref{upperbound} modify the possible range of allowed $n_s$ values obtained from the previous analysis. This can directly shed light on the inflationary model building. Conversely, one can state that for a given inflationary model, CMB can shed light on the possible nature of the dark matter candidate via reheating phase}.

Nevertheless, unifying the dark sector into a single reheating framework is the primary motivation of this section. The basic philosophy is to look into further constraints on the dark-matter parameter space in a more realistic framework of two-phase reheating dynamics and compare with that of the usual perturbative reheating analysis ~ {Maity:2018dgy}. An important outcome is the constraints dut to CMB temperature anisotropy. Particularly, constraints imparted on the dark matter and inflationary parameter space $(\langle\sigma v\rangle-n_s)$ by the CMB anisotropy could enable us to constrain the viable inflationary models though dark matter observable. Conversely, given a viable inflationary model, CMB can potentially shed light on the possible properties of dark matter. keeping this in mind, we study dark matter phenomenology considering two observationally viable inflationary models: Higgs-Starobinsky and minimal plateau models, which are consistent with the current observational bound on $r<0.064$ \cite{Akrami:2018odb}.

\subsection{ Higgs-Starobinsky model  \cite{Bezrukov:2007ep}\cite{Starobinsky:1980te} and dark matter phenomenology}
\begin{figure}[t!]
 	\begin{center}
		\includegraphics[width=005.41cm,height=03.9cm]{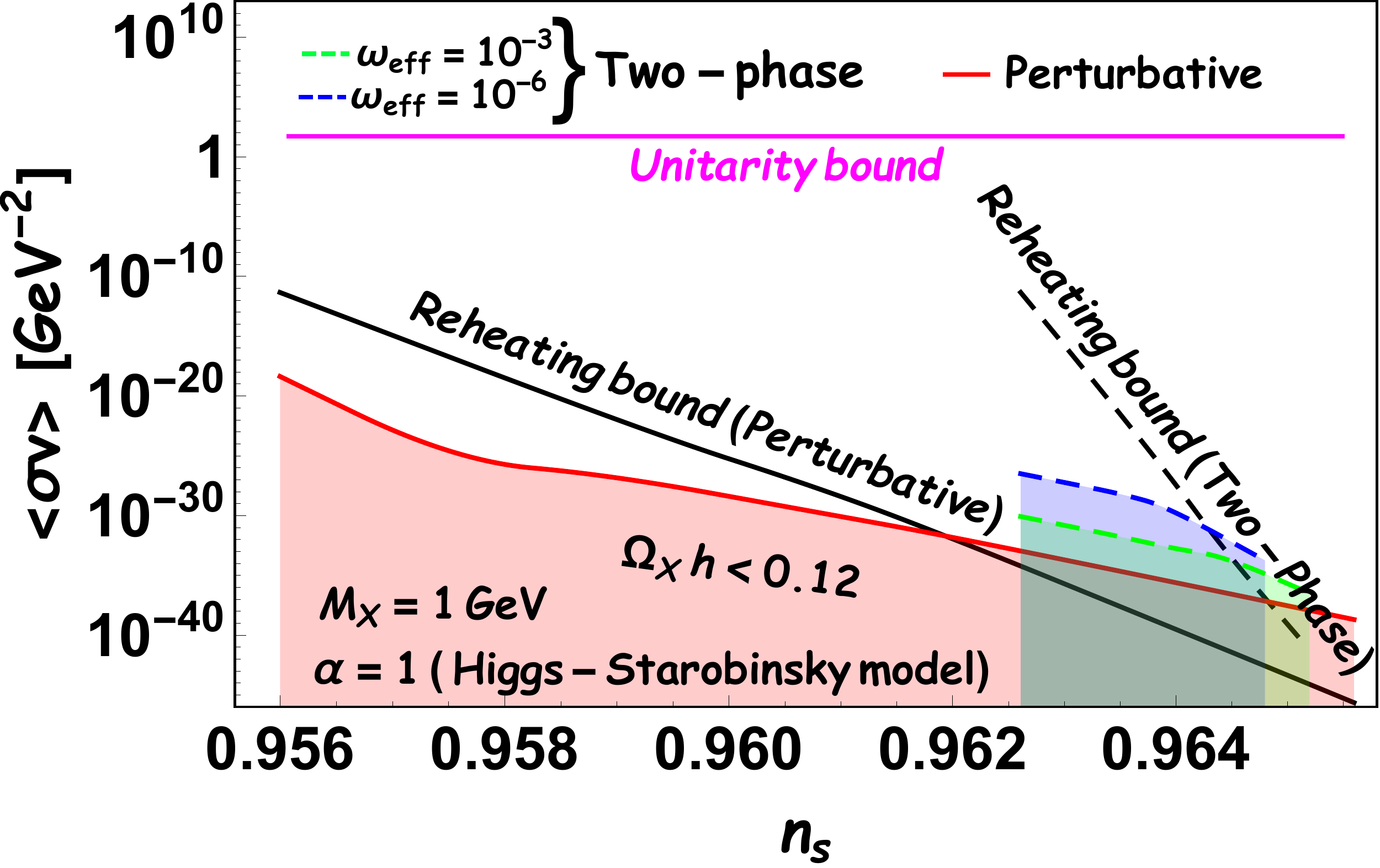}
 		\includegraphics[width=005.41cm,height=03.9cm]{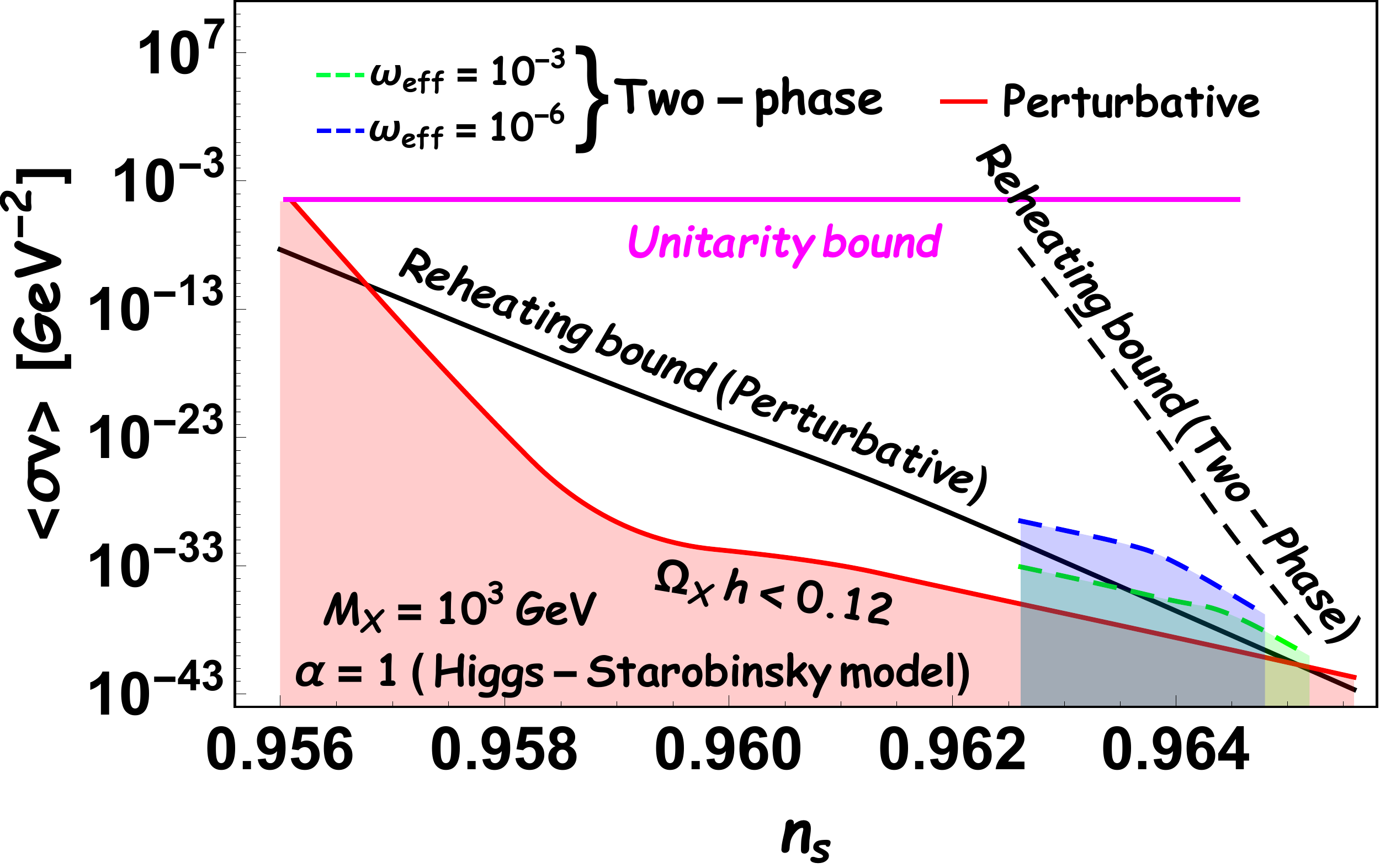}
 \includegraphics[width=005.41cm,height=03.9cm]{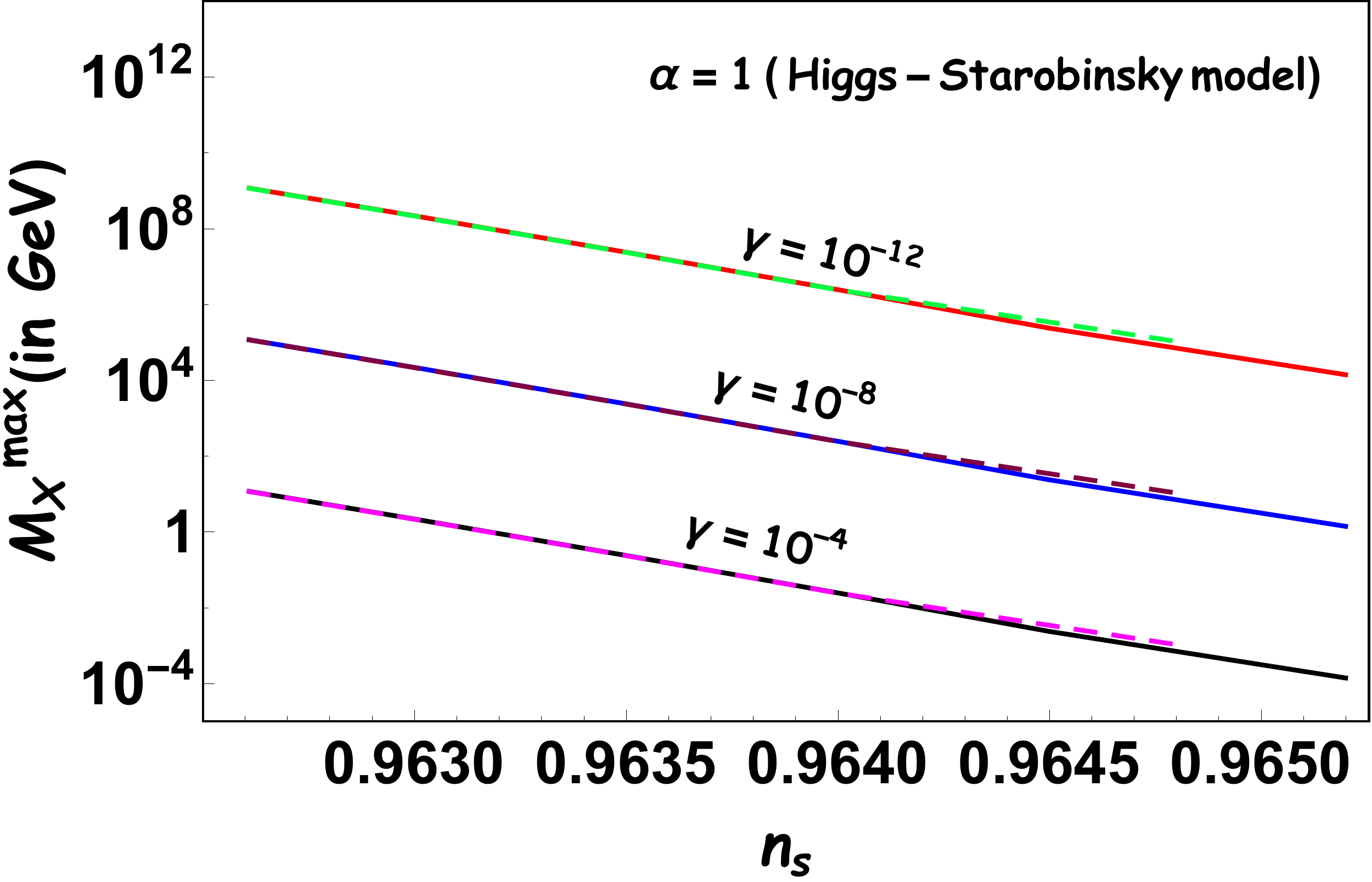}		
 		\caption{\scriptsize
 		 In the first two plots, we have plotted the contour of $\Omega_X h^2=0.12$ in the $n_s-\langle\sigma v\rangle$ plane with a fixed value of dark matter mass within the minimum and maximum values reheating temperature for the Higgs-Satrobinsky model. In the case of two-stage reheating, we have chosen a fixed value of $\gamma=10^{-11}$ (ratio of the dark-matter energy density to the radiation density) during the first stage of reheating. The allowed parameter space is shown by the shaded region below the contour line. The \textcolor{Rhodamine}{\bf{pink}}  horizontal line corresponds to the unitarity bound. The solid and dashed  \textcolor{black}{\bf{black line}}  corresponds to the reheating bound for two different reheating processes. On the right-hand side, we have plotted maximum permitted values of dark matter mass as a function of the spectral index for three different values of $\gamma$. Here the solid and dashed lines are for $\omega_{eff}=(10^{-3},10^{-6})$ respectively.} .
 		\label{alphadarkmatterplot}
 	\end{center}
 \end{figure}
\begin{table}[t!]
\caption{Model parameters and associated constraints on the dark matter parameters for different reheating dynamics:Higgs-Starobinsky model}
$M_X=1$ GeV
\begin{tabular}{|p{2.8cm}|p{2.1cm}|p{2.1cm}|p{2.5cm}|p{2.1cm}|p{2.1cm}|p{2.5cm}|  }

 \hline
 Parameters&\multicolumn{3}{c|}{}&\multicolumn{3}{c|}{Constraints due to \bf{reheating bound}}\\
 \cline{2-7}
 & \multicolumn{2}{c|}{Case-I (Two-phase)}&~Perturbative &\multicolumn{2}{c|}{Case-I (Two-phase)}&~Perturbative \\
\cline{2-7}
&~$\omega_{eff}=10^{-3}$ &~$\omega_{eff}=10^{-6}$&~$\omega_{\phi}=0$& ~$\omega_{eff}=10^{-3}$ &~$\omega_{eff}=10^{-6}$&~$\omega_{\phi}=0$\\
\hline
$n_s^{min}$&~$0.9626$&~$0.9626$&~$0.9560$&~$0.9626$&~$0.9626$&~$0.9560$ \\
$n_s^{max}$&~$0.9652$&~$0.9648$&~$0.9655$&~$0.9645$&~$0.9643$&~$0.9619$ \\
$\langle \sigma v\rangle_{min} \left(\mbox{GeV}^{-2}\right)$&~$3.20\times10^{-37}$&~$1.75\times10^{-34}$&~$2.25\times10^{-39}$&~$1.65\times10^{-34}$&~$7.75\times10^{-32}$ &~$1.66\times10^{-32}$ \\
$\langle \sigma v\rangle_{max} \left(\mbox{GeV}^{-2}\right)$&~$8.15\times10^{-31}$&~$3.10\times10^{-27}$&~$4.20\times10^{-19}$&~$8.15\times10^{-31}$&~$3.10\times10^{-27}$&~$4.20\times10^{-19}$ \\
\hline
\end{tabular}\\[.2cm]
$M_X=10^3$ GeV
\begin{tabular}{|p{2.8cm}|p{2.1cm}|p{2.1cm}|p{2.5cm}|p{2.1cm}|p{2.1cm}|p{2.5cm}|  }

 \hline
 Parameters&\multicolumn{3}{c|}{}&\multicolumn{3}{c|}{Constraints due to \bf{reheating bound}}\\
 \cline{2-7}
 & \multicolumn{2}{c|}{Case-I (Two-phase)}&~Perturbative &\multicolumn{2}{c|}{Case-I (Two-phase)}&~Perturbative \\
\cline{2-7}
&~$\omega_{eff}=10^{-3}$ &~$\omega_{eff}=10^{-6}$&~$\omega_{\phi}=0$& ~$\omega_{eff}=10^{-3}$ &~$\omega_{eff}=10^{-6}$&~$\omega_{\phi}=0$\\
\hline
$n_s^{min}$&~$0.9626$&~$0.9626$&~$0.9561$&~$0.9626$&~$0.9626$&~$0.9568$ \\
$n_s^{max}$&~$0.9652$&~$0.9648$&~$0.9655$&~$0.9652$&~$0.9648$&~$0.9650$ \\
$\langle \sigma v\rangle_{min} \left(\mbox{GeV}^{-2}\right)$&~$9.50\times10^{-41}$&~$1.60\times10^{-37}$&~$2.30\times10^{-42}$&~$9.50\times10^{-41}$&~$1.60\times10^{-37}$&~$1.98\times10^{-41}$ \\
$\langle \sigma v\rangle_{max} \left(\mbox{GeV}^{-2}\right)$&~$8.10\times10^{-34}$&~$3.10\times10^{-30}$&~$2.51\times10^{-5}$&$8.10\times10^{-34}$&~$3.10\times10^{-30}$&~$8.50\times10^{-12}$ \\
 \hline
 \end{tabular} 
 \label{tablehiggsdark}
\end{table}
We have already discussed about the model in the previous section \ref{alpha11}, and the constraints on the reheating parameters $(N_{re}, T_{re})$ in terms of spectral index $(n_s)$. The inclusion of dark matter does not affect much on those parameters. Therefore, the main constraints will be on the thermally averaged cross-section times velocity $(\langle \sigma v\rangle)$, and the dark matter mass $M_X$. 
The first two plots of fig.\ref{alphadarkmatterplot} depicts the variation of annihilation cross-section as a function of the spectral index for two different values of dark-matter mass $M_{X} =(1, 10^3)$ GeV. The range of $n_s$ is taken to be within $(n_s^{min}, n_s^{max})$ depending upon the model of reheating. For comparison, we include  the perturbative reheating scenario \cite{Maity:2018dgy} as well. 
Since the viable range of scalar spectral index $n_s$ is reduced for the two phase reheating than that of the perturbative case, consequently the allowed range of $\langle\sigma v\rangle$ is shrunk as shown by \textcolor{green}{\bf{green dotted}} and \textcolor{blue}{\bf{blue dotted}} lines. Due to larger allowed range of $n_s$ ($n_s^{min} \simeq 0.956, n_s^{max} \simeq  0.9655$), the perturbative reheating \cite{Maity:2018dgy} widens the allowed range of dark matter annihilation cross-section as $2.25 \times 10^{-39}~
\leq\langle \sigma v\rangle\leq4.2 \times 10^{-19}~$ for $M_X=1 ~GeV$ and $2.3 \times 10^{-42}~\leq\langle \sigma v\rangle\leq2.51 \times 10^{-5}~$ for $M_X=10^{3}~$ GeV. Whereas for two phase reheating scenario, for both values of dark matter mass, we can observe the narrower range ($n_s^{min} \simeq 0.9626, n_s^{max} \simeq 0.9652$) for $\omega_{eff} =10^{-3}$ and ($n_s^{min} \simeq 0.9626, n_s^{max} \simeq 0.9648$) for $\omega_{eff}= 10^{-6}$. These ranges of $n_s$ are well within the  $1\sigma$ range of spectral index, $n_s=0.9649\pm0.0042$ (68 \% CL, Planck TT,TE,EE+lowE+lensing) from Planck \cite{Akrami:2018odb}. Detailed constraints on the annihilation cross-section for Higgs inflation model is proved in the table-\ref{tablehiggsdark}.  Therefore, one can observe the significant differences on the allowed range of dark matter annihilation cross-section for two different reheating scenarios (perturbative and two-phase). It is important to note that the dark matter parameter space is constrained by the CMB anisotropy through the inflationary models, or alternatively one can state, {\em how various dark matter experimental observations can have potential to constrain the inflationary model through our unified reheating analysis}.

The inclusion of dark matter dynamics and the associated theoretical constraints discussed in the previous section has put further limits on the range of $n_s$ compatible with the dark matter observation. For example, the perturbative reheating scenario modifies the highest possible value of the spectral index  $n_s^{max}$ as $ 
\rightarrow  0.9619$ , and for two-phase reheating dynamics $n_s^{max}$ shifts as $(0.9652, 0.9648) \rightarrow (0.9645,0.9643)$ with $\omega_{eff} =(10^{-3},10^{-6})$ accordingly for $M_X= 1$ GeV. This modified maximum $n_s$ condition leads to the minimum values of the dark matter cross-section $\langle \sigma v\rangle_{min} \approx 1.66\times 10^{-32} ~\mbox{GeV}^{-2}$ for perturbative case and $\langle \sigma v\rangle_{min}\approx(1.65\times 10^{-34},7.75\times 10^{-32})~ \mbox{GeV}^{-2}$ for two phase reheating case with two different values of $\omega_{eff}=(10^{-3},10^{-6})$. For $M_X=10^3$ GeV instead, the unitary bound put stringent constraints on $n_s^{min}$, only for perturbative process. Further, dynamics during reheating ({\bf reheating bound}) bounds the cross-section within $1.98\times10^{-41}\leq\langle \sigma v\rangle\leq8.5\times10^{-12}$ for perturbative scenario. However, for $M_X=10^3$ GeV, there is no effect of theoretical constraints on the bound of dark matter annihilation cross-section obtained from  two phase reheating analysis. 

From first two plots of fig.\ref{alphadarkmatterplot}, we read the  variation of the cross-section for two different effective equations of state $\omega_{eff}$. As we decrease the value of the $\omega_{eff}$ from $10^{-3} \rightarrow 10^{-6}$, the e-folding number $N_{npre}$, which is nearly independent of inflationary parameter changes from $5.8\to12.2$. Another interesting consequence of the Phase-I dynamics is the maximum possible value of dark matter mass $M_X^{max}$. To understand the underlying reason behind the origin of $M_X^{max}$, we have computed analytic expressions considering relativistic dark matter. The dark matter number density at the point of freeze-out $n_X^f$ (see appendix \ref{relic density}) is expressed as 
\bea\label{darka}
n_X^f x_f^3=n_X^{in}+\langle \sigma v\rangle f(x_f)~~,
 \eea
 where expressions of various symbols are given the appendix. $x_f=A_f/A_{npre}$ and $A_f$ is the normalized scale factor when both comoving dark matter and radiation component become constant. By using the above expression, we can obtain dark matter abundance as
\begin{equation}\label{darkc}
\begin{split}
 \Omega_Xh^2&\simeq \frac{\langle E_X \rangle _f x_f^{-3}}{\rho_R(x_f)}\frac{T(x_f)}{T_{now}}\left(n_X^{in}+\langle \sigma v\rangle f(x_f)\right)\Omega_Rh^2 \\ &=\frac{\sqrt{M_X^2+9T(x_f)^2} x_f^{-3}}{\rho_R(x_f)}\frac{T(x_f)}{T_{now}}\left(n_X^{in}+\langle \sigma v\rangle f(x_f)\right)\Omega_Rh^2~~.
 \end{split}
 \end{equation}
 
The above expression indicates that the dark matter abundance increases with increasing dark matter mass. Moreover, at a particular value of the dark matter mass, the dark matter component's initial number density ($n_X^{in}$) will also play in the final value of the observed dark matter abundance, $\Omega_Xh^2=0.12$. It can be observed from the equation (\ref{darkc}), if $ M_X > M_X^{max}$, then $\Omega_Xh^2$ always $\geq0.12$. Therefore the maximum possible dark matter mass can be obtained from the above equation considering $\Omega_Xh^2 = 0.12$ as
\bea\label{mx1}
M_X^{max}=T(x_f)\sqrt{\left(0.12\frac{\beta}{n_X^{in}} \frac{T_{now}T(x_f)^2}{\Omega_Rh^2 x_f^{-3}}\right)^2-9}~~.
\eea,
which is dependent on the initial dark matter number density for the phase-II evolution,
\bea\label{mx2}
n_X^{in}==\frac{\gamma~ m_\phi}{\langle E_X\rangle}\frac{3\left(\omega_{eff}-\omega_{\phi}\right)}{1-3~\omega_{eff}\left(1+\gamma\right)}\Phi(A_{npre})A_{npre}^{-3}m_\phi^3~~.
\eea
 \begin{figure}[t!]
 	\begin{center}
 \includegraphics[width=005.41cm,height=3.9cm]{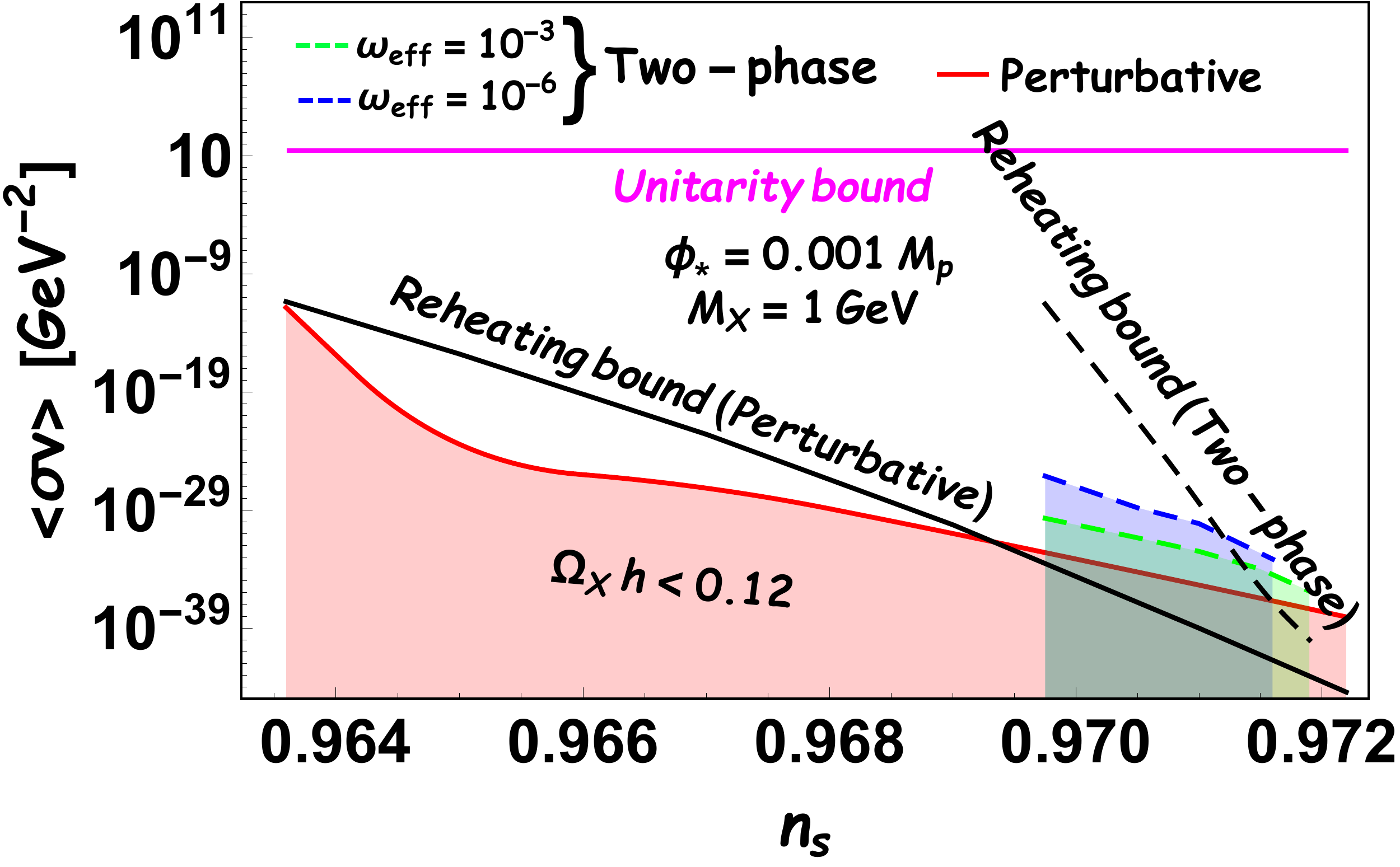}
\includegraphics[width=005.41cm,height=03.9cm]{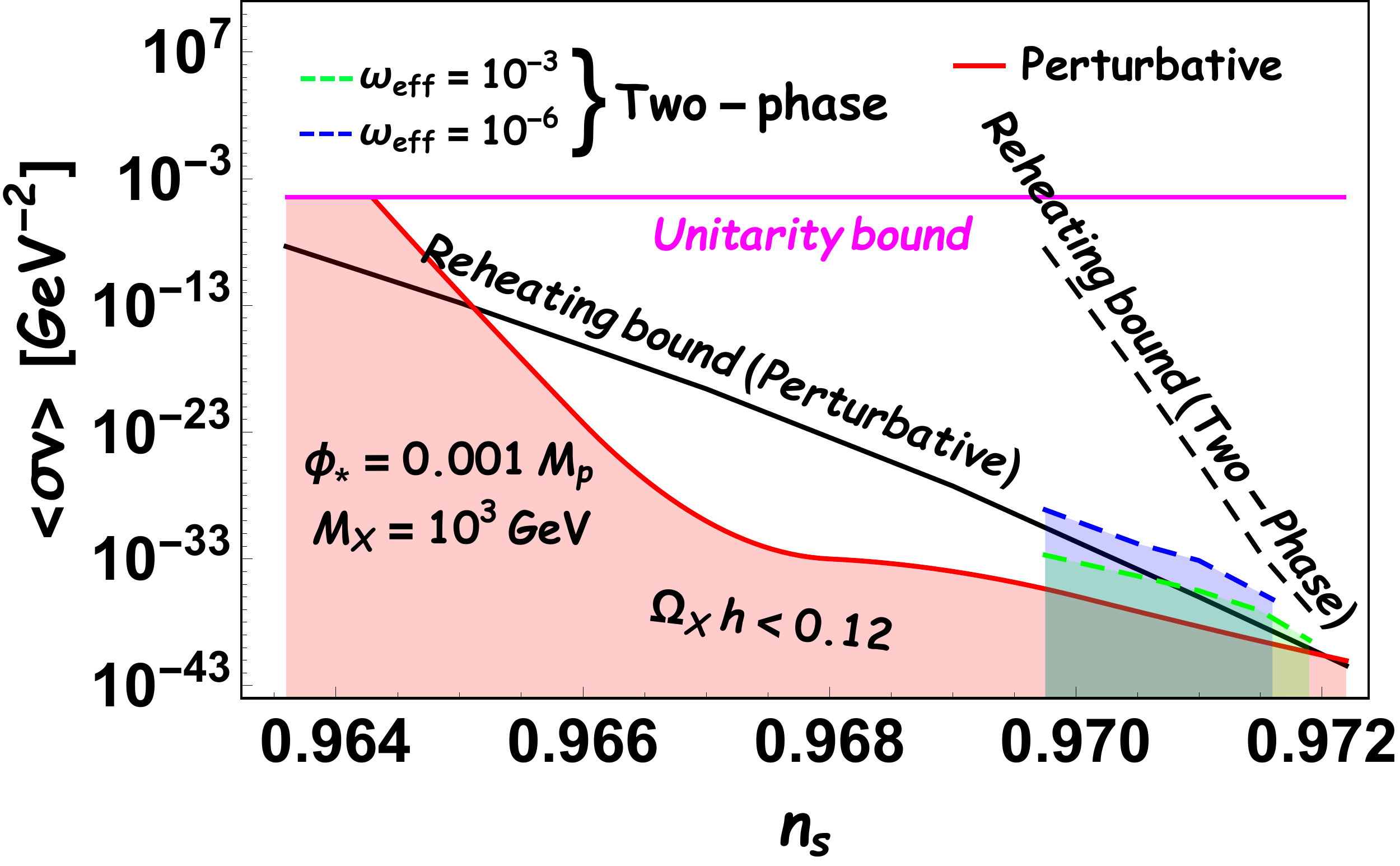}
 \includegraphics[width=005.41cm,height=03.9cm]{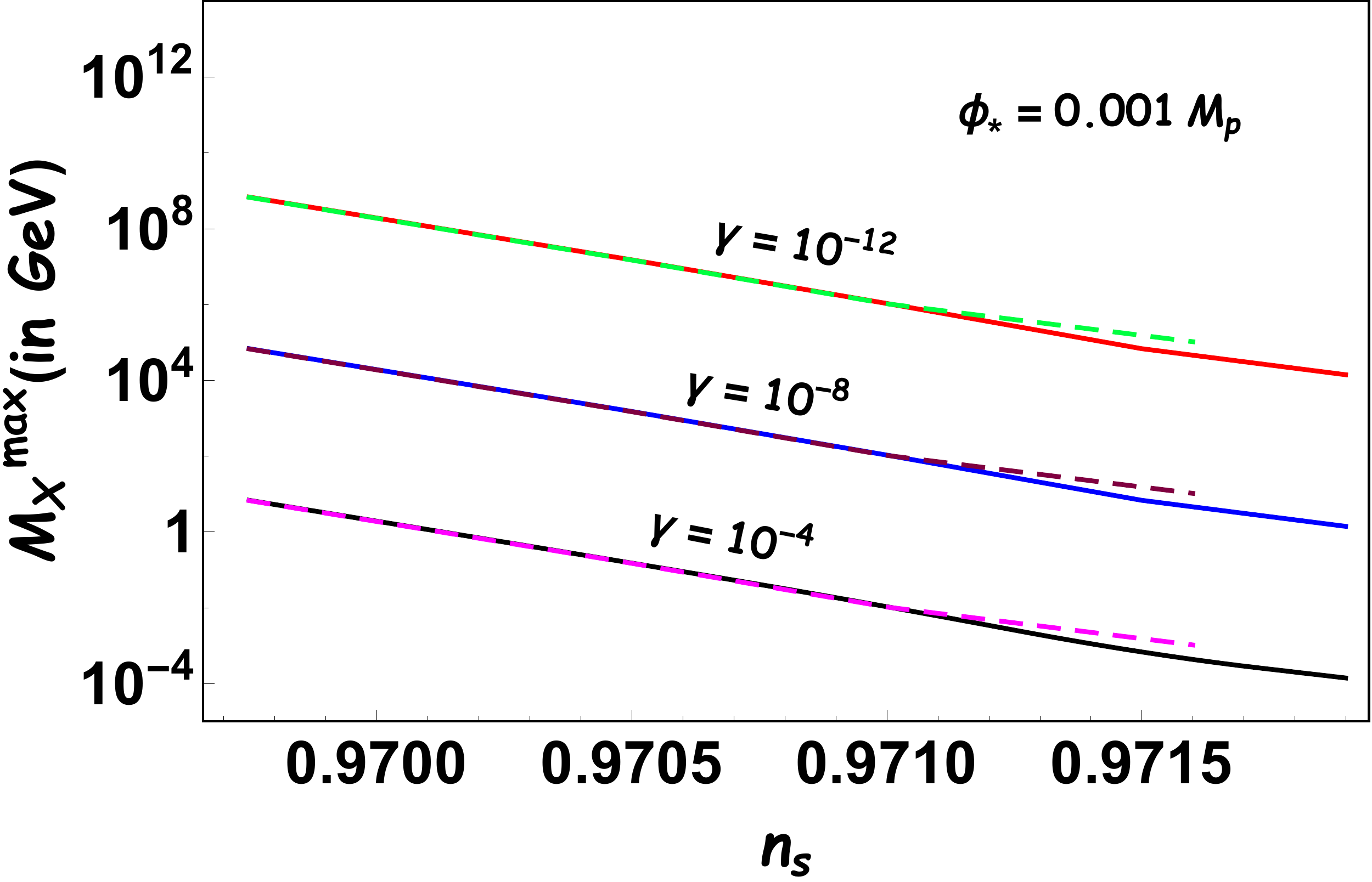}	
 		\caption{\scriptsize All plots are same as in the previous Fig.\ref{alphadarkmatterplot}. The main difference is that, here we have plotted for minimal plateau inflation
model with $\phi_*=0.001 M_p$, $n=2$. }
 		\label{minimaldarkmatterplot}
 	\end{center}
 	\end{figure}
For a given $\gamma$, the initial value of $X$ is clearly set by the value of $\omega_{eff}$. Therefore, for a given value of phase-I dynamics parameters $\omega_{eff}$ and $\gamma$, a particular value of the dark matter mass exits above which present value of the dark matter abundance, $\Omega_Xh^2\approx0.12$ can not be achieved irrespective of the cross-section values. Eqs.(\ref{mx1}) and (\ref{mx2}) illustrate the behavior of $M_X^{max}$, inversely proportional to $ \gamma $ as for a fixed value of the spectral index ($n_s$).  From the third plot of fig.\ref{alphadarkmatterplot}, we can also observe the same. Likewise for $n_s=0.9635$, $M_X^{max}=(2.38\times10^7,2.38\times 10^3,0.238)$ GeV with $\gamma=(10^{-12},10^{-8},10^{-4})$ accordingly once we fixed $\omega_{eff} =10^{-3}$.  Furthermore, from third plot of fig.\ref{alphadarkmatterplot}, one can observe that $M_X^{max}$ is nearly independent of the choice of $\omega_{eff}$ except a small deviation as one approaches towards $n_s^{max}$. The straight forward answer could be that $M_X^{max}$ is proportional to the freeze-out radiation temperature $T(x_f)$, which remains invariant with the choice of $\omega_{eff}$ value.

%
	 
\subsection{Minimal plateau inflation model\cite{Maity:2019ltu}}
\begin{table}[t!]
 	\caption{Model parameters and associated constraints on the dark matter parameters for different reheating dynamics:Minimal plateau model}
 	$\phi_*=0.001 M_p, M_X=1$ GeV
 	\begin{tabular}{|p{2.8cm}|p{2.1cm}|p{2.1cm}|p{2.5cm}|p{2.1cm}|p{2.1cm}|p{2.5cm}|  }
 	
 	\hline
 	Parameters&\multicolumn{3}{c|}{}&\multicolumn{3}{c|}{Constraints from \bf{reheating bound}}\\
 	\cline{2-7}
 	& \multicolumn{2}{c|}{Case-I (Two-phase)}&~Perturbative &\multicolumn{2}{c|}{Case-I (Two-phase)}&~Perturbative \\
 	\cline{2-7}
 	&~$\omega_{eff}=10^{-3}$ &~$\omega_{eff}=10^{-6}$&~$\omega_{\phi}=0$& ~$\omega_{eff}=10^{-3}$ &~$\omega_{eff}=10^{-6}$&~$\omega_{\phi}=0$\\
 	\hline
 	$n_s^{min}$&~$0.96975$&~$0.96975$&~$0.9636$&~$0.96975$&~$0.96975$&~$0.9636$ \\
 	$n_s^{max}$&~$0.9719$&~$0.9716$&~$0.9722$&~$0.97136$&~$0.9712$&~$0.9692$ \\
 	$\langle \sigma v\rangle_{min} \left(\mbox{GeV}^{-2}\right)$&~$1.29\times10^{-36}$&~$6.80\times10^{-34}$&~$8.60\times10^{-39}$&~$3.76\times10^{-34}$&~$7.31\times10^{-32}$ &~$2.89\times10^{-32}$ \\
 	$\langle \sigma v\rangle_{max} \left(\mbox{GeV}^{-2}\right)$&~$1.95\times10^{-30}$&~$7.35\times10^{-27}$&~$1.40\times10^{-12}$&~$1.95\times10^{-30}$&~$7.35\times10^{-27}$&~$1.40\times10^{-12}$ \\
 	\hline
 	\end{tabular}\\[.2cm]
 	$M_X=10^3$ GeV
 	\begin{tabular}{|p{2.8cm}|p{2.1cm}|p{2.1cm}|p{2.5cm}|p{2.1cm}|p{2.1cm}|p{2.5cm}|  }
 	
 	\hline
 	Parameters&\multicolumn{3}{c|}{}&\multicolumn{3}{c|}{Constraints from \bf{reheating bound}}\\
 	\cline{2-7}
 	& \multicolumn{2}{c|}{Case-I (Two-phase)}&~Perturbative &\multicolumn{2}{c|}{Case-I (Two-phase)}&~Perturbative \\
 	\cline{2-7}
 	&~$\omega_{eff}=10^{-3}$ &~$\omega_{eff}=10^{-6}$&~$\omega_{\phi}=0$& ~$\omega_{eff}=10^{-3}$ &~$\omega_{eff}=10^{-6}$&~$\omega_{\phi}=0$\\
 	\hline
 	$n_s^{min}$&~$0.96975$&~$0.96975$&~$0.9643$&~$0.96975$&~$0.96975$&~$0.9651$ \\
 	$n_s^{max}$&~$0.9719$&~$0.9716$&~$0.9722$&~$0.9719$&~$0.9716$&~$0.9720$ \\
 	$\langle \sigma v\rangle_{min} \left(\mbox{GeV}^{-2}\right)$&~$3.80\times10^{-40}$&~$6.30\times10^{-37}$&~$9.10\times10^{-42}$&~$3.80\times10^{-40}$&~$6.30\times10^{-37}$&~$1.82\times10^{-41}$ \\
 	$\langle \sigma v\rangle_{max} \left(\mbox{GeV}^{-2}\right)$&~$1.95\times10^{-33}$&~$7.32\times10^{-30}$&~$2.51\times10^{-5}$&~$1.95\times10^{-33}$&~$7.32\times10^{-30}$&~$7.42\times10^{-14}$ \\
 	\hline
 	\end{tabular}
 	\label{tableminidark}
 	\end{table}	

 The details of this model are discussed in section \ref{minimal}. As was already the case for Higgs's inflation, for the minimal inflation model also the reheating parameters such as $(T_{re}, N_{re})$ will not be modified much because of dark matter dynamics. The reason being, the contribution of dark matter in the background evolution during the reheating phase is insignificant. Throughout this analysis, we consider $\phi_* =0.001M_p$, which satisfies the CMB observation. Another motivation is that as one increases $\phi^*$ value, the models assume simple power law. Details constraints on the dark matter parameter space can be read off from the Fig.\ref{minimaldarkmatterplot}. Furthermore, the numerical values are provided in the table-\ref{tableminidark}.
 From the figure we observed a similar behavior of dark matter annihilation cross-section as a function of the spectral index for two different values of $\omega_{eff}=(10^{-3},10^{-6})$ with dark matter mass $M_X=(1,10^3)~\mbox GeV$. For minimal model we again  identify the maximum allowed values of dark matter mass $M_X^{max}$ followed by the equations (\ref{mx1}) and (\ref{mx2}). Furthermore, for a given value of $n_s$, the $M_X^{max}$ turns out to be linearly varying with $\gamma$ . Like, for $n_s=0.9705$, $M_X^{max}\simeq(1.5\times 10^7,1.5 \times 10^3,0.15)$ GeV with $\gamma=(10^{-12},10^{-8},10^{-4})$.
 

\section{Summary and discussion:}
In this paper, we propose an effective two-phase reheating scenario. After inflation, reheating has been studied extensively in the literature, either through a perturbative or non-perturbative approach. However, it is believed that both approaches independently should not capture the complete picture of the complicated dynamics. In this paper, we, for the first time, study this phase to the best of our knowledge, taking into account both the approaches together motivated by our previous work \cite{Maity:2018qhi}. However, instead of considering explicit non-perturbative decay of the inflaton field through parametric resonance, we model the initial phase by effective dynamics governed by the standard conservation laws and parametrized by a constant effective equation of state ($\omega_{eff}$). The combined form of conservation laws and the initial condition of the reheating dynamics put constraints on the effective equation of state during the effective non-perturbative process calling it as phase-I. However, during perturbative analysis due to explicit decay of the inflaton field into radiation, we obtain the non-trivial time-dependent effective equation of state. At this stage, let us remind the reader that in all the PLANCK analysis \cite{PLANCK2} on constraining the inflationary models $w_{eff}$ is assumed to be a constant free parameter during reheating, which follows from the proposal described in  \cite{martin}. What we argue is that those assumptions should not be correct. After inflation, every inflationary model has its own characteristic oscillatory period, which contributes to the equation of state during reheating.  Therefore, considering $w_{eff}$ as a free parameter loses some of the fundamental characteristic properties of the inflaton potential itself.   Furthermore, if reheating occurs for a longer period of time, the time-dependent $w_{eff}$ during the perturbative process should also be very important to get precise constraints on any inflationary model.  This is where our analysis not only can play an important role in better understanding the inflationary models but also opens up the possibility of understanding the micro-physics of the reheating process through CMB physics. As we can clearly see how the CMB power spectrum constrains the value of inflation-radiation coupling parametrized by $\varGamma_\phi$ through reheating temperature $T_{re}$. The usual connection between $\varGamma_\phi$ and $T_{re}$ will not be correct any more once we consider the decaying inflaton as it is a well-known fact that during the reheating process, even at the end of reheating time, $\varGamma_\phi = H$, inflaton does not decay into radiation completely. Therefore, one certainly needs to take into account this fact while calculating $T_{re}$ and its connection with the scalar power spectrum $n_s$ in the analysis. However, all the previous theoretical as well as in PLANCK analysis, complete decay of inflaton is assumed while relating the cosmological scales exiting and re-entering the horizon at two different time scales. Therefore, based on the two-phase reheating scenario, our prediction of reheating temperature corresponding to the inflationary power-spectrum is more accurate than the previous analysis.\\
At first, we analyzed the viable constraints on the decay width as well as reheating parameters ($N_{re}, T_{re}$) considering the decay of the inflaton field in the perturbative Boltzmann framework. 
Perturbative dynamics have been shown to give rise to a maximum reheating temperature $T_{re}^{max} \simeq 10^{15}$ naturally, which essentially corresponds to almost instantaneous reheating. As long as the decay width is in the perturbative regime, the result from the only perturbative process is trustworthy. However, because of the straightforward relation between $T_{re}$, and $\Gamma_{\phi}$,  high reheating temperature limit can correspond to non-perturbative phenomena. This fact motives us to include non-perturbative aspects of reheating through effective dynamics. In our present scenario, the universe passes through two distinct phases during reheating. Combining the inflation and subsequent standard big-bang evolution with the intermediate two-phase reheating, our approach predicts the critical value of the inflaton decay constant  $\Gamma_{\phi}^{cri}$ depending upon the phase-I equation of state $\omega_{eff}$. 

The critical point naturally defined at  $N_{npre}=N_{pre}$. 
Therefore, if $\Gamma_{\phi} < \Gamma_{\phi}^{cri}$, the reheating phase will be dominated by perturbative one and vice versa. We also compare our numerical results of $\Gamma_{\phi}^{cri}$ with the critical  decay width obtained from the theoretical consideration for different type of inflaton-reheating field interactions $g\phi\chi^2$, $y\phi\chi^3$, $h\phi\psi\bar{\psi}$. It turns out that all the theoretical values of $\Gamma_{cri}(model)$ correspond to an effective phase-I equation of state $\omega_{eff}$ within $10^{-3} \sim 10^{-6}$. Our actual lattice simulation results also appeared to be compatible with this conclusion (see Fig.\ref{compare1}). A summary table-\ref{gammacri} for $\Gamma_\phi^{cri}$ is given for three observationally viable model. 
\begin{table}[t!]
	\caption{Different inflationary models and associated values of $\Gamma_\phi^{cri}$ ($T_{re}^{cri}$), measured in units of GeV}
  \begin{tabular}{|p{3.65cm}|p{1.8cm}|p{1.8cm}|p{1.8cm}|p{1.8cm}|p{2cm}|p{2.15cm}|}
\hline
 &\multicolumn{2}{c|}{$\alpha$-attractor}    &\multicolumn{2}{c|}{Axion} &\multicolumn{2}{c|}{Minimal plateau}\\
\cline{2-7}
 & \quad$\alpha=1$&\quad$\alpha=100$&$f=10M_p$&$f=50M_p$&$\phi_* =0.01M_p$&$\phi _*=0.001M_p$ \\
\hline
$T_{re}^{cri}~\left(\omega_{eff}=10^{-3}\right)$&~$3.5\times10^{10}$&~$7.2\times10^{10}$&~$4.0\times10^{10}$&~$9.0\times10^{10}$&\quad$2.2\times10^{10}$&\quad$8.8\times10^{9}$\\
$T_{re}^{cri}~\left(\omega_{eff}=10^{-6}\right)$&~$2.3\times10^{5}$&~$1.2\times10^{6}$&~$1.8\times10^{5}$&~$1.9\times10^{5}$&\quad$2.8\times10^{5}$&\quad$5.8\times10^{4}$\\
 $\Gamma_\phi^{cri}~\left(\omega_{eff}=10^{-3}\right)$&~$960.0$&~$1.3\times10^{4}$&~$3.7\times10^{4}$&~$2.7\times10^{4}$&\quad$2.3\times10^{3}$&\quad$394.7$\\
 $\Gamma_\phi^{cri}~\left(\omega_{eff}=10^{-6}\right)$&~$1.4\times10^{-7}$&~$3.9\times10^{-6}$&~$1.1\times10^{-7}$&~$1.1\times10^{-7}$&\quad$4.8\times10^{-7}$&\quad$2.7*10^{-8}$\\
 $\Gamma_\phi^{cri}\left(model\right)\left(\phi\to\chi\chi\chi\right)$&~$0.07$&~$0.01$&~$2.8\times10^{-3}$&~$3.2\times10^{-3}$&\quad$0.70$&\quad$15.30$\\
 $\Gamma_\phi^{cri}\left(model\right)\left(\phi\to\chi\chi\right)$&~$5.03$&~$1.60$&~$0.45$&~$0.51$&\quad$34.20$&\quad$231.70$\\
 $\Gamma_\phi^{cri}\left(model\right)\left(\phi\to\bar{\psi}\psi\right)$&~$260.30$&~$42.20$&~$10.80$&~$12.30$&\quad$2.7\times10^3$&\quad$5.8\times10^4$\\
\hline
  \end{tabular}
  \label{gammacri}
\end{table}  


The inclusion of the initial non-perturbative phase naturally changes the maximum reheating temperature value because of its perturbative definition. $T_{re}^{max}$ is no longer defined at the point of instantaneous reheating $N_{re} \simeq 0$, rather is defined at  $N_{re}\approx N_{npre}$, which is equivalent in saying the phase-II e-folding number $N_{pre}\simeq 0$.  At the end of phase-I, approximately $50\%$ of the total comoving energy density remains in the form of the inflaton, which naturally leads to different $T_{re}^{max}$ defined in the perturbative phase-II dynamics. This phase further sets the final equation of the state of the system to $1/3$. All these results have been shown to be crucially dependent upon the phase-I effective equation of state $\omega_{eff}$.
As one changes the value of $\omega_{eff}$ from $10^{-3}\to 10^{-6}$, the phase-I e-folding number $N_{npre}$ changes from $6\to 12$. The maximum reheating temperature $T_{re}^{max}$ accordingly changes from ($10^{13}\to 10^{10}$) GeV. Therefore, the conclusion that can be emphasized upon is that {\it the  value of reheating temperature may encode the information about the non-perturbative phase.} Furthermore, all the inflationary models which are compatible with the observed CMB anisotropy, predict the same maximum reheating temperature ($T_{re}^{max}$) for a given $\omega_{eff}$. This is reminiscent of the maximum reheating temperature $T_{re}^{max}$ obtained for purely perturbative reheating dynamics irrespective of the inflation model. Keeping this point in mind, we have performed a comparative analysis of different existing reheating formalisms such as conventional reheating dynamics (case II) and purely perturbative analysis (case III) with our proposed two-phase (case-I). For both cases II and III, the model-independent maximum value of the reheating temperature turns out to be $T_{re}^{max}\simeq 10^{15}$ GeV, which is reduced to $10^{13} \sim 10^{10}$ GeV when considering two-phase reheating for $\omega_{eff} = 10^{-3} \sim 10^{-6}$. Further, two-phase reheating scenario constraints the inflation model within a very narrow range of allowed scalar spectral index compatible with CMB anisotropy. \\
     Further generalization has been analyzed by including the dark matter component as one of the decay products of the inflaton. Depending upon the mass dark matter annihilation cross-section versus scalar spectral index parameter space has been shown to be reduced because of two-phase reheating as compared to that of standard reheating dynamics, which can be observed from Fig.\ref{alphadarkmatterplot} and \ref{minimaldarkmatterplot}. Details of the allowed parameter space for various models can be obtained from the tables \ref{tablehiggsdark} and \ref{tableminidark}. Because of the non-trivial initial condition for two-phase dynamics, there exists a maximum possible mass $M_X^{max}$ above which dark matter turned out to be overproduced no matter how small the annihilation cross-section is assumed. In the summary table \ref{mmax}, we provide numerical values of maximum possible dark matter mass allowed for different viable models under consideration. summary  As just stated, the value of $M_X^{max}$ is directly connected to $\gamma$ ($\gamma=\frac{\rho_X}{\rho_R}$), which is defined during phase I. Once we fixed the spectral index for a particular inflation model and these values of $M_X^{max}$ is nearly independent of the choice of $\omega_{eff}$.\\
 
 \begin{table}[t!]
	\caption{Models and their associated values of $M_X^{max}$, measured in units of GeV}
	Higgs-Starobinsky model\\[.1cm]
  \begin{tabular}{|p{3cm}|p{2cm}|p{2cm}|p{2cm}|p{2cm}|p{2cm}|p{2cm}|}
\hline
 &\multicolumn{2}{c|}{$\gamma=10^{-12}$} &\multicolumn{2}{c|}{$\gamma=10^{-8}$}    &\multicolumn{2}{c|}{$\gamma=10^{-4}$}\\
\cline{2-7}
 &\multicolumn{1}{c|}{$\omega_{eff}=10^{-3}$}&\multicolumn{1}{c|}{$\omega_{eff}=10^{-6}$}&\multicolumn{1}{c|}{$\omega_{eff}=10^{-3}$}&\multicolumn{1}{c|}{$\omega_{eff}=10^{-6}$}&\multicolumn{1}{c|}{$\omega_{eff}=10^{-3}$}&\multicolumn{1}{c|}{$\omega_{eff}=10^{-6}$} \\
\hline
$M_{X}^{max}~\left(minimum\right)$&\quad$1.4\times10^{4}$&\quad$1.1\times10^{5}$&\quad$1.4$&\quad$10.6$&\quad$1.4\times10^{-4}$&\quad$1.1\times10^{-3}$\\
$M_{X}^{max}~\left(maximum\right)$&\quad$11.7\times10^{8}$&\quad$11.6\times10^{8}$&\quad$11.7\times10^4$&\quad$11.6\times10^{4}$&\quad$11.7$&\quad$11.6$\\
\hline
  \end{tabular}\\[.2cm]
  Minimal plateau model ($\phi_*=0.001M_p$)\\[.1cm]
  \begin{tabular}{|p{3cm}|p{2cm}|p{2cm}|p{2cm}|p{2cm}|p{2cm}|p{2cm}|}
\hline
 &\multicolumn{2}{c|}{$\gamma=10^{-12}$} &\multicolumn{2}{c|}{$\gamma=10^{-8}$}    &\multicolumn{2}{c|}{$\gamma=10^{-4}$}\\
\cline{2-7}
 &\multicolumn{1}{c|}{$\omega_{eff}=10^{-3}$}&\multicolumn{1}{c|}{$\omega_{eff}=10^{-6}$}&\multicolumn{1}{c|}{$\omega_{eff}=10^{-3}$}&\multicolumn{1}{c|}{$\omega_{eff}=10^{-6}$}&\multicolumn{1}{c|}{$\omega_{eff}=10^{-3}$}&\multicolumn{1}{c|}{$\omega_{eff}=10^{-6}$} \\
\hline
$M_{X}^{max}~\left(minimum\right)$&\quad$1.4\times10^{4}$&\quad$1.0\times10^{5}$&\quad$1.4$&\quad$10.4$&\quad$1.4\times10^{-4}$&\quad$1.0\times10^{-3}$\\
$M_{X}^{max}~\left(maximum\right)$&\quad$6.8\times10^{8}$&\quad$6.7\times10^{8}$&\quad$6.8\times10^4$&\quad$6.7\times10^{4}$&\quad$6.8$&\quad$6.7$\\
\hline
  \end{tabular}
  \label{mmax}
\end{table}  
  Nonetheless, one important point we should understand that the existing reheating scenarios, either perturbative or non-perturbative, are not the complete description of this phase. A unified description that connects both non-perturbative and perturbative dynamics is more appropriate. In our present study, we, for the first time, try to construct such a unified description. As a first attempt towards this goal, we describe non-perturbative preheating dynamics by effective dynamics. Our present formalism is particularly suited for the class of inflation models with quadratic potential near its minimum. For inflaton potential with a power greater than two, lattice results generically predict the equation of state $\frac{1}{3}$ after the end of non-perturbative dynamics \cite{Maity:2018qhi}. So for those models, our two-phase reheating is no applicable. We will be considering this case in our future work. Instead of considering an effective non-perturbative approach, actual non-perturbative dynamics integrated with perturbative one would be more appropriate. Recently an interesting approach has been proposed to describe preheating phenomena in the Boltzmann framework \cite{Emond:2018ybc}. In our present two-phase reheating dynamics, the aforementioned non-perturbative Boltzmann framework could be natural to integrate with the perturbative Boltzmann equations. Another important fact we have not considered is the temperature dependency of the effective numbers of relativistic degrees of freedom ($g_*$). Constant effective degrees of freedom is reasonably good approximation for a wide range of temperature \cite{kolb} \cite{gondolo} till  the QCD hadronic transition happens at around $10^2$ MeV scale, around which the vale of effective degrees of freedom changes as $g_* = 100\rightarrow 10$ \cite{Hindmarsh:2005ix}
\cite{Laine:2006cp}.  So our eventual plane in the future to calculate dark matter and reheating parameter space accurately by acknowledging the precise evolution of those degrees of freedom in the thermal bath \cite{Drees:2015exa}-
\cite{olive}.
\acknowledgements
We would like to thank the HEP and Gravity groups at IIT Guwahati for useful discussions. PS would like to thank the cluster computing facility at IIT Madras, where part of the numerical simulations were carried out. The research of PS is partly supported through the Core Research Grant CRG/2018/002200 from Science and Engineering Research Board, Department of Science and Technology, Government of India.

\section{Appendix}
\appendix
\section{Two-phase reheating: Analytic expression of  $T_{max}$ }\label{calculation}
After the end of the effective non-perturbative dynamics, the usual perturbative analysis follows and the  governing Boltzmann equations are
\begin{gather}
\dot{\rho_\phi}+3H(1+\omega_\phi^1)\rho_\phi=-\Gamma_\phi\rho_\phi (1+\omega_\phi^1)\\
\dot{\rho_R}+4H\rho_R=\Gamma_\phi \rho_\phi(1+\omega_\phi^1)+2 \langle E_X\rangle \langle \sigma v\rangle\left(n_X^2-n_{X,eq}^2\right)\\
\dot{n_X}+3Hn_X=-\langle\sigma v\rangle\left(n_X^2-n_{X,eq}^2\right)
\end{gather}
In order to solve analytically, we assume the inflaton energy density to follow the equation,
\bea
\rho_\phi=\rho_\phi^{in}\left(\frac{a}{a_{in}}\right)^{-3(1+\omega_\phi^1)}e^{-\Gamma_\phi(1+\omega_\phi^1) (t-t_i)}\simeq \rho_\phi^{in}\left(\frac{a}{a_{in}}\right)^{-3(1+\omega_\phi^1)}~~.
\eea
Here $\Gamma_\phi$ is the time-independent inflaton decay constant. Notice that the effect of decay constant is being ignored assuming the fact that at the initial stage of perturbative reheating inflaton energy is the dominant one.   $\rho_\phi^i$ and $t_i$ are initial density and initial time during the perturbative era respectively. Using the above equation the radiation energy can be solved as follows,
\begin{equation}
\begin{split}
d\left(\rho_Ra^4\right)&=\left(\Gamma_\phi\rho_\phi (1+\omega_\phi^1) a^4+2\langle E_X\rangle\langle \sigma v\rangle\left(n_X^2-n_{X,eq}^2\right)a^4\right)dt\\
&=\left(\Gamma_\phi\rho_\phi^{in}e^{-\Gamma_\phi (t-t_i)}a_{in}^{3(1+\omega_\phi^1)}a^{1-3\omega_\phi^1}(1+\omega_\phi^1)+2\langle E_X\rangle\langle \sigma v\rangle\left(n_X^2-n_{X,eq}^2\right)a^4\right)dt\\&\simeq\Gamma_\phi\rho_\phi^{in}a_{in}^{3(1+\omega_\phi^1)}a^{-3\omega_\phi^1} \frac{da}{H}+2\langle E_X\rangle\langle \sigma v\rangle\left(n_X^2-n_{X,eq}^2\right)a^3\frac{da}{H}~~.
\end{split}
\label{radan}
\end{equation}
Using the following expression for the Hubble parameter,
\bea
H=\frac{\sqrt{\rho_\phi^{in}\left(\frac{a}{a_{in}}\right)^{-3(1+\omega_\phi^1)}+\rho_R^{in}\left(\frac{a}{a_{in}}\right)^{-4}}}{\sqrt{3}M_p}~~,
\eea
where $\rho_R^{in}$ is the initial radiation density at the beginning of perturbative phase. For the reheating temperature computation we ignore the effect of dark matter whose contribution has been verified to be negligible in our full numerical computation. By solving Eq.\ref{radan} we obtain,
\bea
\begin{split}
\rho_Ra^4 &=\rho_R a_{in}^4+\Gamma_\phi \rho_\phi^{in}a_{in}^3\int\limits_{a_{in}}^a\frac{\left(a/a_{in}\right)^{-3\omega_\phi^1}(1+\omega_\phi^1)da}{\left(\sqrt{3}M_P\right)^{-1}\sqrt{\rho_\phi^{in}\left(\frac{a}{a_{in}}\right)^{-3(1+\omega_\phi^1)}+\rho_R^{in}\left(\frac{a}{a_{in}}\right)^{-4}}}\\\rho_R x^4&=\rho_R^{in}+\Gamma_\phi \rho_\phi^{in}(1+\omega_\phi^1)\int\limits_{1}^x\frac{x^{2-3\omega_\phi^1}dx}{\left(\sqrt{3}M_P\right)^{-1}\sqrt{\rho_\phi^{in}x^{1-3\omega_\phi^1}+\rho_R^{in}}}\\&=\rho_R^{in}+\frac{\Gamma_\phi \rho_\phi^{in}(1+\omega_\phi^1)}{H_{in}}\int\limits_{1}^x\frac{x^{\frac{3-c}{2}}dx}{\sqrt{1+\frac{\rho_R^{in}}{\rho_\phi^{in}}x^{c-1}}}\simeq\rho_R^{in}+\frac{\Gamma_\phi \rho_\phi^{in}(1+\omega_\phi^1)}{H_{in}}\int\limits_{1}^x x^{\frac{3-c}{2}}\left(1-\frac{\rho_R^{in}}{2\rho_\phi^{in}}x^{c-1}\right)dx
\end{split}
\eea
In the above expression, we neglected higher-order terms of ${\rho_R^{in}}/{\rho_\phi^{in}}$. Additionally in terms of radiation temperature  $T_{rad}=\left(\frac{30}{\pi^2 g_*}\rho_R\right)^{1/4}$ the above equation transforms into following expression,
\bea\label{rad}
\frac{\beta T^4 x^4}{\rho_\phi^{in}}=\frac{\Gamma_\phi (1+\omega_\phi^1)}{H_{in}}\left[\frac{2}{5-c}\left(x^{\frac{5-c}{2}}-1\right)+\frac{\rho_R^{in}}{\rho_\phi^{in}}\left(\frac{1-x^{\frac{c+3}{2}}}{c+3}+\frac{H_{in}}{\Gamma_\phi(1+\omega_\phi^1)}\right)\right]~~.
\eea
Here $x$, $\beta$, $c$ and $H_{in}$ defined as
\bea
x=\frac{a}{a_{in}}~,~\beta=\frac{\pi^2 g_*(T)}{30}~,~c=3\omega_\phi^1~,~H_{in}=\frac{\sqrt{\rho_\phi^{in}}}{\sqrt{3}M_P}~~.
\eea
The maximum radiation temperature can be found by taking derivative of the above equation (\ref{rad}) with respect to $x$ and set it to zero
\bea
\frac{4\beta T^3}{\rho_\phi^{in}}\frac{dT}{dx}=\frac{-4\Gamma_\phi(1+\omega_\phi^1)}{H_{in}x^5}\left[\frac{\frac{3+c}{4}x^{\frac{5-c}{2}}-2}{5-c}+\frac{\rho_R^{in}}{\rho_\phi^{in}}\left(\frac{\frac{c-5}{8}x^{\frac{c+3}{2}}+1}{c+3}+\frac{H_{in}}{\Gamma_\phi(1+\omega_\phi^1)}\right)\right]=0~~.
\eea
In the limit of ${\rho_\phi}/{\rho_R}\ll 1$ (perturbative approximation),  the values of $x$ at the point of maximum radiation temperature appears as
\bea
x_{max,p}=\left(\frac{8}{3+c}\right)^{\frac{2}{5-c}}~~.
\eea
In our present analysis, the expression of $x$ associated with maximum radiation temperature leads to the following relation
\bea
x_{max}\simeq\left(\frac{8}{3+c}\right)^{\frac{2}{5-c}}\left[1-\frac{\rho_R^{in}}{\rho_\phi^{in}}\left(\frac{c-5}{8(c+3)}x_{max,p}^{\frac{c+3}{2}}+\frac{1}{c+3}+\frac{H_{in}}{\Gamma_\phi(1+\omega_\phi^1)}\right)\right]=x_{max,p}\left[1-z\right]~~,
\eea
where $z=\frac{\rho_R^{in}}{\rho_\phi^{in}}\left(\frac{c-5}{8(c+3)}x_{max,p}^{\frac{c+3}{2}}+\frac{1}{c+3}+\frac{H_{in}}{\Gamma_\phi(1+\omega_\phi^1)}\right)$. Now after replacing the expression of $x_{max}$ into the above Eq.\ref{rad}, the maximum radiation temperature turns out as
\bea\label{radiationtemp}
T_{max}&\simeq \left(\frac{\Gamma_\phi(1+\omega_\phi^1)\rho_\phi^{in}}{\beta H_{in}x_{max,p}^4}\frac{2}{3+c}\right)^{1/4}\left[1+\frac{3+c}{2}\frac{\rho_R^{in}}{\rho_\phi^{in}}\left(\frac{1-x_{max,p}^{\frac{c+3}{2}}}{c+3}+\frac{H_{in}}{\Gamma_\phi(1+\omega_\phi^1)}\right)\right]^{1/4}\\
&\simeq\left(\frac{\Gamma_\phi(1+\omega_\phi^1)\rho_\phi^{in}}{\beta H_{in}x_{max,p}^4}\frac{2}{3+c}\right)^{1/4}\left[1+\frac{3+c}{8}\frac{\rho_R^{in}}{\rho_\phi^{in}}\left(\frac{1-x_{max,p}^{\frac{c+3}{2}}}{c+3}+\frac{H_{in}}{\Gamma_\phi(1+\omega_\phi^1)}\right)\right]~~.
\eea
In the above expression we have neglected higher order terms of ${\rho_R^{in}}/{\rho_\phi^{in}}$. Next, we will try to express all initial densities in terms of the inflaton energy density at the end of the inflation $\rho_\phi^{end}$. The effective non-perturbative phase-I dynamics solves the radiation and inflaton energy density in terms of $\rho_\phi^{end}$. Therefore, during phase I the dimensionless radiation energy  density $R^I(A)$ can be correlate with inflaton energy density $\Phi^I(A)$ (using (\ref{nonper}), (\ref{nonperturbative}) and (\ref{scale})) as
\bea\label{initial1}
R^I(A)=\frac{3\omega_{eff}}{\left(1-3\omega_{eff}\right)}\Phi^I(A)~A~~.
\eea
 The initial densities during phase II (perturbative era) in terms of dimensionless comoving energy densities are identified as
\bea\label{initial2}
\rho_\phi^{in}=\Phi(A_{npre})A_{npre}^{-3(1+\omega_\phi^1)}m_\phi^4~,~\rho_R^{in}=R(A_{npre})A_{npre}^{-4}m_\phi^4~~.
\eea
Furthermore,  we can relate the $\Phi(A_{npre})$ in terms of $\Phi(A=1)$ as,
\bea\label{initial3}
\Phi(A_{npre})=\left(1-3\omega_{eff}\right)\Phi(A=1)A_{npre}^{-3\omega_{eff}}~~,
\eea
where, $A_{npre}$ is the normalized scale factor at the end of the effective dynamics. $A_{npre}$ is defined when the dimensionless comoving radiation energy density becomes $50\%$ of the total comoving energy density, $\frac{R(A_{npre})}{\Phi(A_{npre})+R(A_{npre})}\simeq\frac{1}{2}\implies\Phi(A_{npre})\simeq R(A_{npre})$. Using Eq.\ref{initial1} one can find $A_{npre}$ and corresponding e-folding number $N_{npre}$ as
\bea\label{initial4}
A_{npre}=\frac{1-3\omega_{eff}}{3\omega_{eff}}~,~N_{npre}=\ln\left(A_{npre}\right)~~.
\eea
From our analytic expression above, we obtain $N_{npre} \sim (5.8, 12.7)$ for two values of $\omega_{eff}=(10^{-3},10^{-6})$ accordingly. These values of the e-folding number during phase I almost exactly match with our numerical result.\\
The final expression for the maximum radiation temperature in terms of comoving energy densities is given by
\bea\label{initial5}
T_{max}\simeq D^{1/4}\left[1+\frac{(3+c)R(A_{npre})}{8\Phi(A_{npre})A_{npre}^{1-c}}\left(\frac{1-x_{max,p}^{\frac{c+3}{2}}}{c+3}+\frac{\sqrt{\Phi(A_{npre})A_{npre}^{-3(1+\omega_\phi^1)}m_\phi^4}}{\sqrt{3}M_p\Gamma_\phi(1+\omega_\phi^1)}\right)\right]~~,
\eea
where
\bea\label{initial6}
D=\left(\frac{2\Gamma_\phi(1+\omega_\phi^1)\sqrt{3M_p^2\Phi(A_{npre})A_{npre}^{-3(1+\omega_\phi^1)}m_\phi^4}}{(3+c)\beta x_{max,p}^4}\right)^{1/4}~~.
\eea
Combining equations from (\ref{initial1}) to (\ref{initial6}), we obtain maximum radiation temperature as a function of $\Phi(A=1)$ (dimensionless comoving inflaton energy density at the end of the inflation).\\
\section{Two phase reheting: analytic expression of inflaton decay width $\Gamma_\phi$ and $T_{re}$}\label{calculation1}
\begin{figure}[t!]
 	\begin{center}
 			
 \includegraphics[width=008.1cm,height=6.3cm]{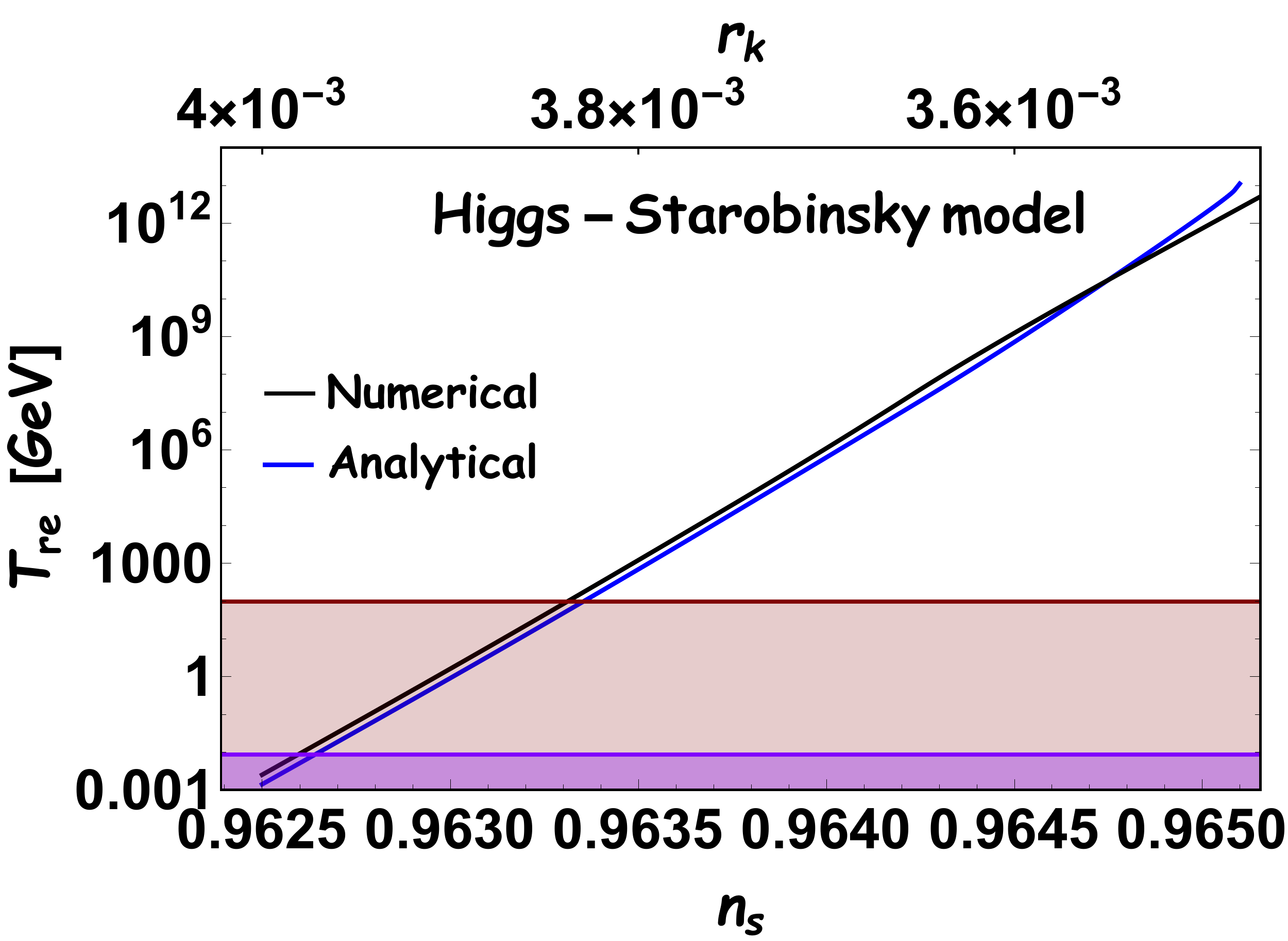}		
 \includegraphics[width=008.1cm,height=6.3cm]{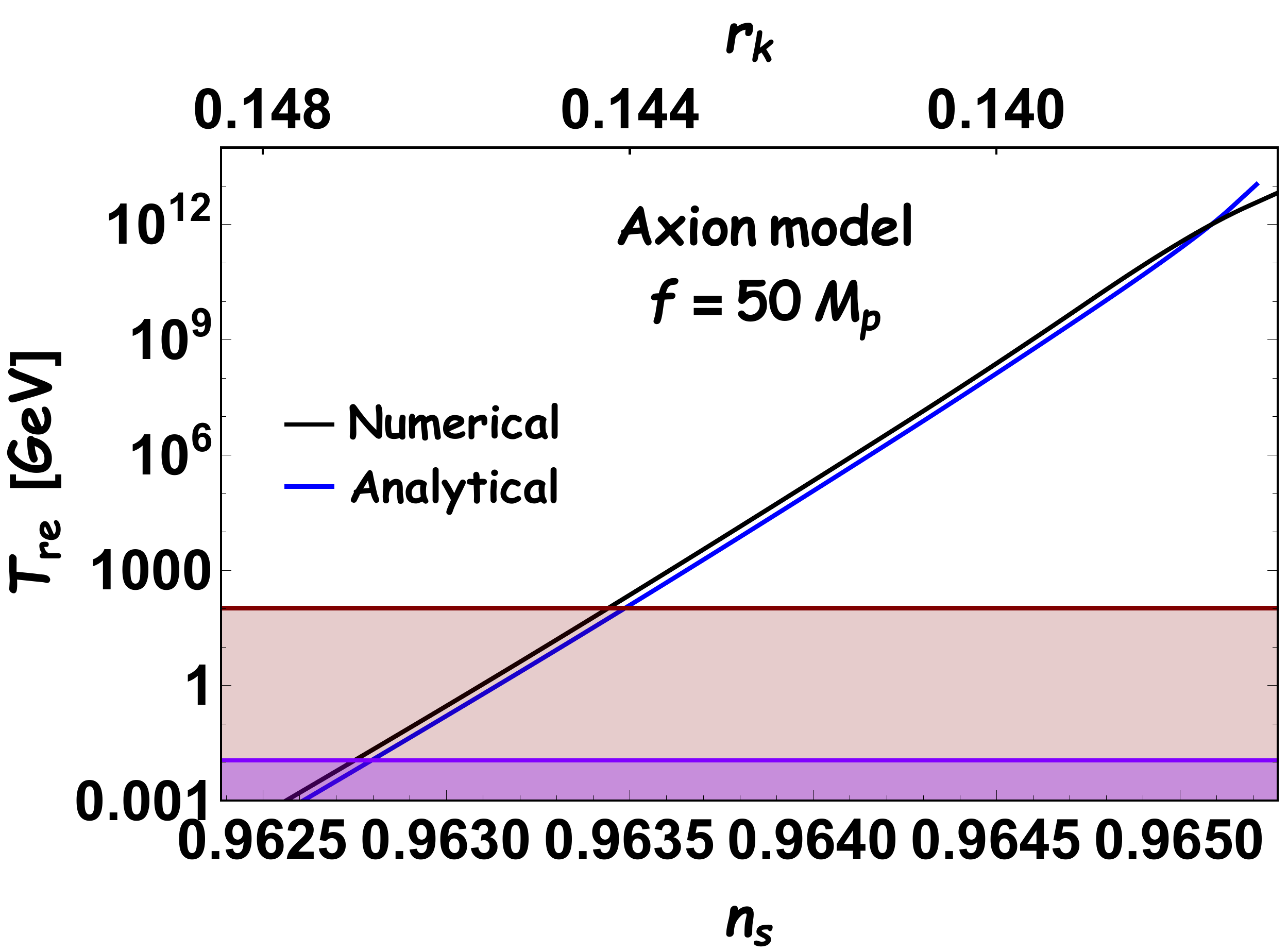}	
 		\caption{\scriptsize Variation of reheating temperature ($T_{re}$) as a function of $n_s$ for Higgs-Starobinsky and axion inflation model with $\omega_{eff}=10^{-3}$ in the framework of two-phase analysis. The \textcolor{blue}{\bf{solid blue line}} indicates the result from approximate analytical expression (equation \ref{reheating10}) whereas, the \textcolor{black}{\bf{solid black line}} shows results from numerical analysis. The \textcolor{brown}{\bf{light brown}} region is below the electroweak scale $T_{ew}\backsim 100 ~GeV$, and the \textcolor{violet}{\bf{violet}} region below $10^{-2} ~GeV$ would ruin the predictions of big bang nucleosynthesis (BBN).}. 
 		\label{compare}
 	\end{center}
 \end{figure}
Assuming the end point of reheating as $x_{re}={a_{re}}/{a_{npre}}$, and considering equation (\ref{rad}), reheating temperature can be obtained as
\bea\label{reheating10}
T_{re}^4 =\frac{\Gamma_\phi \rho_\phi^{in}(1+\omega_\phi^1)x_{re}^{-4}}{\beta H_{in}}\left[\frac{2}{5-c}\left(x_{re}^{\frac{5-c}{2}}-1\right)+\frac{\rho_R^{in}}{\rho_\phi^{in}}\left(\frac{1-x_{re}^{\frac{c+3}{2}}}{c+3}+\frac{H_{in}}{\Gamma_\phi(1+\omega_\phi^1)}\right)\right]~~.
\eea
Using Eq.\ref{reheating 3} (entropy conservation of thermal radiation), one arrives at the following relation
\bea\label{fixgamma2}
T_{re}^4 = \left (\frac{43}{11 g_{re}}\right)^{4/3} \left(\frac{a_0T_0}{k}\right)^4 H_k^4 e^{-4N_k} e^{-4N_{npre}}e^{-4 N_{pre}}=G^4\left(\frac{a_{re}}{a_{npre}}\right)^{-4}=G^4 x_{re}^{-4}~~,
\eea
where
\bea
G=\left (\frac{43}{11 g_{re}}\right)^{1/3} \left(\frac{a_0T_0}{k}\right) H_k e^{-N_k} e^{-N_{npre}}~~.
\eea
Comparing equation (\ref{reheating10}) and (\ref{fixgamma2}), we obtain $\Gamma_\phi$ in terms of $x_{re}$
\begin{equation}\label{decay1}
\begin{split}
\Gamma_\phi&=\left(\frac{G^4\beta}{\rho_\phi^{in}}-\frac{\rho_R^{in}}{\rho_\phi^{in}}\right)\frac{H_{in}}{(1+\omega_\phi^1)}\left[\frac{2}{5-c}\left(x_{re}^{\frac{5-c}{2}}-1\right)+\frac{\rho_R^{in}}{\rho_\phi^{in}}\left(\frac{1-x_{re}^{\frac{c+3}{2}}}{c+3}\right)\right]^{-1},\\
&\simeq\left(\frac{G^4\beta}{\rho_\phi^{in}}-\frac{\rho_R^{in}}{\rho_\phi^{in}}\right)\frac{H_{in}}{(1+\omega_\phi^1)}\frac{5-c}{2}x_{re}^{\frac{c-5}{2}}\left[1+\frac{5-c}{2\left(c+3\right)}\frac{\rho_R^{in}}{\rho_\phi^{in}}x_{re}^{c-1}\right].
\end{split}
\end{equation}
Reheating temperature is defined when the inflaton field comes in thermal equilibrium with the radiation bath at the point, 
\bea\label{fixgamma4}
H(x_{re})^2=\frac{\rho_\phi(x_{re})+\rho_R(x_{re})}{3M_p^2}\simeq\frac{\rho_R(x_{re})}{3M_p^2}=\Gamma_\phi^2~~.
\eea
In the above equation, we ignore the contribution of inflaton energy density to be negligible. Using the expression for the radiation energy density we can obtain the decay width as follows,
\bea
\Gamma_\phi^2\simeq\left(\frac{G^4\beta}{\rho_\phi^{in}}-\frac{\rho_R^{in}}{\rho_\phi^{in}}\right)^2\frac{H_{in}^2}{(1+\omega_\phi^1)^2}\left(\frac{5-c}{2}\right)^2 x_{re}^{c-5}\left[1+\frac{5-c}{c+3}~\frac{\rho_R^{in}}{\rho_\phi^{in}}~x_{re}^{c-1}\right]~~.
\eea
In the earlier expression, we can ignore the second term in the third bracket since $x_{re}\gg1$ for most of the values of the spectral index.  As a result, the $\Gamma_\phi^2$ can now be written as
\bea\label{fixgamma7}
\Gamma_\phi^2\simeq\left(\frac{G^4\beta}{\rho_\phi^{in}}-\frac{\rho_R^{in}}{\rho_\phi^{in}}\right)^2\frac{H_{in}^2}{(1+\omega_\phi^1)^2}\left(\frac{5-c}{2}\right)^2 x_{re}^{c-5}~~.
\eea
Furthermore, the radiation energy density at the ending point of reheating era $\rho_R(x_{re})$ can be expressed as
\bea \label{fixgamma5}
\rho_R(x_{re})\simeq \frac{x_{re}^4\rho_\phi^{in}}{3M_p^2}\left[\frac{2}{5-c}\frac{\Gamma_\phi\left(1+\omega_\phi^1\right)}{H_{in}}x_{re}^{\frac{5-c}{2}}+\frac{\rho_R^{in}}{\rho_\phi^{in}}\left(1-\frac{\Gamma_\phi\left(1+\omega_\phi^1\right)}{H_{in}}\frac{x_{re}^{\frac{c+3}{2}}}{c+3}\right)\right]
\eea
Combining equations (\ref{decay1}) and (\ref{fixgamma5}) one can find
\bea\label{fixgamma6}
\rho_R(x_{re})=\beta T_{re}^4\simeq \frac{x_{re}^4\rho_\phi^{in}}{3M_p^2}\left[\frac{G^4\beta}{\rho_\phi^{in}}+\frac{5-c}{2\left(c+3\right)}\frac{\rho_R^{in}}{\rho_\phi^{in}}\left(\frac{G^4\beta}{\rho_\phi^{in}}-\frac{\rho_R^{in}}{\rho_\phi^{in}}\right)x_{re}^{c-1}\right]~~.
\eea
Now equating this above equation with $\Gamma_\phi^2$ (eqn \ref{fixgamma7}), one arrives at the following expression
\bea\label{are}
x_{re}=\left(\frac{\alpha}{\eta}\right)^{\frac{1}{c-1}}~~,
\eea
Here 
\bea\label{are1}
\alpha=\frac{G^4\beta}{\rho_\phi^{in}}~,~\eta=\frac{5-c}{2}\left(\frac{G^4\beta}{\rho_\phi^{in}}-\frac{\rho_R^{in}}{\rho_\phi^{in}}\right)\left[\frac{\rho_R^{in}}{\left(c+3\right)\rho_\phi^{in}}+\frac{5-c}{2}\frac{3M_p^2H_{in}^2}{\rho_\phi^{in}\left(1+\omega_\phi^{1}\right)^2}\left(\frac{G^4\beta}{\rho_\phi^{in}}-\frac{\rho_R^{in}}{\rho_\phi^{in}}\right)\right]~.
\eea
By utilizing the above equation, we can easily fix decay width (eqn \ref{decay1}) and reheating temperature (eqn \ref{fixgamma6}) as they are the function of $x_{re}$. Besides, the maximum reheating temperature and associated maximum possible value of the spectral index ($n_s^{max}$) can also be defined at the point $x_{re}\to1$ ($N_{pre}\to0$). To check whether our analytical calculations predict the correct result, we plot reheating temperature as a function of the spectral index (fig.\ref{compare}) and compare with our numerical result.

\section{Two phaser reheating: Analytical expression of dark matter abundance and origin of maximum dark matter mass $M_X^{max}$}\label{relic density}
The relevant Boltzmann equation for the evolution of dark matter  during perturbative reheating phase is expressed as 
\bea
d(n_X a^3)=-a^3\langle \sigma v\rangle\left[n_X^2-n_{X,eq}^2\right]dt=-\frac{a^3\langle \sigma v\rangle\left[n_X^2-n_{X,eq}^2\right]da}{aH}~~.
\eea
Through out our calculation we assume dark matter particles are always relativistic and never attain the chemical equilibrium ($n_X \ll n_{X,eq}$) with the radiation bath. Hence in this freeze-in scenario, the dark matter density always remains sub-dominant compared to its thermal equilibrium values. Consequently above dark matter evolution equation can be approximated as,
\bea\label{dark1}
d(n_X a^3)=\frac{a^3\langle \sigma v\rangle n_{X,eq}^2}{aH}da~~.
\eea
In the relativistic limit, the equilibrium distribution is given by
\bea\label{dark2}
n_{X,eq}=\frac{gT^3}{\pi^2}~~,
\eea
where '$g$' is the number of degrees of freedom of the dark matter species.
Furthermore, considering the relativistic dark matter, the dark matter's mass must be less than the reheating temperature. So the freeze-in happens very late after the reheating. Therefore, we can approximate the Hubble parameter as
\bea\label{modhub}
H(a)=\sqrt{\frac{\rho_\phi(a)+\rho_R(a)}{3M_p^2}}\simeq \sqrt{\frac{\rho_R}{3M_p^2}}~~.
\eea
 Connecting equations (\ref{dark1}), (\ref{dark2}), and (\ref{modhub}) one can solve for the dark matter component as
\bea\label{dark3}
\begin{split}
n_X^fa_f^3 &=n_X^{in} a_{in}^3+\int\limits_{a_{in}}^{a_f}\frac{a^2 \langle \sigma v\rangle \frac{g^2}{\pi^4}\beta^{-3/2}\rho_R}{\left(\sqrt{3}M_P\right)^{-1}} da~~,\\n_X^f x_f^3 &\simeq n_X^{in}+\int\limits_{1}^{x_f}\frac{ \langle \sigma v\rangle g^2\sqrt{3}M_P}{\pi^4\beta^{3/2} }\rho_\phi^{in}\left[\frac{2\Gamma_\phi(1+\omega_\phi^1)}{H_{in}(5-c)}x^{\frac{1-c}{2}}+\frac{\rho_R^{in}}{\rho_\phi^{in}}\left(x^{-2}-\frac{\Gamma_\phi(1+\omega_\phi^1)}{H_{in}(c+3)}x^{\frac{c-1}{2}}\right)\right]da~~,
\end{split}
\eea
where $x_f=\frac{a_f}{a_{in}}$ and
\bea\label{dark4}
\rho_R \simeq \rho_\phi^{in}x^{-4}\left[\frac{2\Gamma_\phi(1+\omega_\phi^1)}{H_{in}(5-c)}x^{\frac{5-c}{2}}+\frac{\rho_R^{in}}{\rho_\phi^{in}}\left(1-\frac{\Gamma_\phi(1+\omega_\phi^1)}{H_{in}(c+3)}x^{\frac{c+3}{2}}\right)\right]
\eea
The  scale factor at the point of freeze-in defined as $a_f$, when both comoving dark matter and radiation component become constant. In the preceding expression, we ignore higher-order terms of ${\rho_R^{in}}/{\rho_\phi^{in}}$. With these assumptions the comoving number density $n_X^f$ is found to be
\bea \label{dark7}
n_X^f x_f^3 \simeq n_X^{in}+\langle \sigma v\rangle f(x_f)~~,
\eea
where $f(x_f)$ can be expressed as
\bea \label{dark8}
f(x_f)\simeq \rho_\phi^{in}\left[\frac{4\Gamma_\phi(1+\omega_\phi^1)}{H_{in}(5-c)(3-c)}x_f^{\frac{3-c}{2}}+\frac{\rho_R^{in}}{\rho_\phi^{in}}\left(1-\frac{2\Gamma_\phi(1+\omega_\phi^1)}{H_{in}(c+3)(c+1)}x_f^{\frac{c+1}{2}}\right)\right]~~.
\eea
The dark matter relic can be obtained in terms of radiation abundance $\Omega_R$ ($\Omega_Rh^2=4.3\times10^{-5}$) as
\bea\label{dark5}
\Omega_Xh^2=\frac{\rho_X(x_f)}{\rho_R(x_f)}\frac{T(x_f)}{T_{now}}\Omega_Rh^2=\frac{\langle E_X \rangle _f x_f^{-3}n_X^f(x_f)x_f^3}{\rho_R(x_f)}\frac{T(x_f)}{T_{now}}\Omega_Rh^2=0.12~~.
\eea
Inserting expression of $n_X^f x_f^3$ (equation (\ref{dark7})) into the above equation, one can arrive at the following equation for the dark matter abundance,
\bea\label{dark6}
 \Omega_Xh^2\simeq \frac{\langle E_X \rangle _f x_f^{-3}}{\rho_R(x_f)}\frac{T(x_f)}{T_{now}}\left(n_X^{in}+\langle \sigma v\rangle f(x_f)\right)\Omega_Rh^2~~,
 \eea
The average energy of the single component dark matter at the point of freeze-in can be expressed as 
\bea\label{dark9}
\langle E_X\rangle_f \simeq \sqrt{M_X^2+9T(x_f)^2}\simeq 3T(x_f)\left(1+\frac{M_X^2}{18 T(x_f)^2}\right)~~(\mbox{relativistic ~~approximation})
\eea
Therefore, Connecting the above two equations (\ref{dark6}) and (\ref{dark9}), one arrives at the following expression 
\bea\label{abundance}
 \Omega_Xh^2\simeq \frac{3 x_f^{-3}\rho_R(x_f)^{-1/2}(1+\frac{M_X^2\beta^{1/2}\rho_R(x_f)^{-1/2}}{18})}{\beta^{1/2}T_{now}}\left(n_X^{in}+\langle \sigma v\rangle f(x_f)\right)\Omega_Rh^2~~,
\eea
\subsection{Maximum possible dark matter mass ($M_X^{max}$)}
The approximate analytical expression of dark matter abundance (equation (\ref{abundance})) indicates that the dark matter abundance increases with increasing dark matter mass. Moreover, at a particular value of the dark matter mass, the dark matter component's initial number density ($n_X^{in}$) is sufficient to produce the present observed value of the dark matter abundance $\Omega_Xh^2=0.12$.  We define this particular value of the dark matter mass as $M_X^{max}$. We can clearly see from equation (\ref{abundance}), if the mass of the dark matter $M_X > M_X^{max}$, the abundance $\Omega_Xh^2$ always $\geq0.12$. Therefore the condition for the maximum possible dark matter mass can be written as,
\bea
\Omega_Xh^2\simeq \frac{\sqrt{M_X^2+9T(x_f)^2} x_f^{-3}}{\rho_R(x_f)}\frac{T(x_f)n_X^{in}}{T_{now}}\Omega_Rh^2=0.12~~.
\eea
The outcome of this equation is the maximum possible mass, $M_X^{max}$, which is determined to be
\bea
M_X^{max}=T(x_f)\sqrt{\left(0.12\frac{\beta}{n_X^{in}} \frac{T_{now}T(x_f)^2}{\Omega_Rh^2 x_f^{-3}}\right)^2-9}~~.
\eea

  \hspace{0.5cm}

\end{document}